\documentclass{JFM-FLM_Au}
\usepackage{float}

\usepackage[export]{adjustbox}
\usepackage{subcaption}
\usepackage{comment}
\usepackage{svg}
\usepackage{graphicx}
\usepackage{subcaption}
\usepackage{tikz}
\usepackage{pgfplots}
\usepackage[percent]{overpic}
\usepackage{cleveref}
\usepackage{xcolor}
\usepackage{soul}

\newcommand{\tcr}[1]{\textcolor{red}{#1}}
\newcommand{\tcb}[1]{\textcolor{blue}{#1}}

\lefttitle{A. N. Lagwankar, A. Chiarini, F. Gallaire and E. Boujo}
\righttitle{Journal of Fluid Mechanics}

\title{Imperfect bifurcations in a laminar 3D bluff body wake: effect of pitch and yaw}

\author{A. N. Lagwankar\aff{1}, A. Chiarini\aff{2}, F. Gallaire\aff{1} \and E. Boujo\aff{1}}

\affiliation{
\aff{1}Laboratory of Fluid Mechanics and Instabilities, \'Ecole Polytechnique F\'ed\'erale de Lausanne, CH-1015 Lausanne, Switzerland
\aff{2}Dipartimento di Scienze e Tecnologie Aerospaziali, Politecnico di Milano, via La Masa 34, 20156 Milano, Italy}

\corresau{A. N. Lagwankar, atharva.lagwankar@epfl.ch}

\begin{document}
\maketitle

\begin{abstract}
We study the effect of pitch and yaw on the bifurcations of the laminar flow past an Ahmed body, a bluff body of width-to-height ratio $W/H = 1.2$ and length-to-height ratio $L/H = 3$. When perfectly aligned with incoming flow, its wake first undergoes pitchfork bifurcations leading to a static vertical or horizontal deflection, and becomes oscillatory via a secondary Hopf bifurcation at larger Reynolds number $Re$. As commonly observed for imperfect pitchfork bifurcations, any small misalignment preselects one of the steady states, referred to as the ``primary'' branch, at low $Re$. At larger $Re$, a disconnected ``secondary'' branch is born via a saddle-node bifurcation. Here, we investigate the effect of misalignment using three-dimensional computations. We first focus on pure pitch and pure yaw. In both cases, the secondary branch becomes unstable as soon as the incidence exceeds a small value. In the pure yaw case, the primary branch becomes unstable via a horizontal Hopf bifurcation, destabilised by yaw. In the pure pitch case, the primary branch sees the crossover of a vertical Hopf bifurcation and a horizontal pitchfork bifurcation, respectively destabilised and stabilised by pitch; this motivates use of DNS to study this nonlinear competition, already observed for aligned Ahmed bodies in ground proximity. We then study the effect of simultaneous yaw and pitch with a weakly nonlinear analysis. The competition between the two stationary modes gives rise to rich bifurcation diagrams, allowing us to predict the number of stable solutions and phase space trajectories between different states.

\end{abstract}

{\section{Introduction}
The circular cylinder and the sphere are arguably the simplest and most studied bluff bodies in two and three dimensions, respectively. 
One reason for their ubiquity in fundamental studies lies in the invariance of their shape under rotation, which makes them insensitive to the direction of the incoming flow. 
In practice, however, invariant geometries are rather the exception than the rule, so that most flows past bluff body do depend on the flow direction. 
For example, two-dimensional (2D) square cylinders may have some pitch angle relative to the free stream, three-dimensional (3D) circular cylinders some yaw angle, and generic 3D bluff bodies any combination of pitch, yaw and roll angles, as shown in Figure \ref{fig:geometry}. In this study, we focus on a non-axisymmetric 3D bluff body,  the Ahmed body (\cite{ahmed}), extensively studied in the literature as a generic, simplified model of ground vehicle allowing for a fundamental understanding of flow phenomena relevant to engineering such as wake deflection and bistability.
As briefly reviewed below,  misalignment or ``incidence'' can lead to significant qualitative and quantitative modifications of the flow, both in 2D and 3D configurations, and in laminar and turbulent regimes. 
\subsection{Laminar wakes of 2D bluff bodies}
In the laminar regime, most steady flows past 2D bluff bodies first lose stability via a Hopf bifurcation, which leads to vortex shedding in the wake. 
At slightly larger Reynolds number, the periodic flow undergoes a secondary bifurcation that breaks invariance in the third direction.
For example, the flow past a 2D circular cylinder becomes unstable to an oscillatory eigenmode at $Re_c^{2D} \simeq 47$, where the Reynolds number $Re=U_\infty D/\nu$ is based on the free stream velocity $U_\infty$, the cylinder diameter $D$ and the fluid kinematic viscosity $\nu$. 
Secondary instabilities occur via 3D eigenmodes called modes $A$ and $B$ at $Re_c^A\simeq 190$ and $Re_c^B\simeq 260$, with transverse wavelengths $\lambda_z^A \simeq 4D$ and $\lambda_z^B \simeq 1D$, as shown experimentally, in unsteady numerical simulations and with Floquet stability analyses (\cite{THOMPSON1996190, Williamson_1996, barkley_three-dimensional_1996}).
At zero pitch angle, $\alpha=0^\circ$, the bifurcation sequence is similar for a 2D square cylinder, with a primary Reynolds number $Re_c^{2D} \simeq 45-50$,  secondary Reynolds numbers $Re_c^A\simeq 160$, $Re_c^B\simeq 190$ and $Re_c^S\simeq 200$ and transverse wavelengths $\lambda_z^A \simeq 5D$, $\lambda_z^B \simeq 1D$ and $\lambda_z^S \simeq 3D$ respectively (\cite{Robichaux1999}). 

At incidence, however, the picture is modified: $Re_c^{2D}$ decreases  with $\alpha$, and
properties of the 2D limit cycle are modified too (e.g. Strouhal number and mean drag coefficient); at larger $Re$, the 3D mode $A$ is strongly stabilised for intermediate pitch angles, such that for $12^\circ \leq \alpha \leq 26^\circ$ the  first 3D bifurcation occurs via a new subharmonic mode of wavelength $\lambda_z \simeq 2D$, (\cite{Sohankar_1998, TONG2008994, SHEARD_2009_square}).
Incidence affects the flow past many other 2D bluff bodies. 
For example, \cite{Jackson_1987} used linear stability analysis to show the effect of inclination on the onset of the Hopf bifurcation for elliptic cylinders of aspect ratio 2 and flat plates.
\cite{SOHANKAR_1997} performed numerical simulations of the wake of rectangular cylinders of aspect ratio $2$ and 4 at $Re=200$ for a wide range of pitch angle, revealing variations in Strouhal number and aerodynamic coefficients, some of them strong and non-monotonic.

\subsection{Laminar wakes of 3D bluff bodies}

\subsubsection{Axisymmetric bodies}

Most steady flows past axisymmetric bodies aligned with the free stream first lose stability via a pitchfork bifurcation, which leads to a static deflection of the wake, followed by a Hopf bifurcation at larger Reynolds number. 
Both bifurcations are characterised by an azimuthal wavenumber $m=1$.
For example, linear stability analysis of steady axisymmetric flows yields the following critical Reynolds numbers for the pitchfork and Hopf bifurcations: $Re_c\simeq 115$ and 125 for a thin disc, $Re_c\simeq 215$ and 280 for a sphere, and 
$Re_c \in [215,435]$ and $[285,845]$ for bullet-shaped bodies of length $L \in [1D,6D]$, respectively
(see e.g. \cite{Natarajan_Acrivos_1993, Fabre_PoF_2008, MELIGA_JFS_2009, bohorquez2011}).

Other axisymmetric bluff bodies such as ellipsoids, cones and bicones of various lengths have been investigated in  \cite{Chiarini_Gauthier_Boujo_2025}.
Linear stability analysis predicts an increase of $Re_c$ with $L/D$ for both the pitchfork and Hopf bifurcations, in good agreement with experiments and unsteady numerical simulations.
Secondary stability analysis of the steady deflected wake past a sphere obtained $Re_c \simeq 270-275$ (\cite{Citro_AIAA_2017, Frantz_2025}), close to the $Re_c$ of the second unstable mode of the axisymmetric flow. 
The torus is an exception to the initial sequence of two bifurcations ($m=1$ pitchfork followed by $m=1$ Hopf) described above. 
Indeed, depending on the ratio $D/d$ of the centerline diameter to the cross-section diameter, the geometry continuously varies from a sphere ($D/d=0$) to a 2D circular cylinder (infinitely thin ring, $D/d=\infty$). 
As a result, when oriented such that the axis of symmetry is aligned with the flow, the first bifurcation shifts at $D/d \simeq 4$ from an $m=1$ pitchfork to an axisymmetric or non-axisymmetric Hopf  (\cite{sheard2003, SHEARD_2004}).
In a series of experimental studies motivated by the absence of end effects, parallel and oblique vortex shedding regimes have been extensively characterised in the flow past thin toruses ($D/d \simeq 10-30$) (e.g. \cite{Leweke_PRL_1993} and \cite{Leweke_JFM_1995}).

\cite{YU_2018_torus} numerically investigated the low-$Re$ steady flow of an inclined torus of aspect ratio $2 \leq D/d \leq 3$. They observed a monotonic decrease of the drag coefficient with inclination $\alpha$ (angle between the axis of symmetry and the free stream), and a non-monotonic variation of the lift coefficient, from $C_L=0$ at $\alpha=0^\circ$ and $90^\circ$ (as expected by symmetry) to a maximum of $|C_L|$ at intermediate inclination.

\cite{Inoue_1999_torus} experimentally studied the vortex shedding in an inclined torus wake at two aspect ratios, $D/d = 3$ and 5, and two Reynolds numbers, $Re=600$ and 1500. 
While the wake of the upstream section of the torus remains periodic for all inclinations, that of the downstream section exhibits three regimes: (i)~periodic but unsynchronised with the wake of the upstream section for $\alpha \lesssim 30^\circ$;
(ii)~aperiodic for $30^\circ \lesssim  \alpha \lesssim 60^\circ$;
(iii)~periodic and synchronised for $\alpha \gtrsim  60^\circ$. 

\cite{Chrust_et_al_2015} studied the effect of inclination on the flow past a thin disc.
Their numerical and experimental work showed that 
small inclinations $\alpha \lesssim 35^\circ$ have a destabilising effect on the Hopf bifurcation while, conversely, larger inclinations have a stabilising effect.
In numerical simulations at the larger Reynolds number  $Re=500$, \cite{TIAN2017152} observed that inclination gradually modifies the flow from chaotic ($\alpha\lesssim 35^\circ$) to periodic ($\alpha \gtrsim 55^\circ$), 
while the mean drag and lift coefficients monotonically decrease and increase, respectively.

\subsubsection{Non-axisymmetric bodies}

Simple non-axisymmetric bodies include rectangular flat plates. 
\cite{marquet2015global} studied the linear stability of thin plates perpendicular to the flow and found that the first instability depends on the width-to-height ratio: pitchfork bifurcation for small $W/H \leq 2$ leading to a static wake deflection (like a thin disc), or Hopf bifurcation for large $W/H$, leading to vortex shedding either across the height for  $W/H \geq 2.5$ (like a 2D cylinder) or across the width for intermediate values of $W/H$.

Recently, other non-axisymmetric 3D bluff bodies perfectly aligned with the free stream have been investigated. 
\cite{Zampogna_Boujo_2023} characterised the linear stability of 3D rectangular prisms of width-to-height ratio $W/H=1.2$ and various length-to-height ratio $L/H$ from thin plates to elongated bodies, with sharp and rounded leading edges. 
They found two pitchfork bifurcations followed by two Hopf bifurcations, each eigenmode breaking either the vertical or horizontal symmetry. 
For thin plates, these four modes become unstable at similar Reynolds numbers. 
For longer bodies, the two oscillatory modes are much more stable than the two stationary ones. 
Focusing on geometries typical of the Ahmed body (a simplified car model (\cite{ahmed}) with $W/H \simeq 1.2-1.4$, $L/H \simeq 3-4$ and rounded leading edges, widely studied in the turbulent regime), a weakly nonlinear analysis clarified the nonlinear interaction between the two stationary modes: although the first bifurcation leads to a steady vertical  deflection of the wake (along the smallest dimension of the base), this state becomes unstable at larger $Re$ such that the steady wake deflection  eventually becomes  horizontal (along the largest dimension of the base). 

\cite{Chiarini_Boujo_2025} extended this study to 3D rectangular prisms of various width-to-height ratios. 
They showed that the first bifurcation is always stationary for long enough bodies (large enough $L/W$)  and oscillatory for wide enough bodies  (large enough $W/L$). 
For $L/H=5$ and selected values of $W/H$ corresponding to narrow, intermediate and wide prisms, fully nonlinear simulations revealed rich sequences of secondary and subsequent bifurcations leading to unsteady periodic, aperiodic and chaotic regimes at larger $Re$.

To the best of our knowledge, there is no experimental or numerical study on the laminar wake of 3D rectangular prisms or Ahmed-like bodies at incidence.

\subsection{Turbulent wakes}

While turbulent flows past bluff bodies are fully 3D at all times, temporal averaging may restore some symmetry, but it does so differently depending on the body geometry. 
For nominally 2D bluff bodies, 3D structures such as oblique or spanwise-modulated vortex shedding  disappear under temporal averaging, and  the time-averaged flow is 2D.
For 3D bluff bodies of cross-section aspect ratio close to one, a steady symmetry breaking (SSB) is observed in the flow, but random transitions allow the wake to visit multiple states, leading to a time-averaged flow with the same symmetries as the body, e.g. axisymmetry or planar symmetry  (\cite{Grandemange_PoF_2012,Grandemange_PoF_2013}). 
Many studies have specifically investigated the effect of incidence on the turbulent flow past 2D bluff bodies.For example, \cite{VanOudheusden_1995}, \cite{CHEN_1999}, \cite{Dutta_2003}, \cite{VANOUDHEUSDEN_2008}, \cite{HUANG_2010}, \cite{YEN_2011} and others have experimentally varied the pitch angle $\alpha$ of  2D square cylinders and studied variations in Strouhal number, mean drag, mean lift, lift fluctuations and flow topology, while \cite{LOU_2016} studied the effect of the yaw angle $\beta$. 

Turbulent flows past 3D bluff bodies with a blunt base of aspect ratio close to one exhibit SSB: even if the body is symmetric with respect to one or multiple planes and is perfectly aligned with the free stream,  the wake is  most often deflected. 
However, transitions between multiple coexisting deflected states occur randomly (with no preferred frequency) as rare events (with a typical mean switching time of the order of $10^3$  convective times). 
These transitions allow the system to explore the phase-space parameter and restore symmetry in the time-averaged flow.
For bodies with a rectangular base (e.g. Ahmed body, Windsor body, simple frigate model), the SSB is usually along the large dimension of the base, i.e. horizontal for wider-than-tall bodies and vertical for taller-than-wide bodies \citep{Legeai20}, consistent with the laminar bifurcations \citep{Grandemange12PRE, Evstafyeva17, Zampogna_Boujo_2023, Chiarini_Boujo_2025}.
However, strong ground proximity suppresses bistability and leads to a vertical deflection \citep{Grandemange_PoF_2013}. 

For axisymmetric bodies, the SSB corresponds to a $m=1$ modification of the flow, with infinitely many possible azimuthal orientations, which  again is similar to the first laminar bifurcation ($m=1$ pitchfork), while the time-averaged flow is axisymmetric ($m=0$) as the wake eventually visits all azimuthal orientations \citep{Grandemange_PoF_2012, Rigas2014}.

The SSB has a twofold aerodynamic impact: 
(i)~a non-zero cross-flow force (side force or lift force), which rapidly changes sign during transitions and is detrimental to handling/manoeuvrability; 
(ii)~an increased drag compared to the symmetric flow under the same conditions \citep{Perry_2016, EVRARD2016, bonnavion_cadot_2018, Pavia_et_al_2018_ExpFluids, Plumejeau_2019, Fan_2020b, Haffner_et_al_2020a, Fan_2022, Khan_2022a, LIPPERT_2022}.
Incidence has a strong effect on the SSB and the multistability described above.
For bodies with a rectangular base, rotation about an axis perpendicular to the direction of the SSB (yaw $\beta$ for wide bodies, pitch $\alpha$ for tall bodies) tends to select one of the two possible states and to suppress bistability;
rotation about an axis aligned with the direction of the SSB (pitch $\alpha$ for wide bodies, yaw $\beta$ for tall bodies) induces a static deflection in the other direction (along the short dimension of the base), while the SSB gradually weakens and eventually disappears/vanishes
\citep{bonnavion_cadot_2018, Fan_2022}.
Combined rotations (simultaneous pitch and yaw) were investigated by \cite{Fan_2022}, who obtained a non-trivial map of vertical and horizontal deflections in the $(\alpha,\beta)$ plane.
For elongated axisymmetric bodies, misalignment selects a preferred azimuthal orientation, and the wake becomes increasingly non-axisymmetric with $\alpha$ \citep{Gentile_2017}.
Incidence also affects the switching rate for both rectangular and axisymmetric bodies (see e.g. \cite{Fan_2023} and \cite{Gentile_2017}).
Finally, it should be noted that the effect of pitch and yaw is rather dramatic, with significant changes occurring within a few degrees. 
\subsection{Present work}
As summarised in the previous subsections $1.1-1.3$, both incidence and ground proximity significantly affect the bifurcations of bluff-body wakes.
In this study, we investigate the effect of incidence on the laminar wake of a square-back Ahmed body. 
When perfectly aligned with the free stream, and in the absence of ground, this body has two planes of symmetry and its wake undergoes two pitchfork bifurcations at nearly the same $Re$ (\cite{Zampogna_Boujo_2023}).
We expect incidence, especially pitch and yaw, to affect the linear stability and nonlinear interaction of these two stationary modes, and possibly of other oscillatory modes.
We first compute the fully nonlinear steady base flows in the presence of either pitch or yaw, including multiple solutions that coexist for some values of  pitch angle $\alpha$, yaw angle $\beta$, and Reynolds number $Re$. 
We then systematically characterise the linear stability of these base flows. 
Using direct numerical simulations, we validate the linear stability analysis and gain insight into the nonlinear interaction between competing unstable modes.
Finally, we use a weakly nonlinear analysis to efficiently tackle ``oblique'' incidences, for which both pitch and yaw angles are non-zero. 

The paper is organised as follows. 
Section~\ref{NumMeth} presents the body geometry and flow configuration and gives details on the numerical methods. 
Section~\ref{BaseFlows}  describes the base flows for a range of Reynolds number and incidence.
Section~\ref{LinStab} gives linear stability results under pure pitch and pure yaw, and discusses some fully nonlinear scenarios investigated via direct numerical simulations.
Section~\ref{WNLSection} explores the combined effect of pitch and yaw via a weakly nonlinear analysis.
Finally, the conclusions are provided in section~\ref{Conclusion}. 
\section{Flow configuration and numerical methods}\label{NumMeth}
In this study, we consider the incompressible, laminar flow of a Newtonian fluid past a standard flat-back Ahmed body, as described by \cite{ahmed}. 
\begin{figure}
    \centering
    \begin{center}
        \begin{subfigure}{0.49\textwidth}
        \begin{overpic}[width=\linewidth]{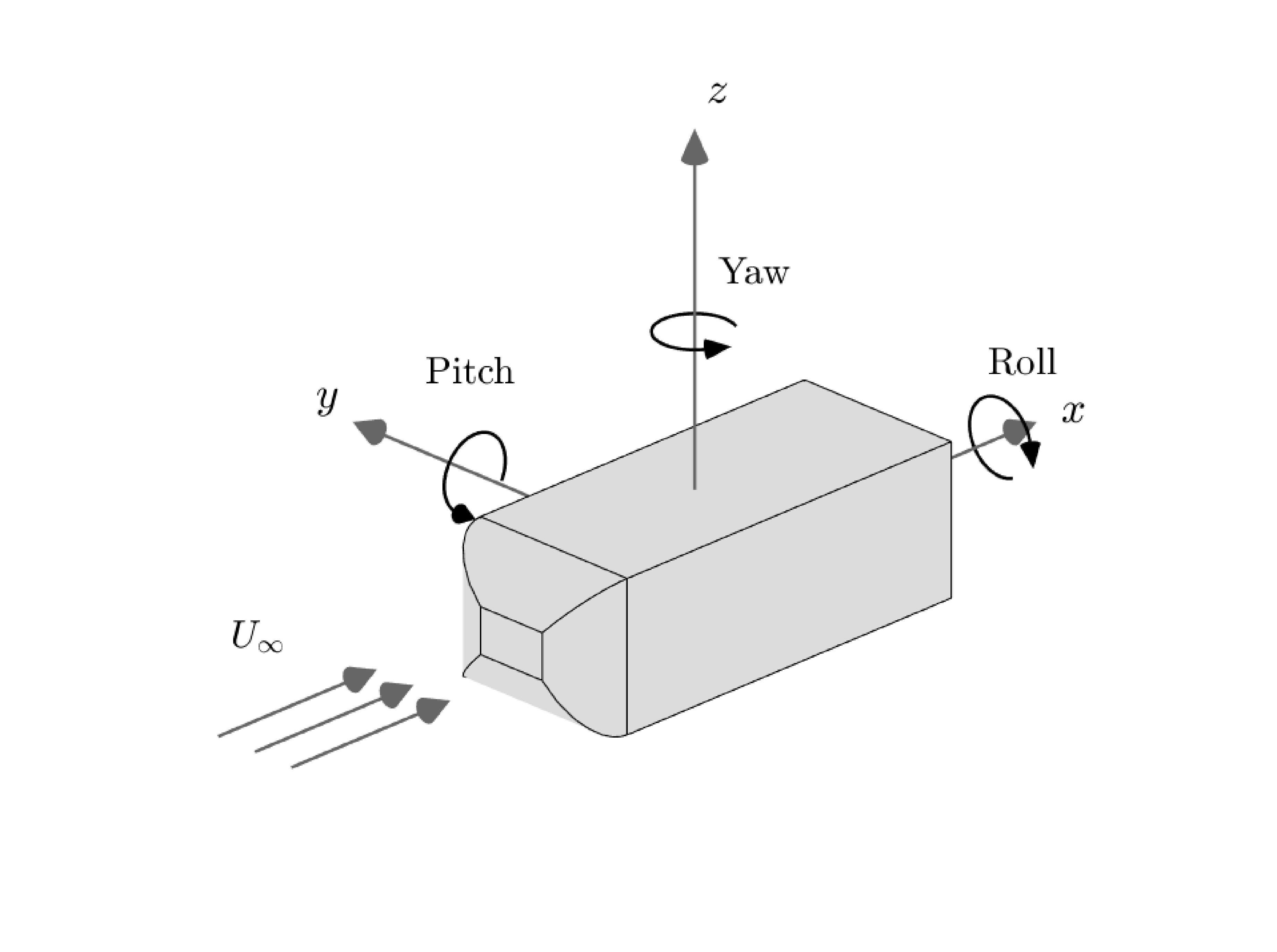}
        \put(-2,65){\small (a)}
        \end{overpic}
        \end{subfigure}
        \begin{subfigure}{0.49\textwidth}
        \begin{overpic}[width=\linewidth]{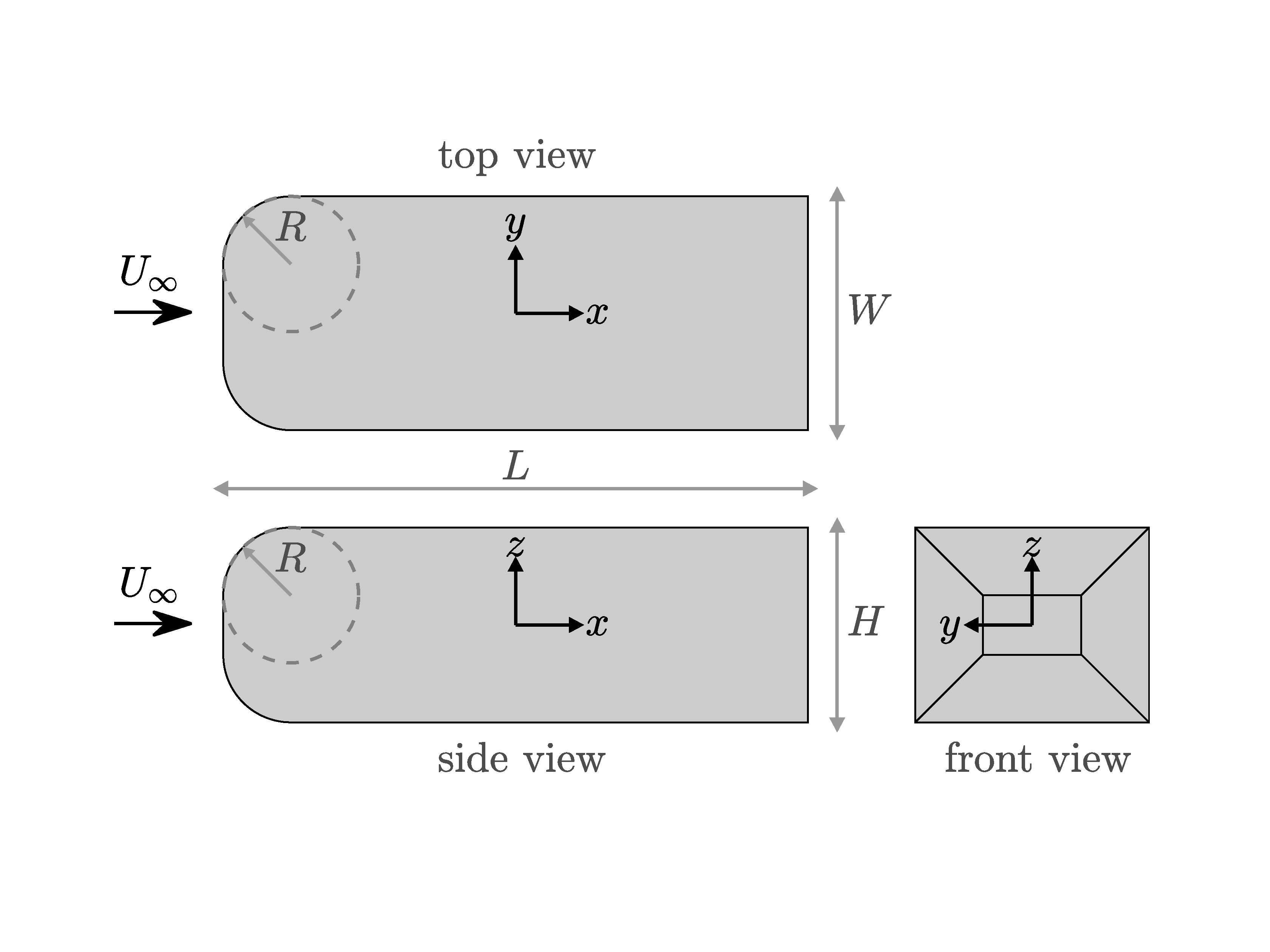}
        \put(-2,65){\small (b)}
        \end{overpic}
        \end{subfigure}
    \end{center}
    \vspace{-9mm}
    \begin{subfigure}{0.49\textwidth}
        \begin{overpic}[width=1\linewidth]{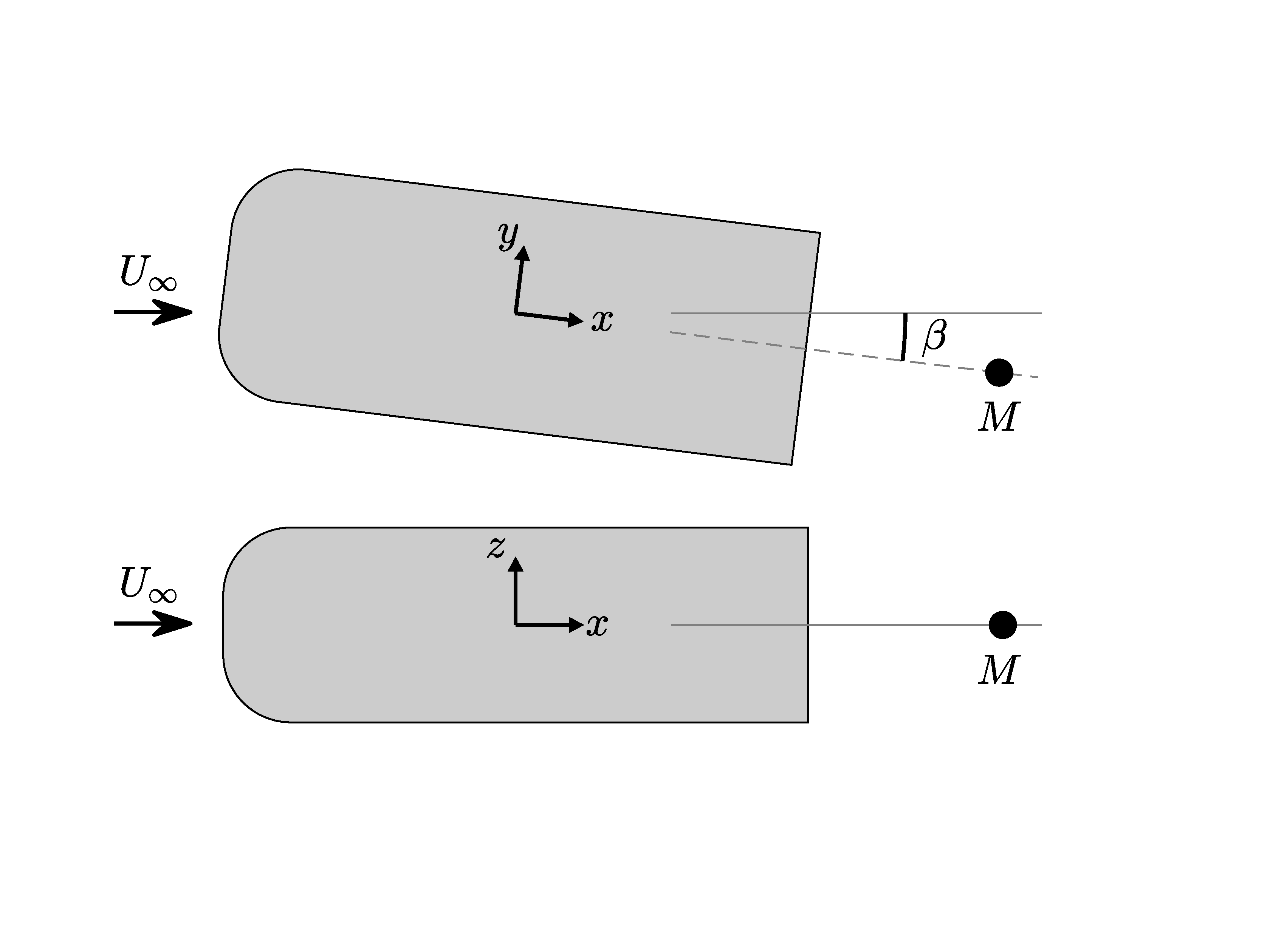}
        \put(-2,65){\small (c)}
        \end{overpic}
    \end{subfigure}
    \begin{subfigure}{0.49\textwidth}
        \begin{overpic}[width=1\linewidth]{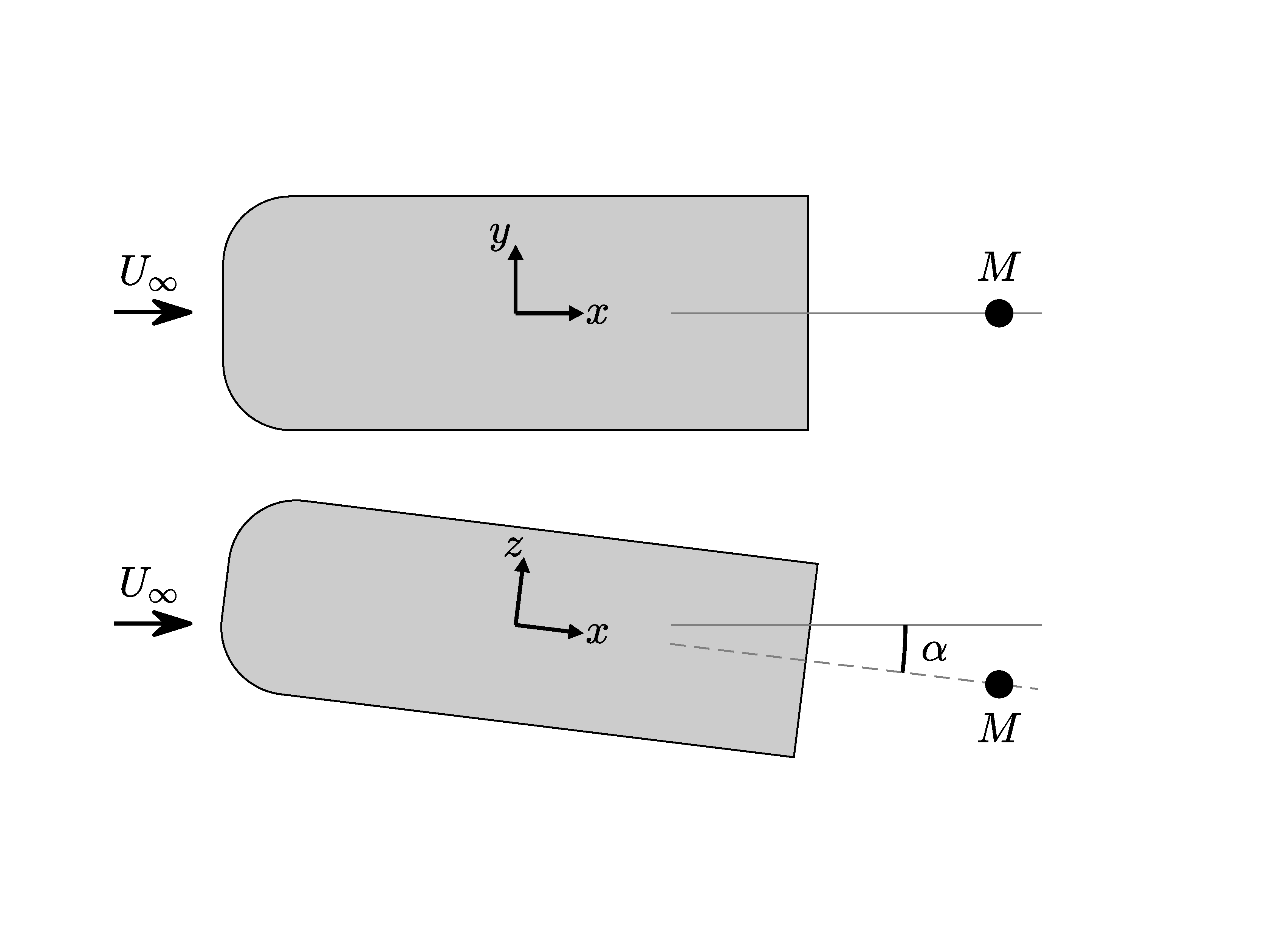}
        \put(-2,65){\small (d)}
        \end{overpic}
    \end{subfigure}
    \caption{Sketch of the geometry and flow configurations. Body dimensions: $W = 1.2H, L = 3H, R = 0.3472H$. (a) 3-D visualization of pitch ($\alpha$), yaw ($\beta$) and roll ($\gamma$), (b) body perfectly aligned with the incoming flow, (c) yaw misalignment and (d) pitch misalignment. The sensor $M$ has coordinates $(2.5H,0,0)$ in the body frame.}
    \label{fig:geometry}
\end{figure} 
The $x,y$ and $z$ axes are the streamwise, lateral and vertical directions respectively. 
We choose the body dimensions as height $H$, length $L = 3H$ and width $W = 1.2H$. 
The fillet radius is selected according to the original geometry of \cite{ahmed} as $R = 100H/288 \approx 0.347H$. The nomenclature for rotations follows standard definitions: pitch, yaw, and roll about the lateral, vertical and streamwise axes, respectively, as shown in Figure \ref{fig:geometry}. The coordinate axis is positioned at the centre of the body, defined as the point equidistant from all the planar faces.
In the absence of ground and lateral walls, positive and negative rotations about a single axis are equivalent. We hence perform base flow calculations and linear stability analysis in half of the numerical domain $\{x,y,z \, | \, -10 \leq x \leq20 ; \, -10 \leq y,z \leq 10 \}$: $y \geq 0$ for pure pitch and $z \geq 0$ for pure yaw. The velocity and pressure fields, $\boldsymbol{u}(\boldsymbol{x},t) = (u,v,w)^T$ and $p(\boldsymbol{x},t)$, are governed by the Navier\textendash Stokes equations,
\begin{equation}
    \nabla\cdot\boldsymbol{u} = 0, \quad \partial_t \boldsymbol{u} + (\boldsymbol{u}\cdot\nabla)\boldsymbol{u} = -\nabla p + \frac{1}{Re}\nabla^2\boldsymbol{u},
\end{equation}
where the Reynolds number $Re = U_\infty H/\nu$ is defined with the freestream velocity $U_\infty$, body height $H$ and  fluid kinematic viscosity $\nu$.
All  quantities are made dimensionless with $H$ and $U_\infty$.
The geometry and mesh are created using the 3D finite element mesh generator Gmsh (\cite{gmsh}). We use the finite element software FreeFEM++ (\cite{hecht:hal-01476313}) to 
perform  the nonlinear base flow calculations (section \ref{BaseFlows}), linear stability analysis (section \ref{LinStab}) and weakly nonlinear (WNL) analysis (section \ref{WNLSection}). 
The DNS  (section~\ref{sec:DSN}) are performed using an in-house finite difference solver introduced by \cite{LUCHINI2016340} and based on the immersed boundary method. The numerical domain used for DNS calculations is given by $\{x,y,z \, | \, -20 \leq x \leq60 ; \, -25 \leq y,z \leq 25 \}$. Table~\ref{tab:methods} summarises information about the domains, meshes and computational methods.

\begin{table}
  \begin{center}
\def~{\hphantom{0}}
  \begin{tabular}{p{4.cm} p{3.5cm} p{2.5cm} p{3.5cm}}
       Calculation & Domain & Mesh size & Software  
       \\ [15pt]
       Base flow and linear stability \newline for pure pitch case & Half domain: $y\geq0$ & $\approx2.94M$ elements & FreeFEM++ \newline (finite elements) 
       \\
       \\ 
       Base flow and linear stability \newline for pure yaw case & Half domain: $z\geq0$ & $\approx2.91M$ elements & FreeFEM++ \newline (finite elements) 
       \\
       \\
       Weakly nonlinear analysis & Quarter domain: $y,z\geq0$ & $\approx1.5M$ elements & FreeFEM++ \newline (finite elements) 
       \\
       \\
       Direct numerical simulation & Full domain & $\approx400M$ points & In-house CPL code \newline (immersed boundary,\newline finite differences) \\ 
  \end{tabular}
  \caption{Summary of the domains, meshes,  and computational methods.}
  \label{tab:methods}
  \end{center}
\end{table}

\section{Base flow}\label{BaseFlows}

In this section, we take a look at the base flow $\textbf{q}_{0}(\boldsymbol{x}) = (\boldsymbol{u}_{0},p_{0})^T$ given by the steady Navier\textendash Stokes (NS) equations, 
\begin{equation} \label{NS}
    \nabla\cdot\boldsymbol{u}_0 = 0, \quad (\boldsymbol{u}_0\cdot\nabla)\boldsymbol{u}_0 = -\nabla p_0 + \frac{1}{Re}\nabla^2\boldsymbol{u}_0.
\end{equation}

We compute the base flow for cases of either pure pitch or pure yaw. The inherent symmetries of the problem allow us to use a half domain. The body is kept fixed and the inlet boundary conditions are adjusted to introduce the misalignment.
The inlet boundary conditions depend on the configuration: the freestream velocity is $\boldsymbol{U}_{\infty} = U_\infty(\cos\alpha, 0, \sin\alpha)^T$ for the pure pitch case, and $\boldsymbol{U}_{\infty} = U_\infty(\cos\beta, \sin\beta, 0)^T$ for the pure yaw case, where $\alpha, \beta$ are the  pitch and yaw angles, respectively.
A sufficiently large domain was chosen based on a domain-dependence study to implement far-field boundary conditions on the lateral surfaces of the domain (\cite{Zampogna_Boujo_2023}). A symmetry boundary condition is applied on the symmetry plane.

\begin{figure}[!h]
\centering
\begin{subfigure}{0.495\textwidth}
\begin{overpic}[width=1\linewidth]{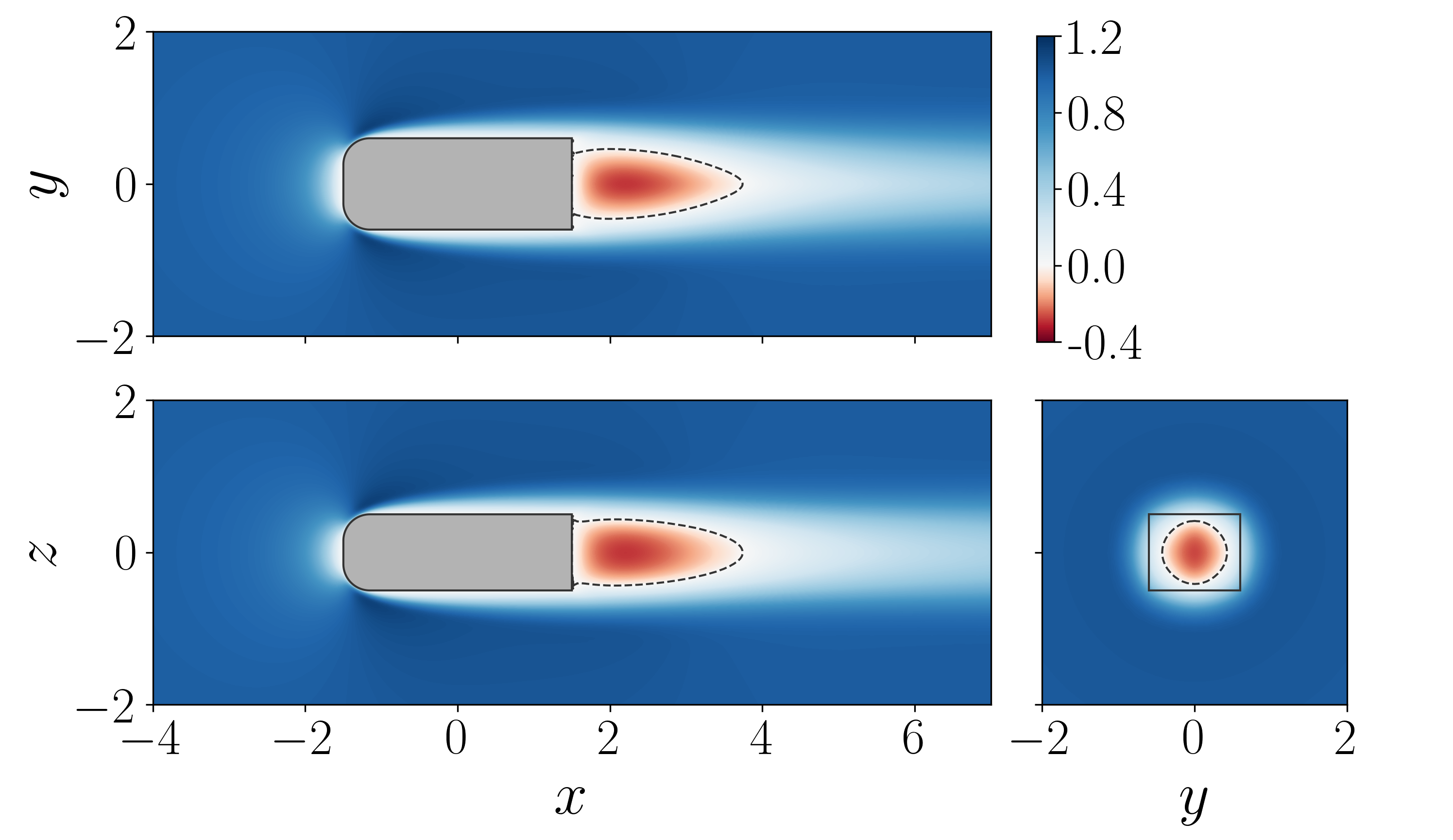} 
\label{fig:subim1}
\put(-5,55){\small (a)}
\end{overpic}
\end{subfigure}
\begin{subfigure}{0.495\textwidth}
\begin{overpic}[height=0.6\linewidth]{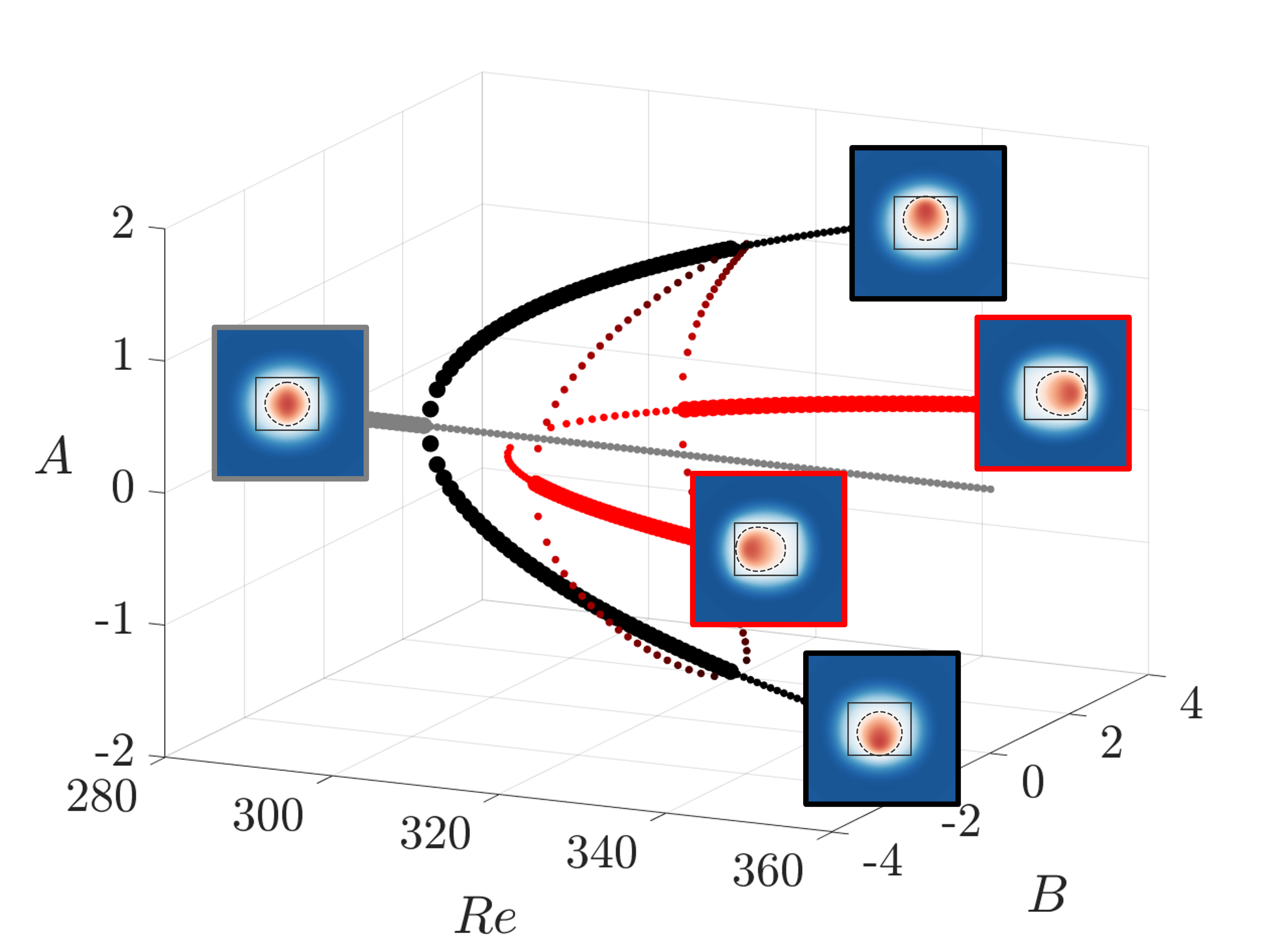} 
\label{fig:subim1}
\put(-5,55){\small (b)}
\end{overpic}
\end{subfigure}
\begin{subfigure}{0.495\textwidth}
\begin{overpic}[width=1\linewidth]{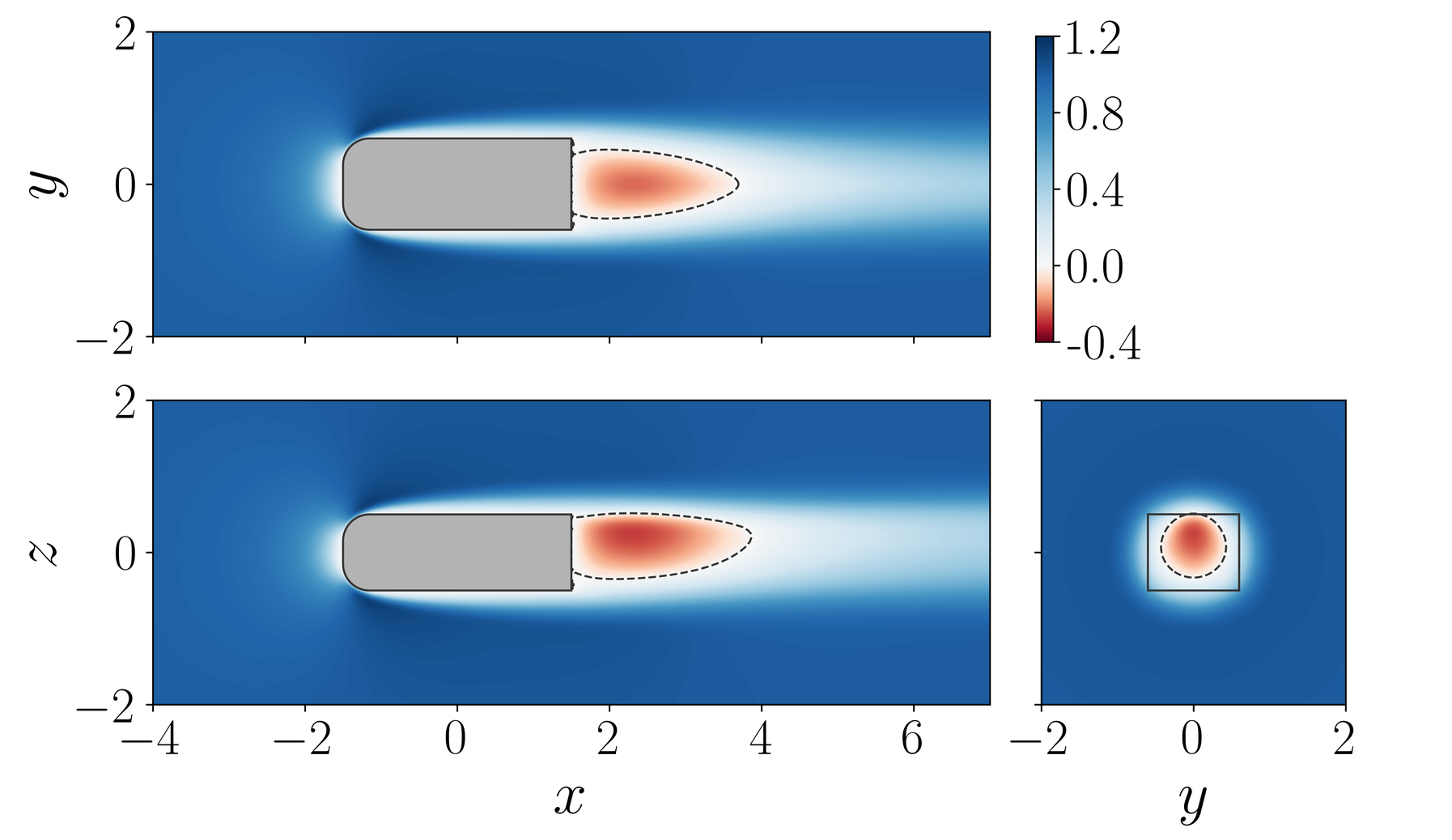} 
\label{fig:subim1}
\put(-5,55){\small (c)}
\end{overpic}
\end{subfigure}
\begin{subfigure}{0.495\textwidth}
\begin{overpic}[width=1\linewidth]{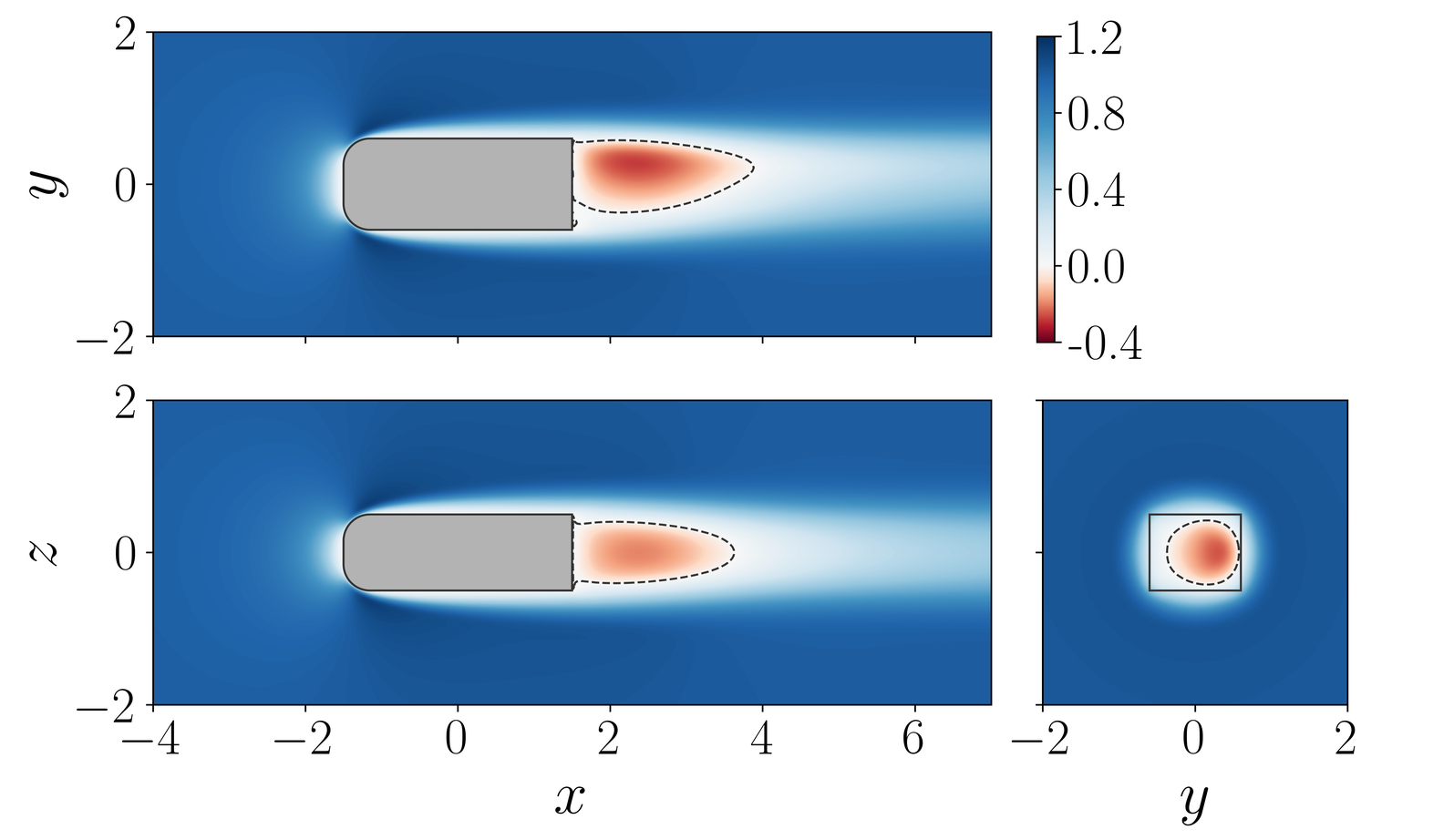}
\label{fig:subim2}
\put(-5,55){\small (d)}
\end{overpic}
\end{subfigure}
\caption{Streamwise velocity $u$ past a perfectly aligned Ahmed body $(H = 1, W = 1.2, L = 3)$, in the planes $z=0$  (top view), $y=0$ (side view) and $x=2.5$ (rear view). 
Dashed line: isocontour $u=0$.
$(a)$ Doubly symmetric state at $Re = 300$.
$(b)$ Complete bifurcation scenario.
$(c)$ State A: vertically deflected wake at $Re = 340$ . 
$(d)$ State B: horizontally deflected wake at $Re = 340$.}
\label{fig:BF_P0Y0}
\end{figure} 
For a perfectly aligned body ($\alpha=\beta=0^\circ$), \cite{Zampogna_Boujo_2023} showed that the doubly symmetric base flow shown in figure \ref{fig:BF_P0Y0}(a) undergoes two pitchfork bifurcations at approximately $Re=300$, each breaking one of the two planar symmetries and leading to a static deflection of the wake. The complete bifurcation sequence is shown in figure \ref{fig:BF_P0Y0}(b). 
In state A, the wake deflection is vertical, for example upward, as shown in figure \ref{fig:BF_P0Y0}(c). 
In state B, the wake deflection is horizontal, for example to the right, as shown in figure \ref{fig:BF_P0Y0}(d). 
As expected for perfect pitchfork bifurcations of a perfectly symmetric base flow, there exists another state A, symmetric to that shown in figure \ref{fig:BF_P0Y0}(c) with respect to the horizontal plane $z=0$, and another state B, symmetric to that shown in figure \ref{fig:BF_P0Y0}(d) with respect to the vertical plane $y=0$. 
In the presence of pitch (resp. yaw), the base flow preselects one of the two states $A$ (resp. $B$) already at $Re=0$. 
This leads to an imperfect pitchfork bifurcation, as seen in figure~\ref{fig:BF_P2Y0}(a), for $\alpha=2^\circ$. 
Here, we use $w_M$ as a measure of the asymmetry, i.e. the vertical velocity measured at the sensor $M$ of coordinates $(2.5,0,0)$ in the body frame.
For a positive pitch angle, the low-$Re$ wake is deflected upward and $w_M$ is positive (figure~\ref{fig:BF_P2Y0}(b)).  This branch, which can be followed continuously from $Re=0$, will be referred to as the primary branch.
When $Re$ becomes large enough, another branch appears via a saddle-node bifurcation. 
This disconnected branch will be referred to as the secondary branch.
In the configuration considered here, the wake is deflected downward on the outer secondary branch (figure~\ref{fig:BF_P2Y0}(c)), while it is barely deflected on the inner secondary branch (figure~\ref{fig:BF_P2Y0}(d)).
The qualitative behaviour of the base flow is similar in the case of pure yaw.

\begin{figure}[h]
        \centering
\begin{subfigure}{0.495\textwidth}
\begin{overpic}[width=\linewidth]{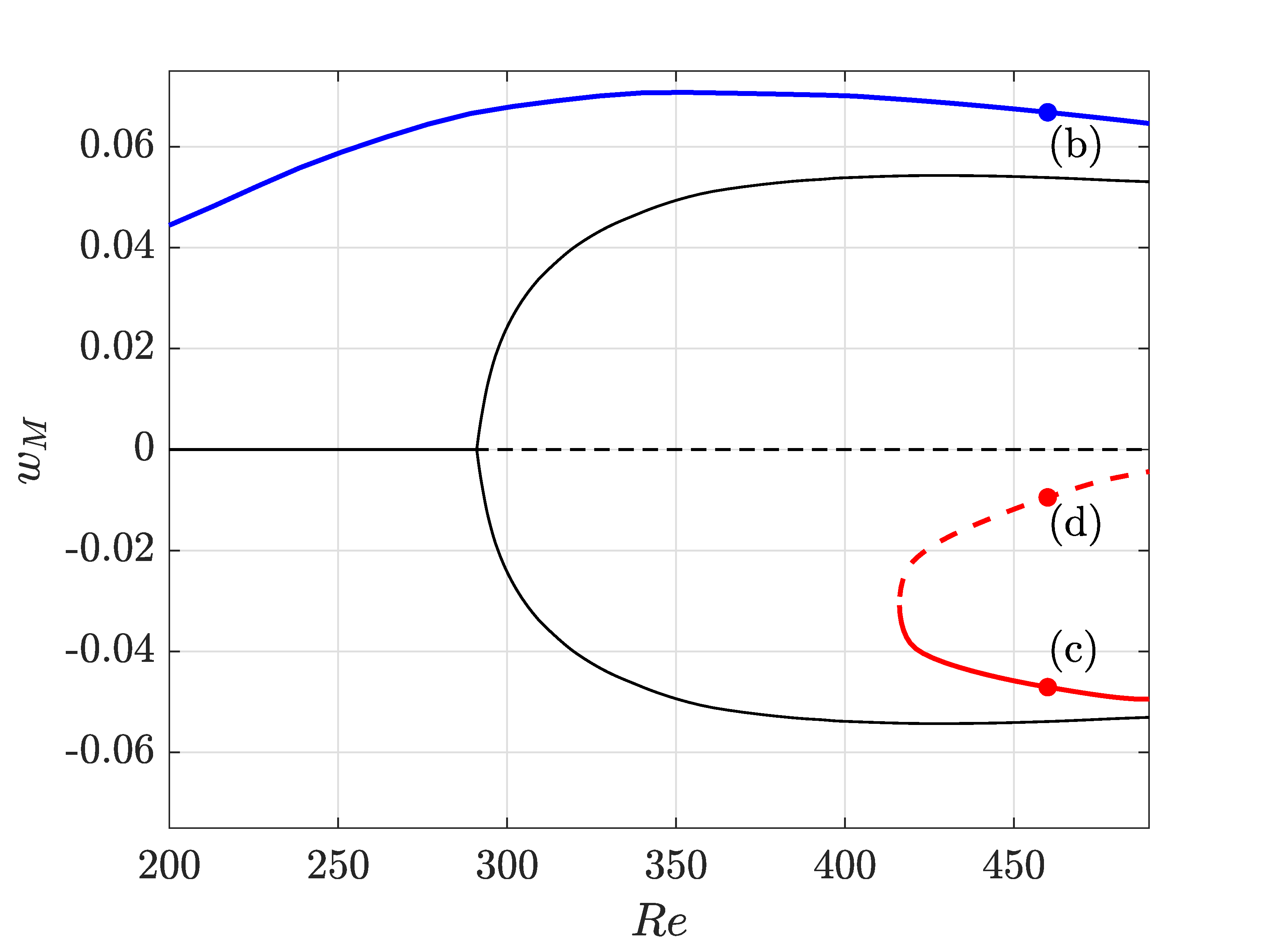} 
\put(-2,55){\small (a)}
\put(37,62){\small \tcb{primary branch}}
\put(50,30){\small \tcr{secondary}}
\put(54,24){\small \tcr{branch}}
\end{overpic}
\label{fig:subim1}
\end{subfigure}
\begin{subfigure}{0.495\textwidth}
\begin{overpic}[width=\linewidth]{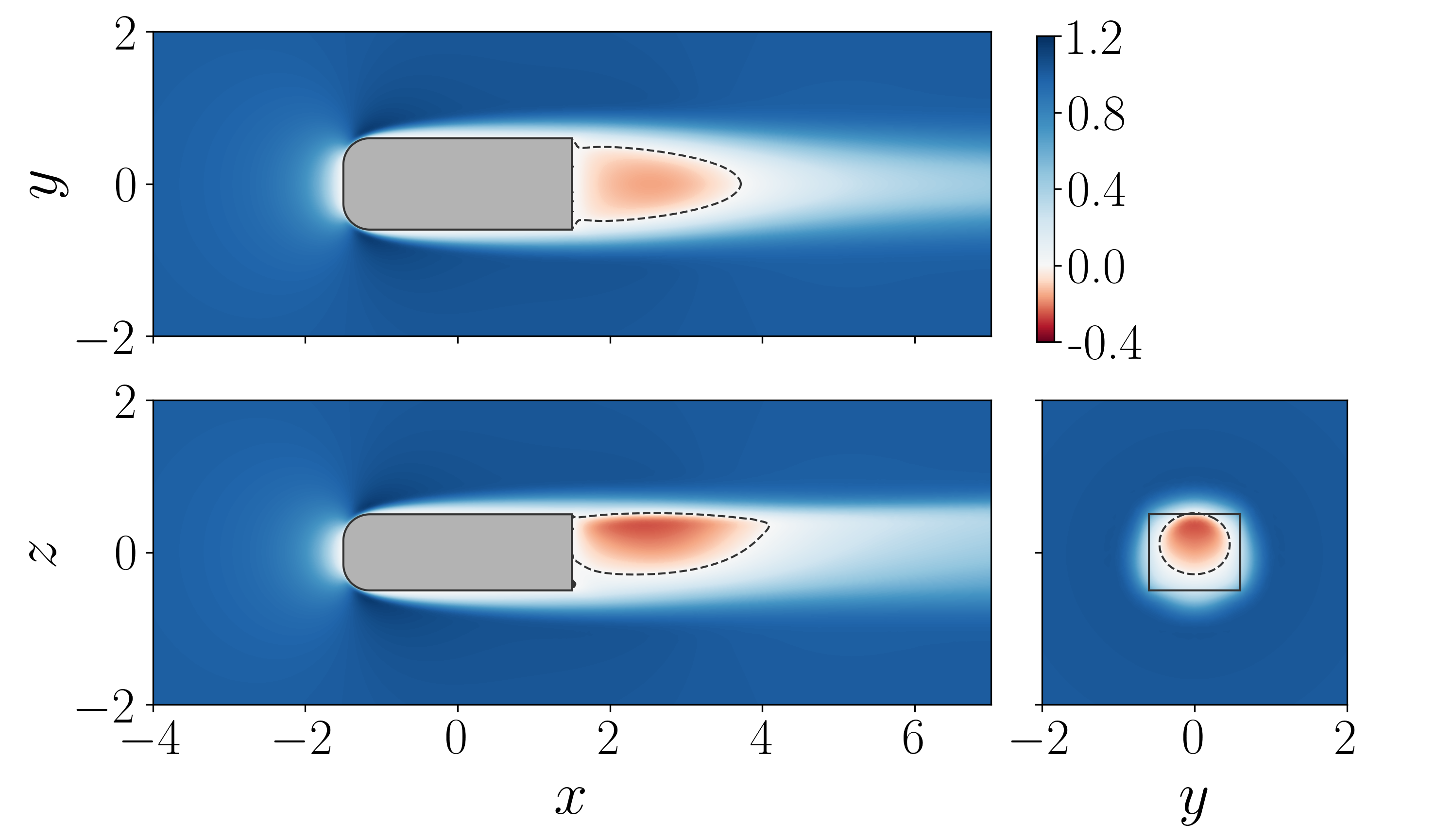}
\put(-2,55){\small (b)}
\end{overpic}
\label{fig:subim2}
\end{subfigure}
\begin{subfigure}{0.495\textwidth}
\begin{overpic}[width=\linewidth]{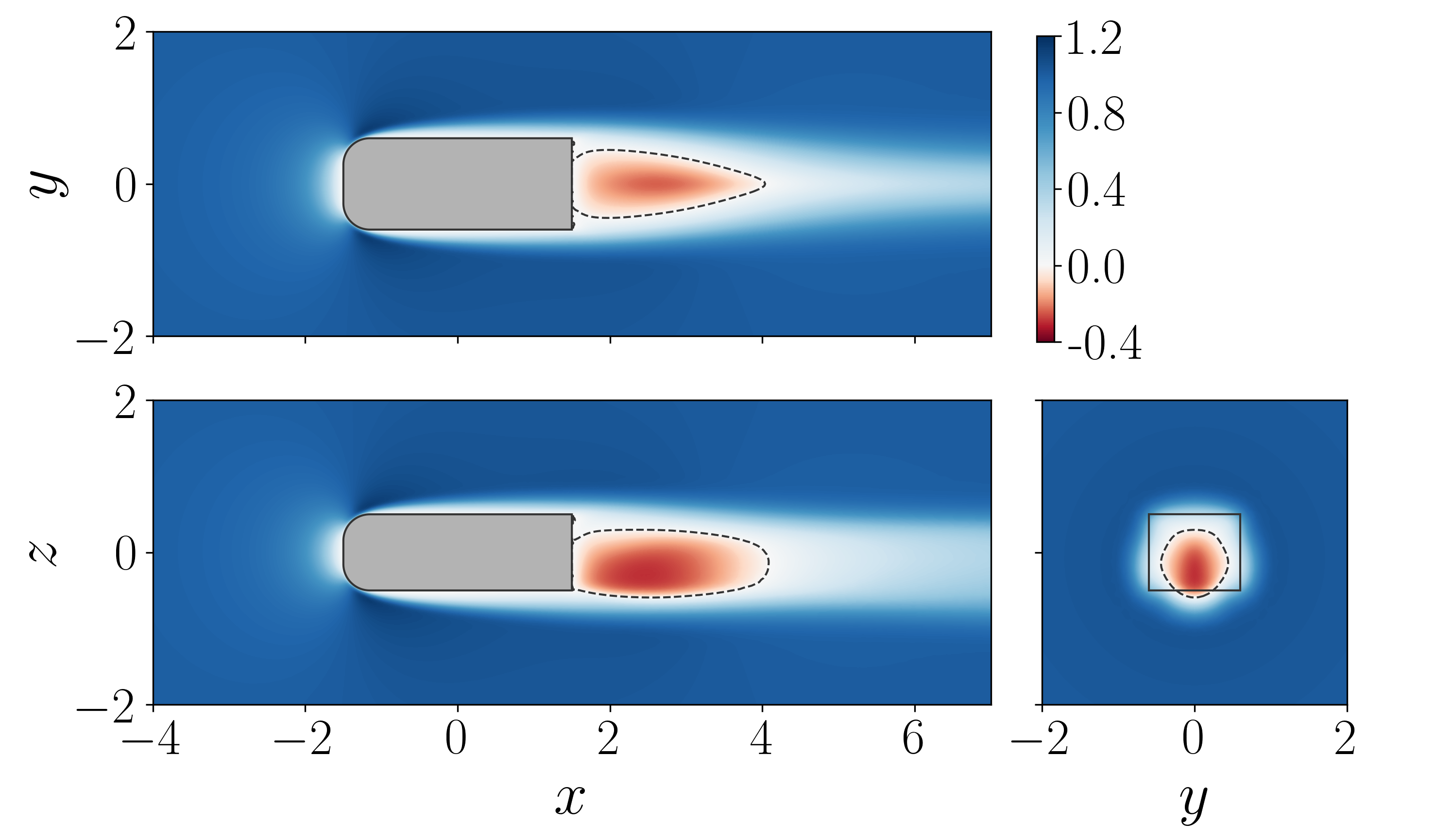}
\put(-2,55){\small (c)}
\end{overpic}
\label{fig:subim2}
\end{subfigure}
\begin{subfigure}{0.495\textwidth}
\begin{overpic}[width=\linewidth]{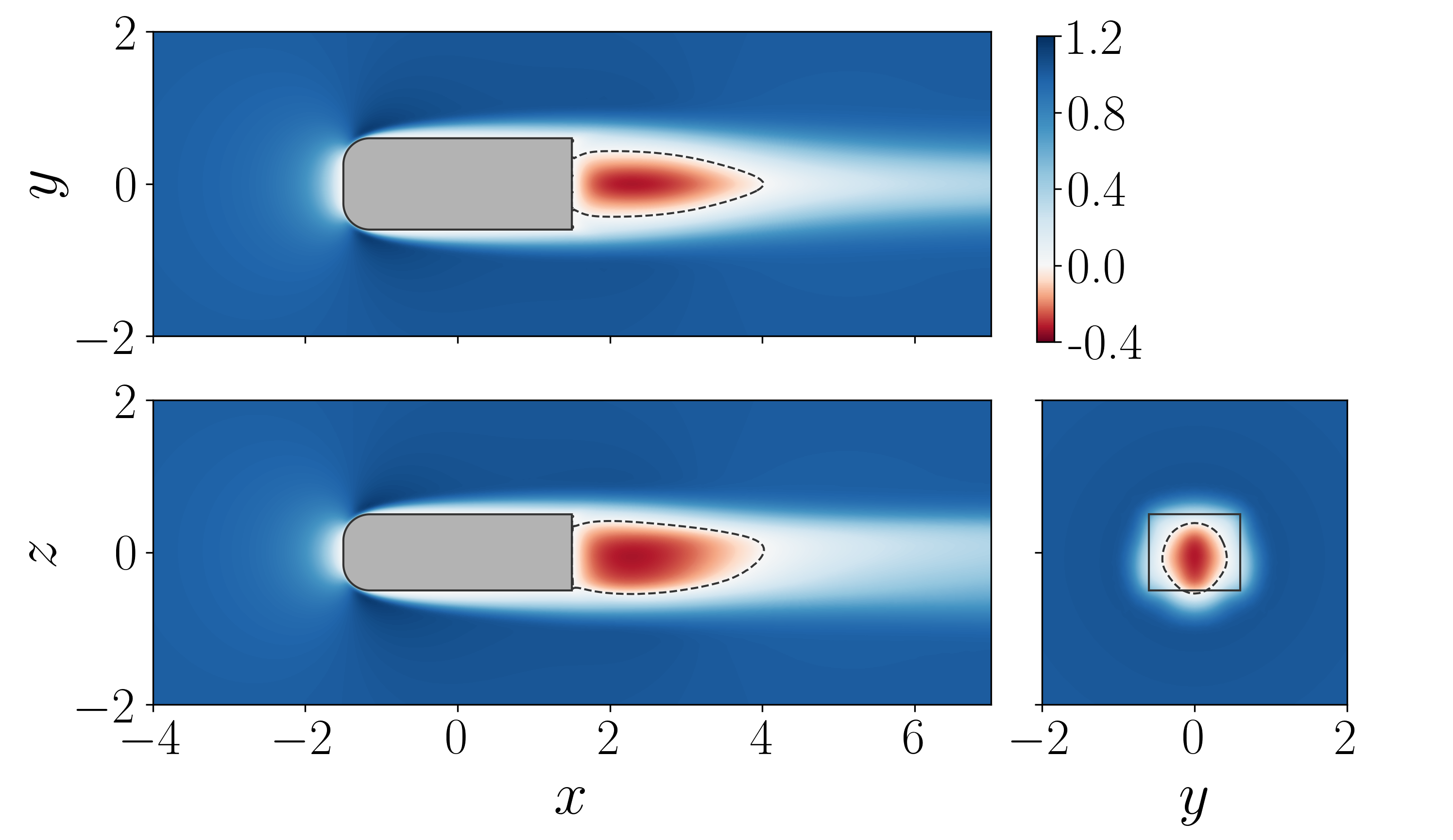}
\put(-2,55){\small (d)}
\end{overpic}
\label{fig:subim2}
\end{subfigure}
\caption{(a) Bifurcation diagram demonstrating the imperfect pitchfork bifurcation as a result of asymmetry (pure pitch  $\alpha=2^\circ$). 
The solid lines correspond to stable solutions while the dashed lines correspond to unstable solutions. The thin black lines correspond the bifurcation diagram in a perfectly aligned case.
(b),(c),(d) Streamwise velocity $u$ of the base flow at $Re = 460$, on the primary, outer secondary and the inner secondary branches respectively.}
\label{fig:BF_P2Y0}
\end{figure}
\begin{figure}[h!]
        \centering
\begin{subfigure}{0.495\textwidth}
\begin{overpic}[width=1\linewidth]{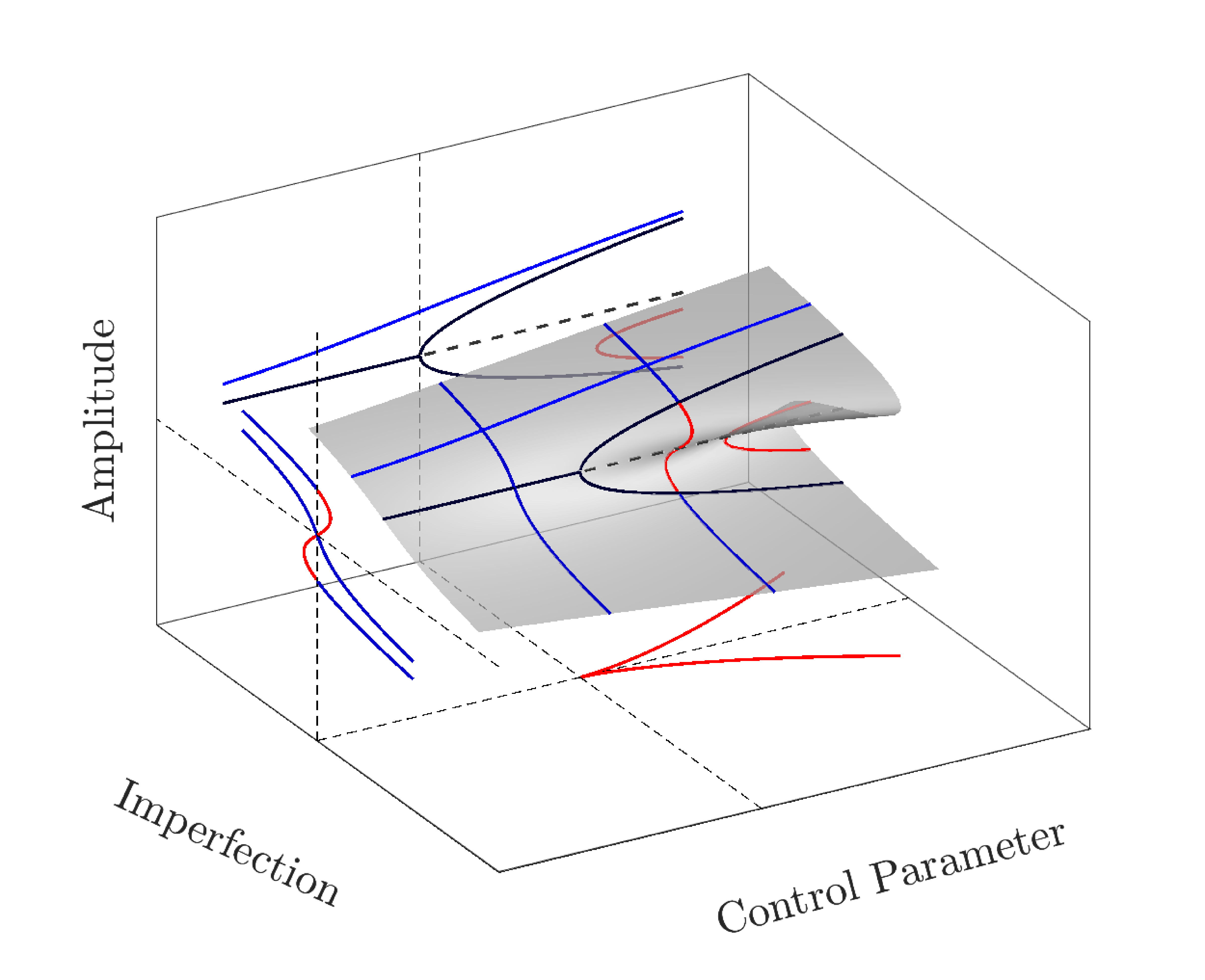} 
\put(-2,65){\small (a)}
\end{overpic}
\end{subfigure}
\begin{subfigure}{0.495\textwidth}
\begin{overpic}[width=1\linewidth]{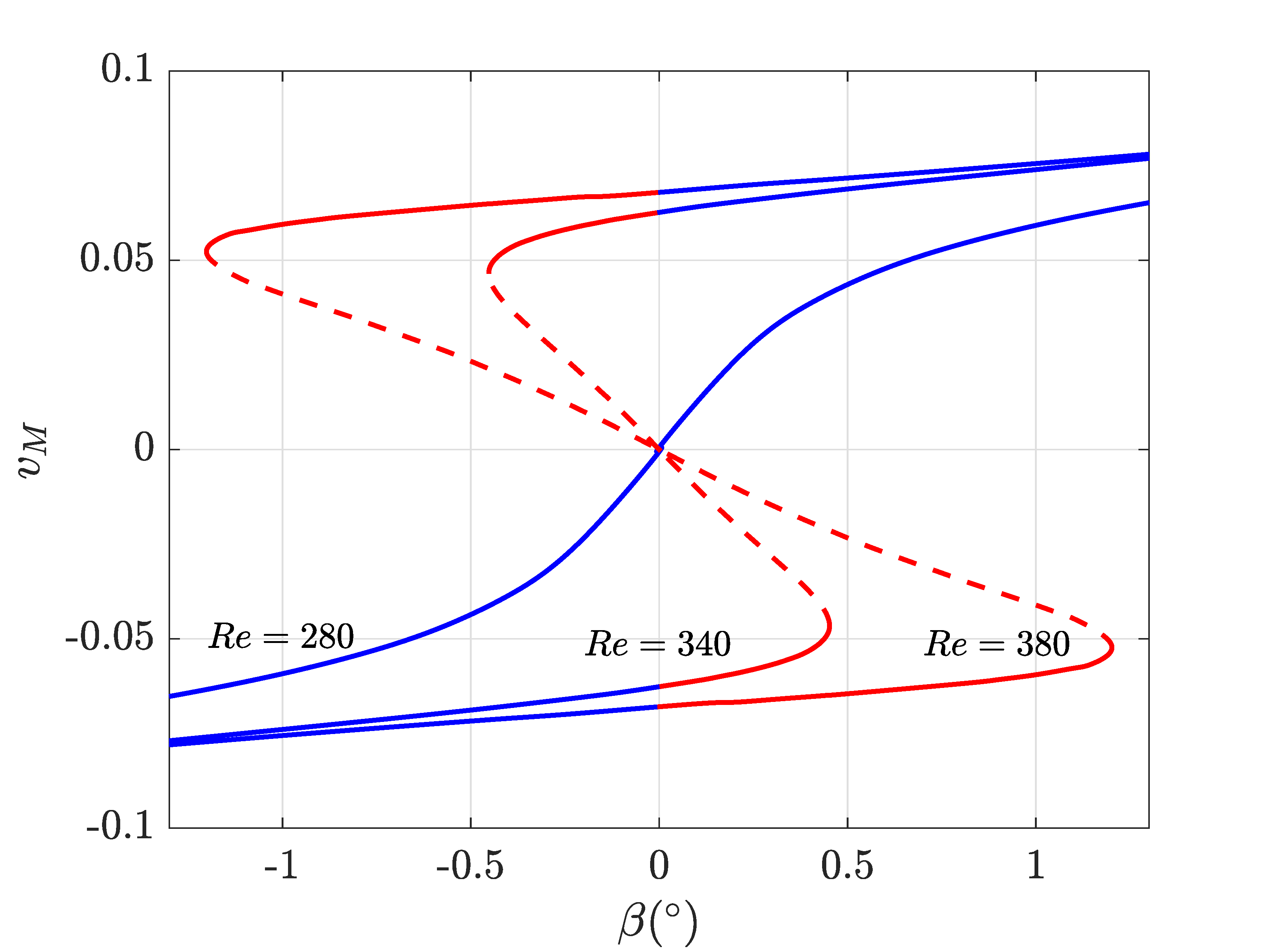}
\put(-2,65){\small (b)}
\end{overpic}
\end{subfigure}
\caption{(a) 3D sketch of an imperfection pitchfork bifurcation (cusp catastrophe). 
(b) Velocity measure at sensor $M$ for pure yaw at $Re =280$, $Re =340$ and $Re =380$.}
\label{splots}
\end{figure} 

To help understand the effect of $Re$ and $\alpha,\beta$, figure \ref{splots}(a) shows a sketch of a generic imperfect pitchfork bifurcation in the 3D space of control parameter, imperfection and amplitude.  
In this study, the control parameter is $Re$, the imperfection is the pitch angle $\alpha$ or yaw angle $\beta$, and the amplitude is the vertical velocity $w_M$ or the horizontal velocity $v_M$.
Solutions lie on the grey surface. Without imperfection, the system undergoes a perfect pitchfork as the control parameter increases (black lines). 
For a fixed non-zero imperfection, the pitchfork bifurcation becomes imperfect, with primary (blue) and secondary (red) branches, as already shown in figure \ref{fig:BF_P2Y0}. These branches are disconnected for all values of the control parameter. However, they can be connected by varying the imperfection parameter, as shown in figure \ref{splots}(b).
Note also that projecting the surface onto the plane of control parameter and imperfection shows the location of the saddle-nodes, which delimits two regions with either one or three solutions. This boundary has the shape of a cusp (hence the other name ``cusp catastrophe'' for imperfect pitchfork bifurcations) and will be discussed again in section \ref{WNLSection} when investigating the number of solutions when varying $Re$ and $\alpha,\beta$.
\section{Linear Stability}\label{LinStab}

In this section, we  study the linear stability of the steady base flows $\boldsymbol{q}_0$ computed in the previous section. Considering small-amplitude perturbations added to the base flow,
\begin{equation}
    \boldsymbol{q}_0(\boldsymbol{x}) = \boldsymbol{q}_0(\boldsymbol{x}) + \varepsilon \boldsymbol{q}_1(\boldsymbol{x}),
\end{equation}
with $0<\varepsilon \ll 1$, and injecting  into the steady NS equations yields at order $\varepsilon^1$ the linearised NS equations,
\begin{equation}
    \nabla\cdot\boldsymbol{u}_1 = 0, \quad  \mathcal{C}(\boldsymbol{u}_0,\boldsymbol{u}_1) +\nabla p_1 - \frac{1}{Re}\nabla^2 \boldsymbol{u}_1 = \boldsymbol{0},
    \label{eq:LNS}
\end{equation}
where $\mathcal{C}(\boldsymbol{a},\boldsymbol{b}) = (\boldsymbol{a}\cdot\nabla)\boldsymbol{b}+(\boldsymbol{b}\cdot\nabla)\boldsymbol{a}$ is the symmetric convection operator. 
We shall rewrite these equations as
\begin{equation}
    \mathcal{B} \partial_t \boldsymbol{q}_1 + \mathcal{L}\boldsymbol{q}_1 = \boldsymbol{0},
\end{equation}
where
\begin{equation}
    \mathcal{L} = \begin{pmatrix}
        \mathcal{C}(\boldsymbol{u}_0,\cdot) -Re^{-1}\nabla^2(\cdot)& \nabla(\cdot)\\
        \nabla\cdot(\cdot) & 0
    \end{pmatrix}, \quad
    \mathcal{B} = \begin{pmatrix}
        \mathcal{I}&0\\
        0& 0
    \end{pmatrix},
\end{equation}
and where $\mathcal{I}$ is the identity operator. 
Introducing the normal mode ansatz $\boldsymbol{q}_1 = \hat{\boldsymbol{q}}_1(\boldsymbol{x}) \text{e}^{\lambda t} + c.c.$, where $c.c.$ is the complex conjugate, yields the eigenvalue problem given by:
\begin{equation}
    \lambda\mathcal{B} \hat{\boldsymbol{q}}_1 + \mathcal{L}\hat{\boldsymbol{q}}_1  = \boldsymbol{0}.
\end{equation}
Eigenmodes are defined up to a complex-valued factor. In this study, we  normalise the modes as $\langle \mathcal{B} \hat{\boldsymbol{q}}_1,\hat{\boldsymbol{q}}_1 \rangle = \langle\hat{\boldsymbol{u}}_1,\hat{\boldsymbol{u}}_1\rangle = 1$. We define the inner product as $\langle \boldsymbol{a},\boldsymbol{b}\rangle = \int_\Omega (\boldsymbol{a}^{*}\cdot \boldsymbol{b}) \mathrm{d}\boldsymbol{x}$, where * stands for the complex conjugate.
We will also compute adjoint modes $\boldsymbol{q}^\text{\textdagger}_1$, which are a solution of
\begin{equation}
    \lambda\mathcal{B} \hat{\boldsymbol{q}}^\text{\textdagger}_1 + \mathcal{L}^\text{\textdagger}\hat{\boldsymbol{q}}^\text{\textdagger}_1  = \boldsymbol{0},
\end{equation}
where the adjoint NS operator $\mathcal{L}^\text{\textdagger}$ is defined by $\langle\mathcal{L}\textbf{a},\textbf{b}\rangle = \langle\textbf{a},\mathcal{L}^\text{\textdagger}\textbf{b}\rangle$, which yields
\begin{equation}
    \mathcal{L}^\text{\textdagger} = \begin{pmatrix}
        \mathcal{C}^\text{\textdagger}(\cdot,\boldsymbol{u}_0) -Re^{-1}\nabla^2(\cdot)& \nabla(\cdot)\\
        \nabla\cdot(\cdot) & 0,
    \end{pmatrix}
\end{equation}
where $\mathcal{C}^\text{\textdagger}(\boldsymbol{a},\boldsymbol{b}) = \nabla \boldsymbol{b}^T\cdot \boldsymbol{a} - (\boldsymbol{b}\cdot\nabla) \boldsymbol{a}$, or in index notation $\mathcal{C}_i^\text{\textdagger} = \partial_ib_j a_j - b_j\partial_j a_i$, is the (non-symmetric) adjoint convection operator. 

As we perform calculations in a half domain, for each base flow there are two families of eigenmodes with opposite symmetry properties. 
For example, for non-zero pitch, the vertical plane $y=0$ is a symmetry plane of the base flow, and eigenmodes can be either symmetric ($S_y$) or antisymmetric ($A_y$) with respect to that plane. The boundary conditions for the direct and adjoint modes in the case of non-zero pitch are hence given by
\begin{equation}
 \begin{split}
     S_y :& \quad \partial_y u_1 =  v_1 = \partial_y w_1 = 0 \quad  \text{on the $y=0$ plane}, 
     \\
     A_y :& \quad  u_1 =  \partial_y v_1 =  w_1 = p_1= 0 \quad  \text{on the $y=0$ plane}. \\
 \end{split}   
\end{equation}
Similarly, for non-zero yaw, the horizontal plane $z=0$ is a symmetry plane of the base flow, and eigenmodes can be either symmetric ($S_z$) or antisymmetric ($A_z$) with respect to that plane, in which case the boundary conditions are given by
\begin{equation}
 \begin{split}
     S_z :& \quad \partial_z u_1 =  \partial_z v_1 =  w_1 = 0 \quad \text{on the $z=0$ plane}, \\
     A_z :& \quad  u_1 =   v_1 = \partial_z w_1 = p_1 = 0 \quad \text{on the $z=0$ plane}.
 \end{split}   
\end{equation}
In both cases, the remaining boundary conditions are derived directly from the base flow. 
A homogeneous Dirichlet boundary condition, $\boldsymbol{u}_1 = \boldsymbol{0}$, is imposed at the inlet and lateral surfaces of the domain; 
no-slip boundary condition, $\boldsymbol{u}_1 = \boldsymbol{0}$, is imposed on the body surface and stress-free boundary condition $-p_1\boldsymbol{n} + Re^{-1}\nabla\boldsymbol{u}_1 \cdot \boldsymbol{n} = \boldsymbol{0}$ on the outlet plane.

The stationary modes $S_yA_z$ (mode $A$) and $A_yS_z$ (mode $B$) that become unstable near $Re \simeq 300$ for a perfectly aligned Ahmed body's doubly symmetric base flow are shown in Figure~\ref{fig:eigst_0_0}.

As mentioned in section \ref{BaseFlows}, adding pitch (resp. yaw) introduces a vertical (resp. horizontal) asymmetry 
and preselects one of the two branches of the (now imperfect) pitchfork bifurcation of mode $A$ (resp. $B$).
The deflected steady flow on the two branches may itself become unstable to perturbations. For example, figures~\ref{fig:eig_0_0}(a) and (b) show the unstable oscillatory eigenmode that lives on the primary branch at $Re=440$ for  $(\alpha,\beta)=(2^\circ,0^\circ)$ and  $(\alpha,\beta)=(0^\circ,2^\circ)$, respectively.

\begin{figure}[ht]
        \centering
\begin{subfigure}{0.49\textwidth}
\begin{overpic}[width=0.95\linewidth]{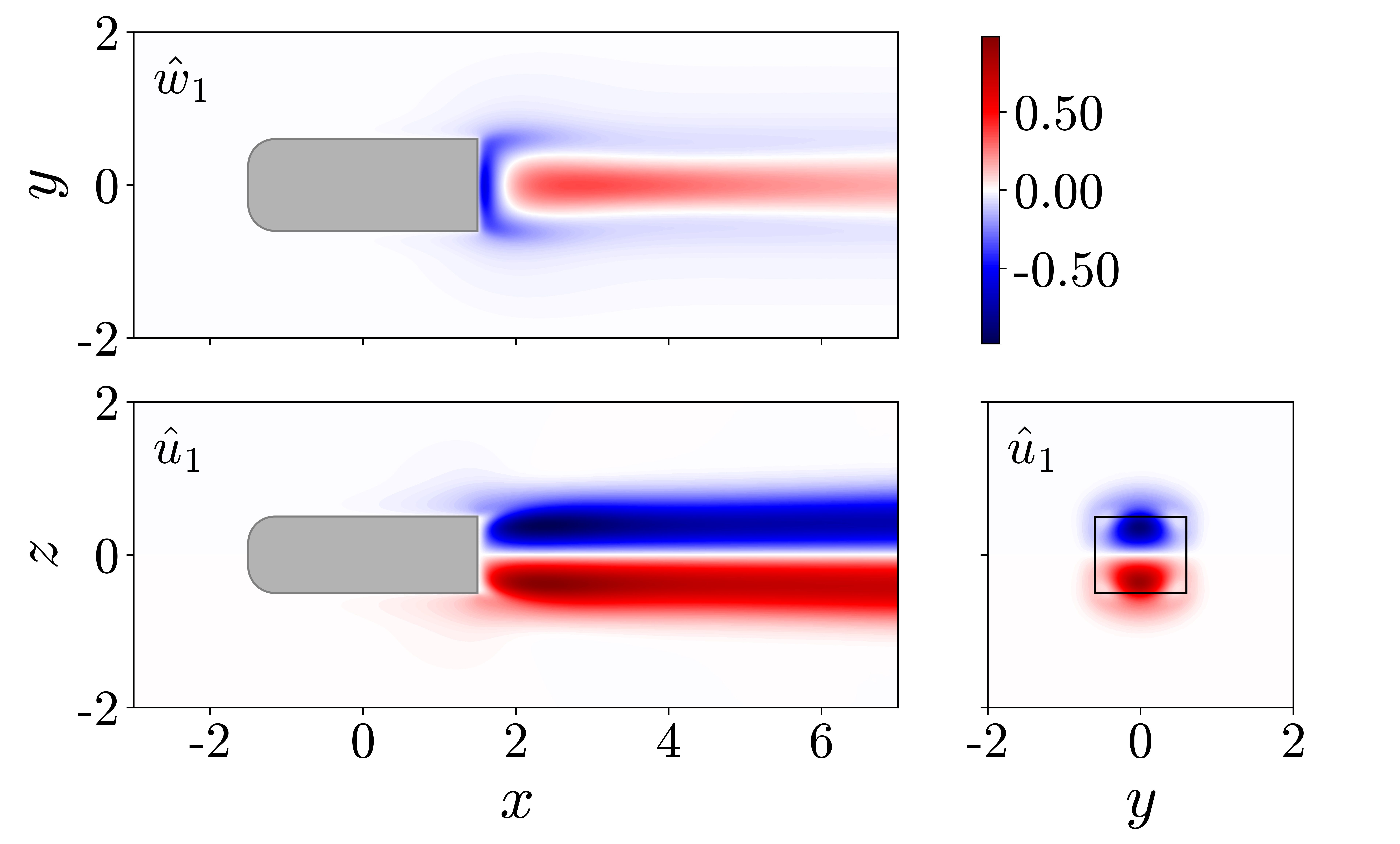} 
\put(-5,55){\small (a)}
\end{overpic}
\label{fig:subim1}
\end{subfigure}
\begin{subfigure}{0.49\textwidth}
\begin{overpic}[width=0.95\linewidth]{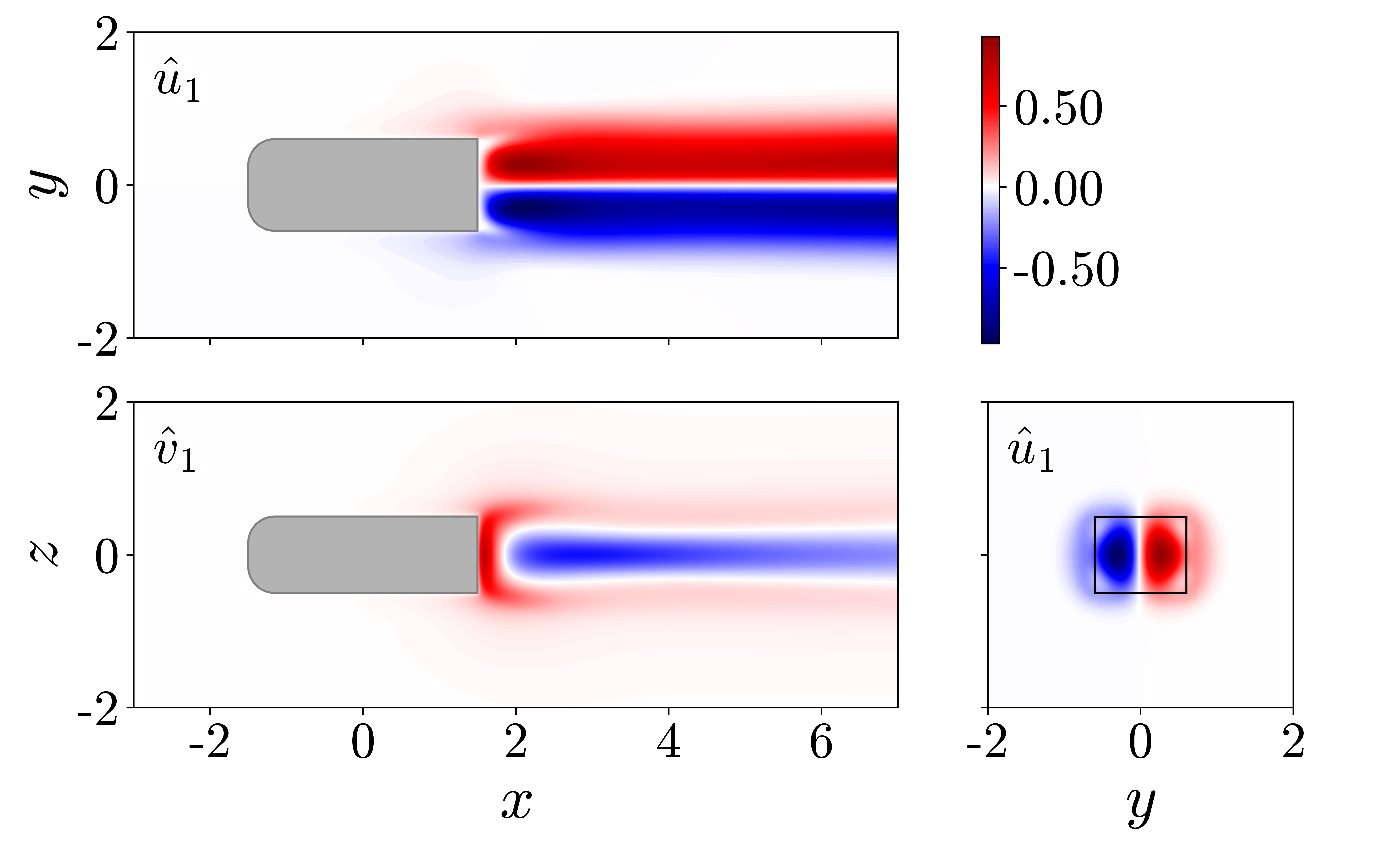}
\put(-5,55){\small (b)}
\end{overpic}
\label{fig:subim2}
\end{subfigure}
\caption{Stationary eigenmodes (a) $A$ and (b) $B$ for perfectly aligned fully-symmetric ($\alpha = 0^{\circ}, \beta = 0^{\circ}$) Ahmed body at $Re = 300$, in $z = 0$ plane (top view), $y=0$ plane (side view) and $x=2.5$ plane (rear view)}
\label{fig:eigst_0_0}
\end{figure}
\begin{figure}[ht]
        \centering
\begin{subfigure}{0.49\textwidth}
\begin{overpic}[width=0.95\linewidth]{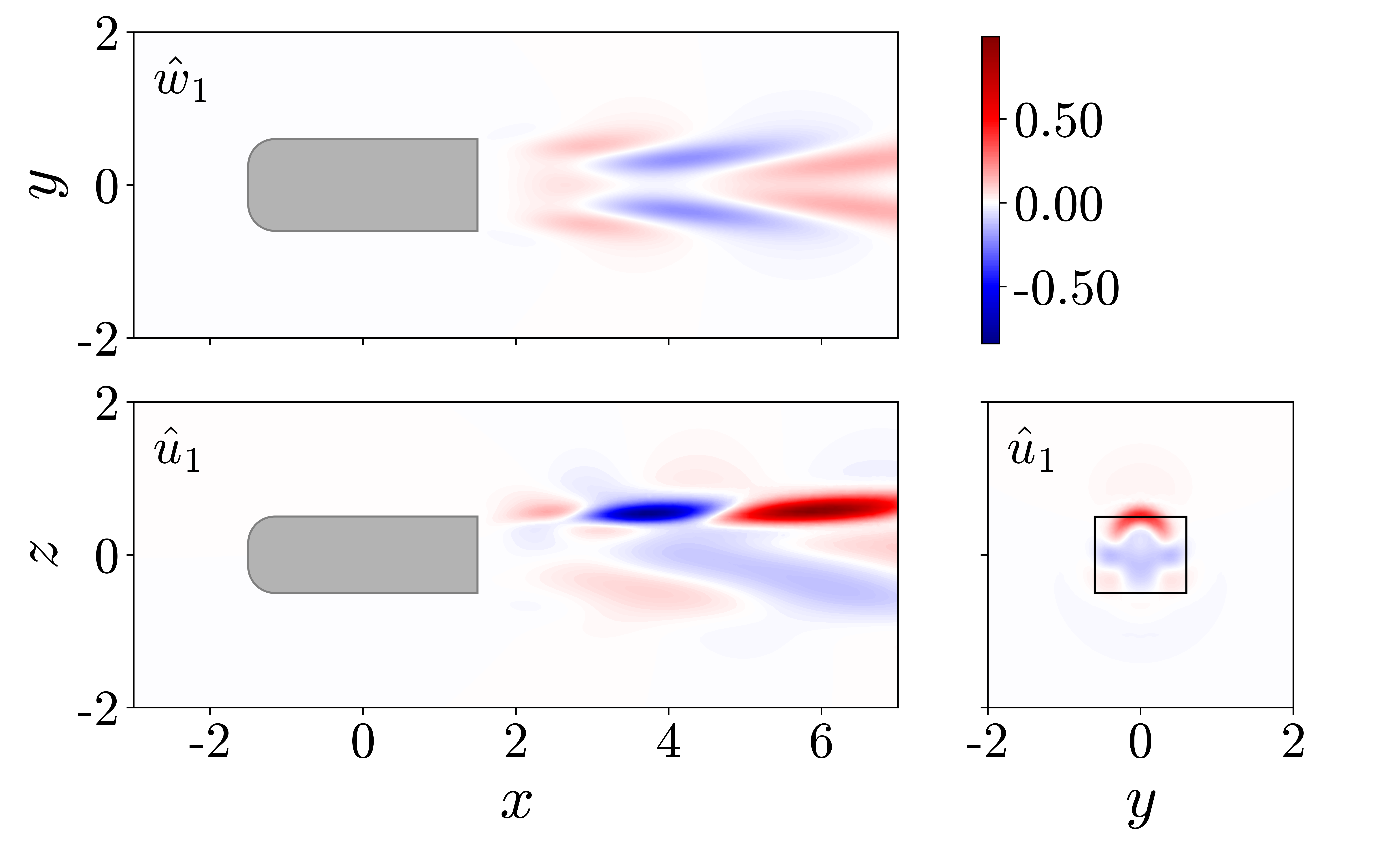} 
\put(-5,55){\small (a)}
\end{overpic}
\label{fig:subim1}
\end{subfigure}
\begin{subfigure}{0.49\textwidth}
\begin{overpic}[width=0.95\linewidth]{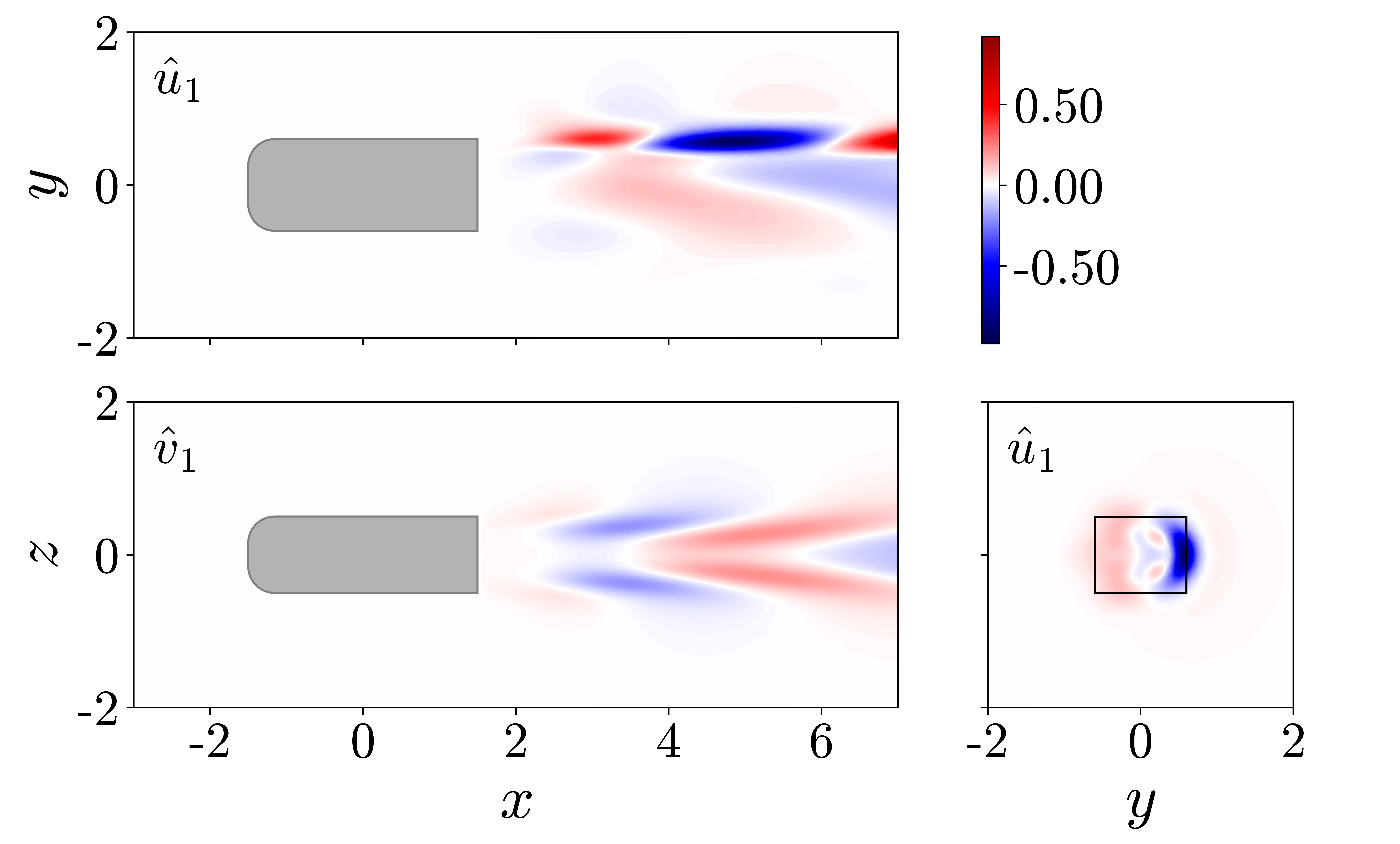}
\put(-5,55){\small (b)}
\end{overpic}
\label{fig:subim2}
\end{subfigure}
\caption{Oscillatory eigenmodes at $Re = 440$, in $z = 0$ plane (top view), $y=0$ plane (side view) and $x=5$ plane (rear view) for (a) $\alpha = 2^{\circ}, \beta = 0^{\circ}$, (b) $\alpha = 0^{\circ}, \beta = 2^{\circ}$}
\label{fig:eig_0_0}
\end{figure} 

We now consider either pure pitch or pure yaw and study the linear stability of the deflected flow on the primary and secondary branches of the resulting imperfect bifurcation. 

\begin{figure}[ht]
\centering
\begin{subfigure}{0.48\textwidth}
\begin{overpic}[height=0.75\linewidth,width=\linewidth]{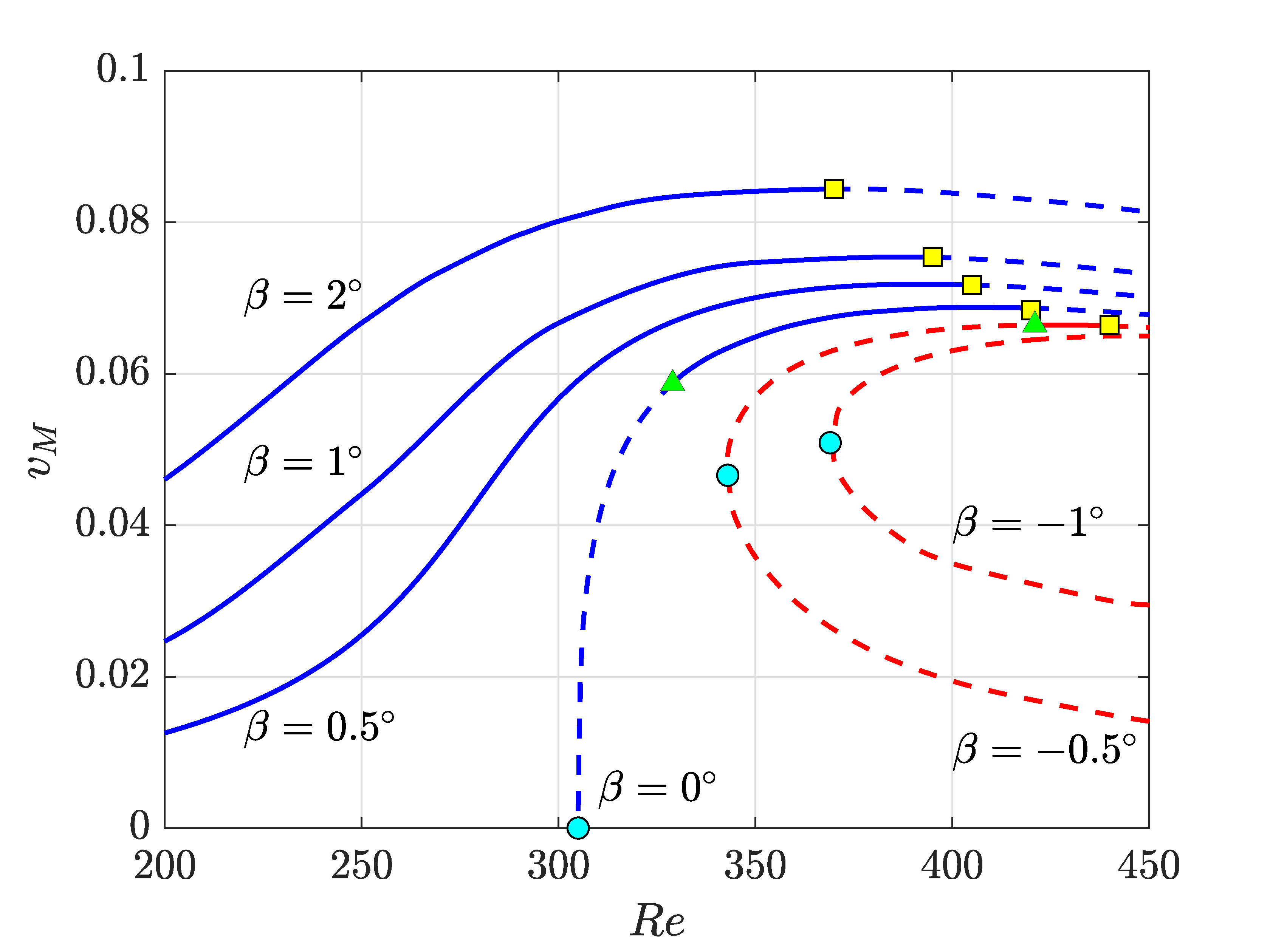}
\put(-2,65){\small (a)}
\end{overpic}
\end{subfigure}
\begin{subfigure}{0.48\textwidth}
\begin{overpic}[height=0.75\linewidth,width=\linewidth]{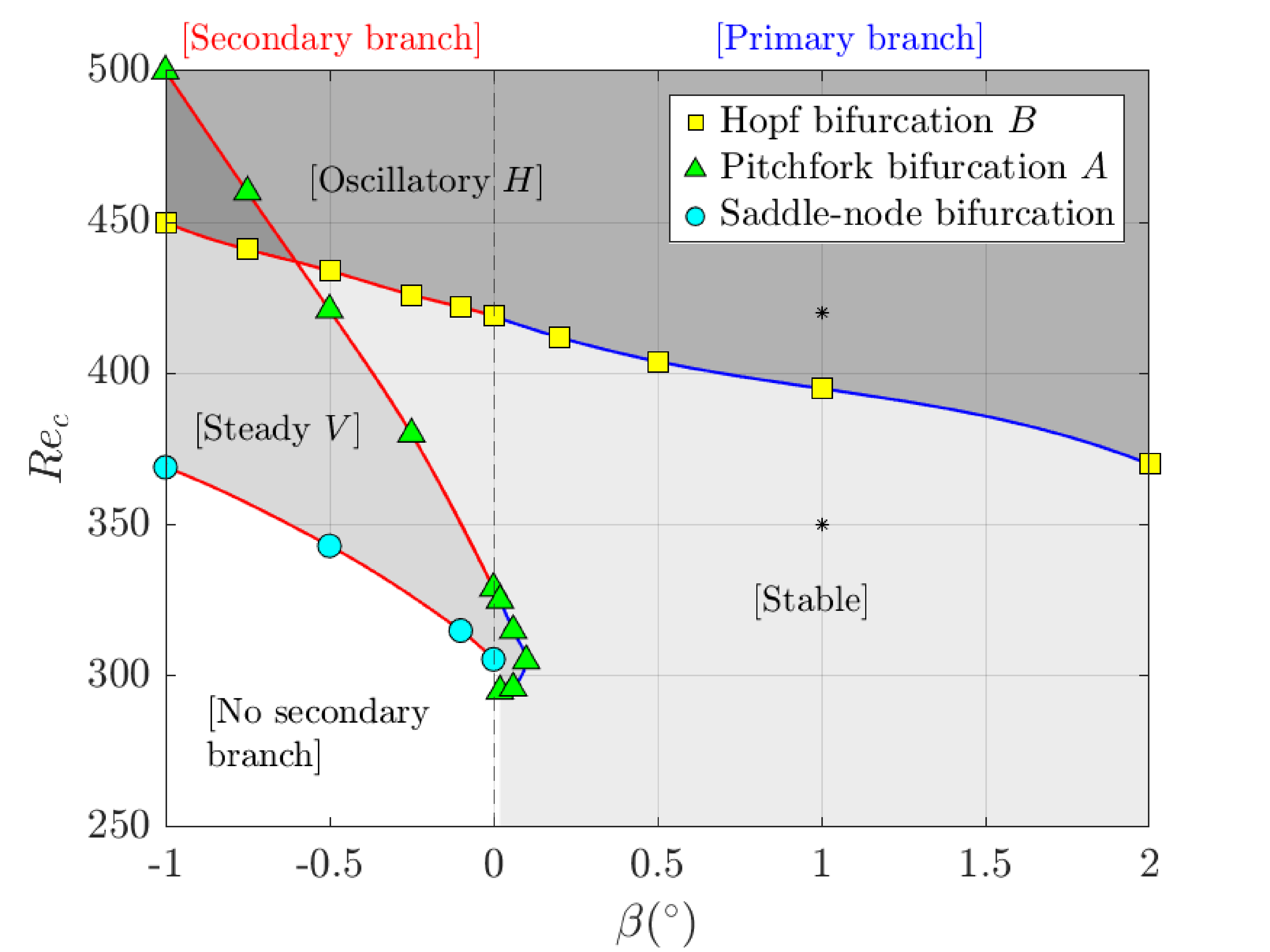}
\put(-2,65){\small (b)}
\end{overpic}
\end{subfigure}
\caption{Bifurcation diagram for pure yaw.
(a)~Horizontal velocity at sensor $v_M$ as function of Reynolds number, for several yaw angles. 
Bifurcations of the primary and secondary branches are marked with symbols (see legend in panel b). 
(b) Stability diagram. Red (left) and blue (right) correspond to the secondary and primary branches, as in panel a. $A$ and $B$ denote instability to perturbations preserving the horizontal and vertical symmetries, respectively. $V$ and $H$ denote vertical and horizontal states for the wake. The asterisks denote the parameters where the DNS calculations were performed, as compiled in table \ref{tab:dns}. 
}
\label{linstabyaw}
\end{figure}

\begin{figure}[ht]
\centering
\begin{subfigure}{0.48\textwidth}
\begin{overpic}[height=0.75\linewidth,width=\linewidth]{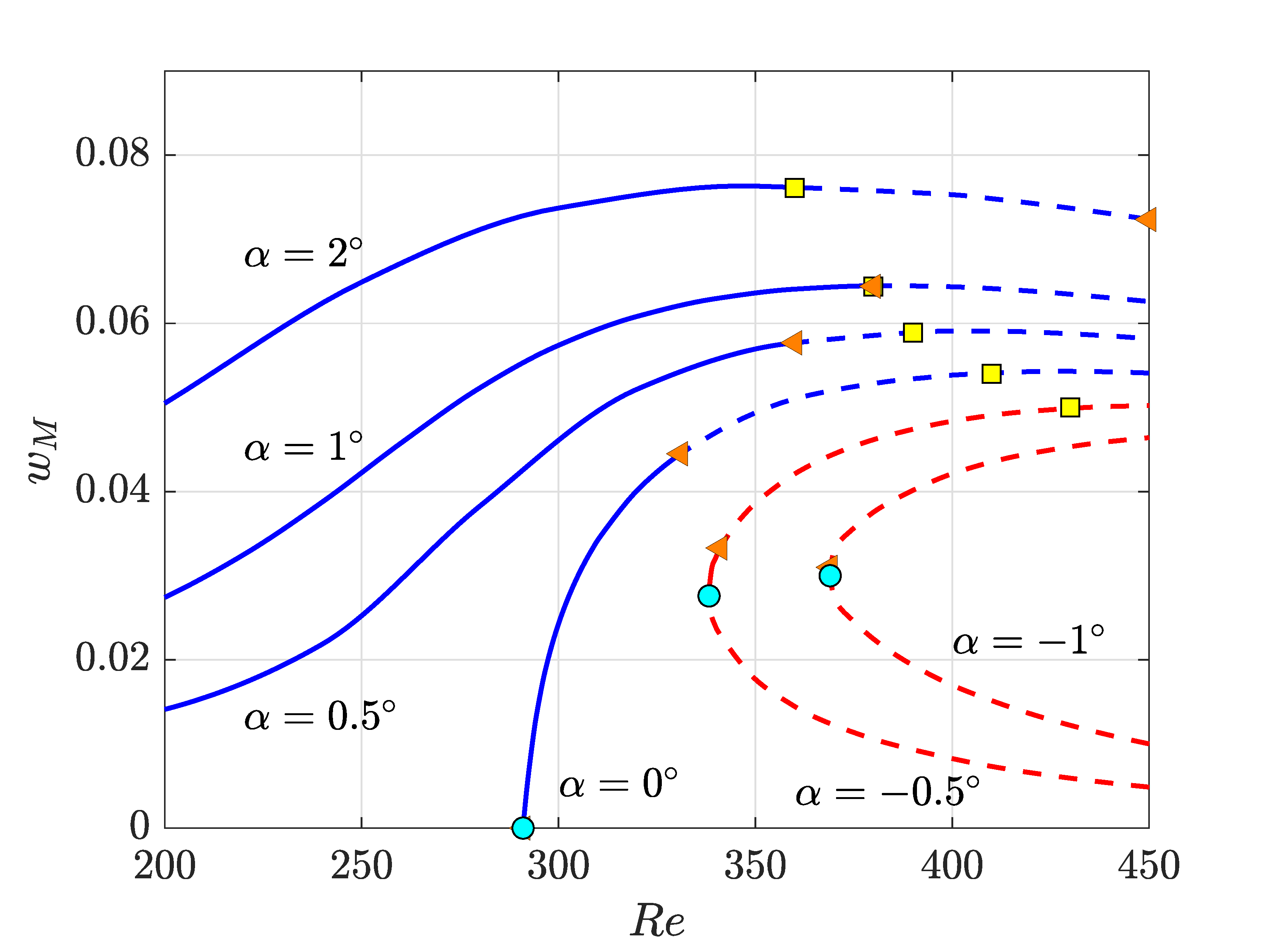}
\put(-2,65){\small (a)}
\end{overpic}
\end{subfigure}
\begin{subfigure}{0.48\textwidth}
\begin{overpic}[height=0.75\linewidth,width=\linewidth]{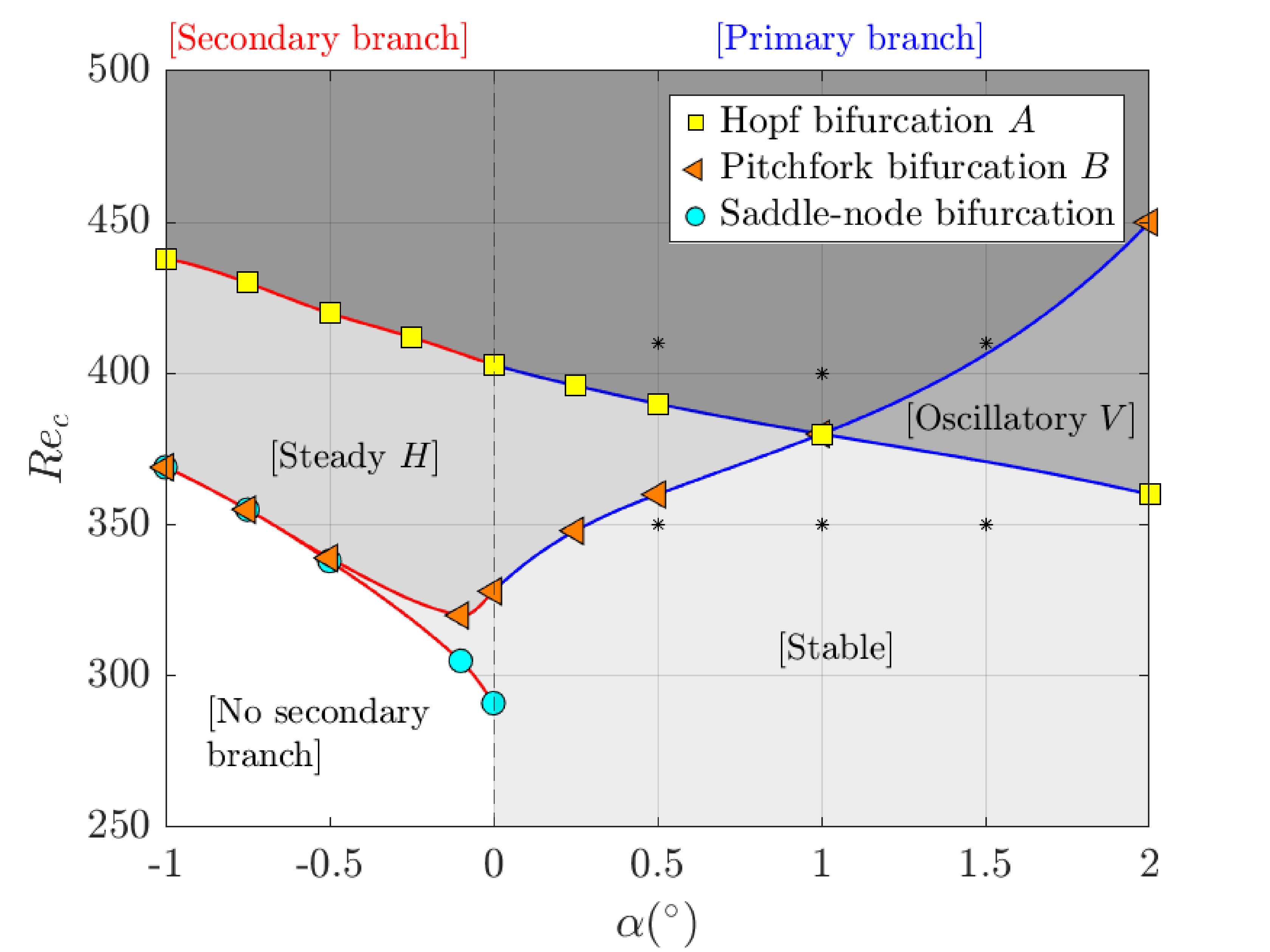}
\put(-2,65){\small (b)}
\end{overpic}
\end{subfigure}
\caption{
Bifurcation diagram for pure pitch.
(a)~Vertical velocity at sensor $w_M$ as function of Reynolds number, for several pitch angles. 
Bifurcations of the primary and secondary branches are marked with symbols (see legend in panel b). 
(b) Stability diagram. Red (left) and blue (right) correspond to the secondary and primary branches, as in panel a. $A$ and $B$ denote instability to perturbations preserving the horizontal and vertical symmetries, respectively. $V$ and $H$ denote vertical and horizontal states for the wake. The asterisks denote the parameters where the DNS calculations were performed, as compiled in table \ref{tab:dns}. 
}
\label{linstabpitch}
\end{figure}

\subsection{Pure yaw scenario: $\alpha = 0^{\circ}, \beta \neq 0^{\circ}$}

Figure 7 shows the results of the linear stability analysis for pure yaw.
Figure 7(a) shows the primary (blue) and secondary (red) branches of the imperfect pitchfork as a function of $Re$, for several yaw angles. To obtain a compact representation with $v_M>0$, we report the primary branch for $\beta>0$ and the secondary branch for $\beta<0$. 
As expected, as $|\beta|$ increases, the deflection becomes stronger on the primary branch, and the saddle-node of the secondary branch moves to larger $Re$. We recall figure \ref{splots} to better understand the primary and secondary branches.
Symbols show the critical Reynolds number of the different instabilities found along these branches.
Figure 7(b) summarises the variation of $Re_c$ with $\beta$ in a stability diagram. 

For the perfectly aligned case (\cite{Zampogna_Boujo_2023}), we recover the horizontal-symmetry-breaking pitchfork bifurcation of mode $B$ at $Re\simeq305$ (blue circle). Initially, the resulting horizontally deflected wake is itself unstable to stationary vertical-symmetry-breaking perturbations, because mode $A$ is unstable too, but becomes stable at $Re\simeq330$ (green triangle). Further increasing $Re$  up to $Re \simeq 420$, the horizontally deflected wake becomes unstable to a horizontal vortex shedding mode (yellow square).
When yaw is introduced, the pitchfork bifurcation of mode $B$ disappears on the primary branch, as the flow is already horizontally asymmetric, and becomes the saddle-node of the secondary branch.
The horizontal Hopf bifurcation can be followed continuously on both primary and secondary branches through $\beta=0$. Increasing yaw destabilises this oscillatory mode, which becomes the first unstable mode encountered on the primary branch as soon as $\beta>0.1^\circ$.
Finally, the outer secondary branch is always unstable to a stationary vertical eigenmode which, similar to the $\beta=0$ case, restabilises via a pitchfork bifurcation. The $Re_c$ of this restabilisation increases strongly with $|\beta|$. As a result, for $-0.6^\circ<\beta<0^\circ$ the outer secondary branch is stable in some interval of $Re$ delimited by pitchfork and Hopf bifurcations, while it becomes unconditionally unstable for $\beta<-0.6^\circ$. The inner secondary branch is unconditionally unstable at any yaw angle, similar to standard imperfect pitchfork bifurcations.

\subsection{Pure pitch scenario: $\alpha \neq 0^{\circ}, \beta = 0^{\circ}$}

We now turn our attention to pure pitch configurations.
Figure 8(a) shows the primary (blue) and secondary (red) branches of the imperfect pitchfork as a function of $Re$, for several pitch angles. Similar to sec. 4.1, to obtain a compact representation with $w_M>0$, we report the primary branch for $\alpha>0$ and the secondary branch for $\alpha<0$. 
Again, as $|\alpha|$ increases, the deflection become stronger on the primary branch, and the saddle-node of the secondary branch moves to larger $Re$.
Figure 8(b) summarises the variation with $\alpha$ of the critical Reynolds number of the different instabilities found along these branches. 

For the perfectly aligned case (\cite{Zampogna_Boujo_2023}), we recover the vertical-symmetry-breaking pitchfork bifurcation of mode $A$ at $Re\simeq 295$ (blue circle). Initially, the resulting vertically deflected wake is linearly stable, but becomes unstable to stationary horizontal-symmetry-breaking perturbations at $Re\simeq330$ (orange triangle), once mode $B$ is sufficiently unstable. 
This unstable state undergoes an additional instability at $Re \simeq 410$: a vertical vortex shedding mode (yellow square).

When pitch is introduced, the pitchfork bifurcation of mode $A$ disappears on the primary branch, as the flow is already vertically asymmetric, and becomes the saddle-node of the secondary branch. 
Both the vertical pitchfork and vertical Hopf bifurcations can be followed continuously on both primary and secondary branches through $\beta=0^\circ$. Decreasing pitch ($\alpha<0$) stabilises both bifurcations on the outer secondary branch, but the critical Reynolds number of the pitchfork bifurcation quickly approaches the $Re$ of the saddle-node, so this branch is unconditionally unstable for $|\alpha|>0.5^\circ$.
Increasing pitch ($\alpha>0$) stabilises the pitchfork bifurcation and destabilises the Hopf bifurcation, so the  vertical vortex shedding mode becomes the first unstable mode encountered on the primary branch when $\alpha>1^\circ$. 
Finally, the inner secondary branch is unconditionally unstable at any pitch angle, similar to standard imperfect pitchfork bifurcations. 

\subsection{DNS results}
\label{sec:DSN}

Fully nonlinear direct numerical simulations were carried out using the same numerical method as in \cite{Chiarini_Boujo_2025}, with the aim of (i)~validating the linear stability results, and (ii)~investigating the crossover of the stability curves for the pure pitch case.

\subsubsection{Validation of linear stability analysis}
\label{sec:DSN_valid}
Table~\ref{tab:dns} summarises the results of the DNS for different values of pitch, yaw and Reynolds number.
In all cases, these DNS confirm that the wake is steady at low Reynolds number $Re=350$, with a vertical or horizontal deflection in the pure yaw and pure pitch cases, respectively.
For $(\alpha,\beta)=(0.5^\circ,0^\circ)$ and $(1^\circ,0^\circ)$, increasing $Re$ leads to a steady wake that is now deflected in both vertical and horizontal directions. This is in line with the critical $Re$ of the horizontal pitchfork bifurcation predicted by linear stability analysis (fig.~\ref{linstabpitch}). As $Re$ is above the predicted Hopf bifurcation too, this also suggests that the nonlinear interaction between the stationary and oscillatory modes promotes the former.
By contrast, for $(\alpha,\beta)=(1.5^\circ,0^\circ)$, increasing $Re$ to $410$ leads to an unsteady periodic wake, with both the oscillations and the mean position being non-zero in the vertical and horizontal directions.
This confirms that both the horizontal stationary mode and vertical oscillatory mode are unstable, in line with the linear stability analysis, and that nonlinear interactions lead to a superposition of the two. 
The Strouhal number predicted by linear stability at $\alpha = 1.5^\circ$, $\beta = 0^\circ$, $Re=410$ is $St_{LSA} = 0.111$ and matches well that obtained with the DNS, $St_{DNS} = 0.102$.
Finally, simulations for $(\alpha,\beta)=(0^\circ,1^\circ)$ lead to an unsteady periodic wake with horizontal oscillations about a mean horizontal deflection, confirming that a horizontal oscillatory mode becomes unstable as predicted by linear stability (fig.~\ref{linstabyaw}). 
The Strouhal number predicted by linear stability at $\alpha = 0^\circ$, $\beta = 1^\circ$, $Re=420$  is $St_{LSA} = 0.0943$ and matches well that obtained with the DNS, $St_{DNS} = 0.0902$.

\subsubsection{Crossover of Hopf and pitchfork bifurcations for $(\alpha,\beta)=(1^\circ,0^\circ)$}

We now take a closer look at the competition between the vertical oscillatory and horizontal stationary modes in the immediate vicinity of their crossover at $(\alpha,\beta)=(1^\circ,0^\circ)$ and $Re \simeq 380$.
As mentioned in section~\ref{sec:DSN_valid}, a first simulation at $Re=350$ (yellow circle in figure~\ref{dns2}(a)) converges to a stationary flow, as expected from linear stability results.
This flow has some vertical asymmetry but preserves the horizontal symmetry: $w_M\neq0$, $v_M=0$ in figure~\ref{dns2}(b).
Next, using this flow as an initial condition, the Reynolds number is increased to $Re=400$ (green circle in figure~\ref{dns2}(a)). Initially, the wake  starts to oscillate vertically and remains horizontally symmetric, as shown in figure~\ref{dns2}(c) for $t \leq 500$. This is consistent with the vertical Hopf bifurcation described earlier.
For $t \geq 600$, however, vertical oscillations  disappear very slowly while a horizontal deflection $v_M\neq 0$ appears, and the flow finally settles to a static horizontal wake deflection. 
This, in turn, is consistent with the horizontal pitchfork bifurcation described earlier. 
It also shows that, while both modes are unstable, their nonlinear interaction unfolds into slow dynamics and, eventually, the stationary mode alone prevails.

\begin{table}
  \begin{center}
\def~{\hphantom{0}}
  \begin{tabular}{cc c c c cc cl}
      $\alpha(^\circ)$ & $\beta(^\circ)$ && $Re$ && $w_M$ & $v_M$ && Flow state \\ \\[3pt]
       0.5 & 0 && 350 && -0.053  & 0~        && Steady V deflection due to pitch\\
           &   && 410 && -0.018  & 0.067~    && Additional steady H deflection post bifurcation
       \\ 
       \\ 
       1.0 & 0 && 350 && -0.058  & 0~        && Steady V deflection due to pitch\\
           &   && 400 && -0.036  &  0.0534~  && Additional steady H deflection post bifurcation\\ 
       \\ 
       1.5 & 0 && 350 && -0.066~  & 0~        && Steady V deflection due to pitch\\
           &   && 410 && -0.044* & 0.065*   && V and H oscillations about mean V and H deflection\\
       \\ 
       0 & 1.0 && 350 && 0       & -0.068~   && Steady H deflection due to yaw\\
           &   && 420 && 0       & -0.072*  && H oscillations about mean H deflection\\
  \end{tabular}
  \caption{Summary of the DNS study conducted for pure pitch and pure yaw cases (see also asterisks in figures \ref{linstabyaw}(b) and \ref{linstabpitch}(b)).
  V and H denote vertical and horizontal directions, respectively.
  Stars indicate mean values.}
  \label{tab:dns}
  \end{center}
\end{table}
The same scenario has been observed in the laminar experiments of \cite{Grandemange12PRE} and numerical simulations by \cite{Evstafyeva17}, without pitch but with a fixed ground below the Ahmed body: increasing the Reynolds number to $Re=365$, the wake started to oscillate vertically and remained centred horizontally, before evolving very slowly to a statically deflected state. 
This very slow evolution took place over approximately $500~s$, i.e. several hundreds of convective time units $H/U_\infty$ (in their experiment in water, $H=26$~mm, $U_\infty \simeq 1.4$~cm/s, so  $H/U_\infty \simeq 1.9$~s). 
This suggests that pitch and ground have a qualitatively similar effect on the bifurcations of the Ahmed body wake: by introducing a vertical imperfection, they stabilise the horizontal pitchfork bifurcation and destabilise the vertical Hopf bifurcation (compared to the no-ground, perfectly aligned case studied in \cite{Zampogna_Boujo_2023}); this can lead to a crossover and to a slow competition between the two unstable eigenmodes.
Of course, it remains to be seen whether this generic scenario is robust to the flow configuration (ground clearance, thickness of the incoming ground boundary layer, fixed / moving ground).

\begin{figure}
    \centering
    \begin{overpic}[width=1\linewidth]{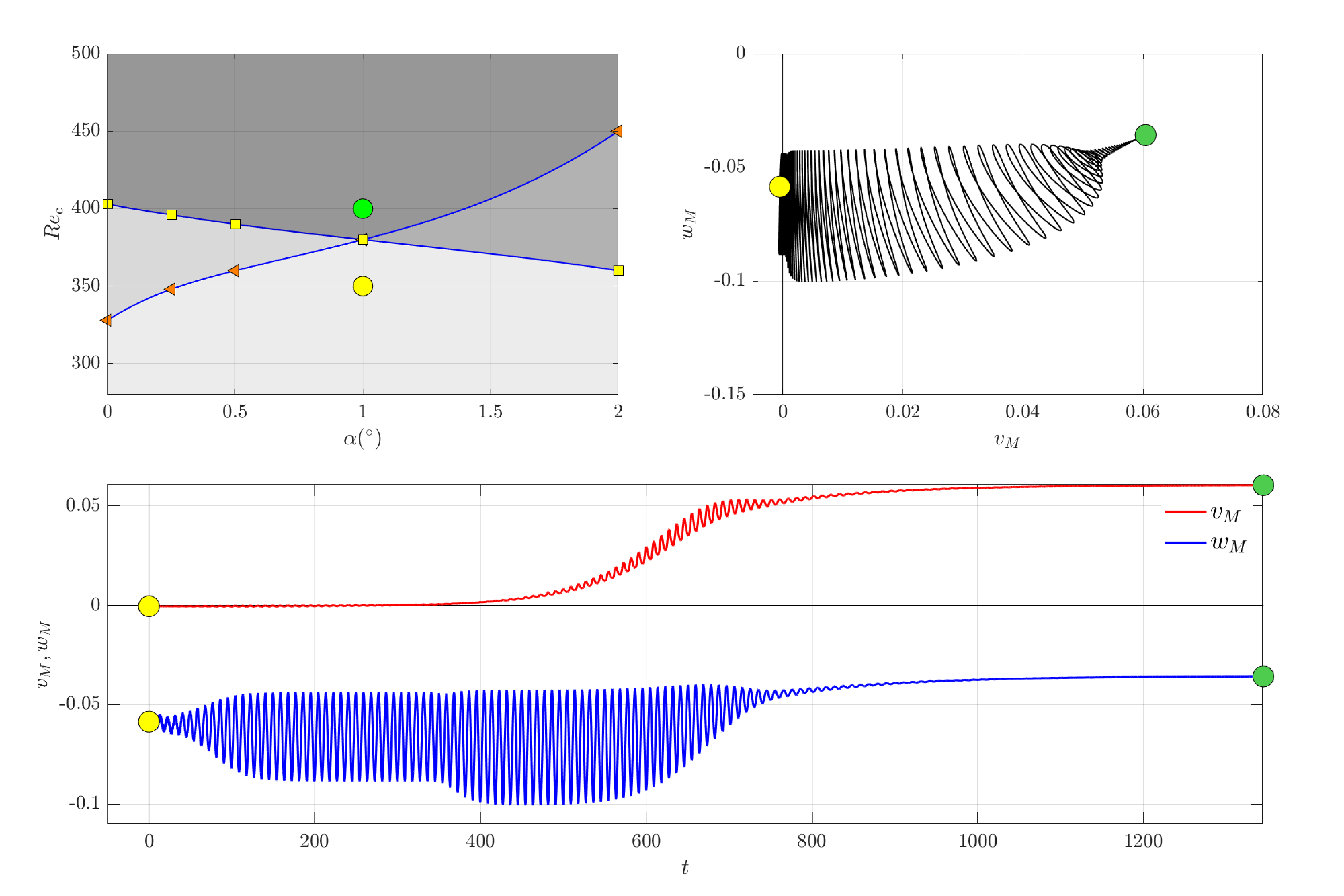}
     \put(-2,60){\small (a)}
     \put(50,60){\small (b)}
     \put(-2,30){\small (c)}
    \end{overpic}
    \caption{
    Competition between vertical vortex shedding and horizontal static deflection for $\alpha=1^\circ$ and $Re \simeq 380$.
    (a)~Linear stability diagram (zoom of figure~\ref{linstabpitch}(b)) highlighting the two Reynolds numbers $Re=350$ and $Re=400$ used in the DNS. 
    (b)~DNS evolution of the horizontal velocity $v_M$ and vertical velocity $w_M$ at sensor $M$ for $Re=400$, starting from the flow at $Re=350$.
    (c)~Corresponding time signals $v_M(t)$ and $w_M(t)$.}   
    \label{dns2}
\end{figure}

\section{Weakly nonlinear analysis}\label{WNLSection}
In general, when the flow has multiple bifurcations, the competition between the unstable modes is of interest. Fully nonlinear computations require the use of the complete computation domain and subsequently are computationally heavy, especially when both pitch and yaw are non-zero. Even after obtaining the base flow, linear stability analysis cannot predict the flow when multiple modes are unstable. 
We are hence motivated to develop a weakly nonlinear (WNL) method to understand the flow behaviour in presence of both pitch and yaw and multiple unstable modes.

Previous studies have used a weakly nonlinear analysis to study one or multiple bifurcations in wake flows.
\cite{sipp_lebedev_2007} used the multiple-scale method to perform a WNL analysis of the Hopf bifurcation in the wake of a 2D circular cylinder. 
\cite{MELIGA_CHOMAZ_SIPP_2009} described the nonlinear interactions between a stationary $m=1$ mode and an oscillatory $m=1$ mode in the wake of a thin disk, and considered the imperfection resulting from a steady volume forcing.
\cite{Carini_Auteri_Giannetti_2015} used the center-manifold approach to describe the nonlinear interaction between a steady  mode and an oscillatory mode in the wake of two side-by-side 2D circular cylinders.
\cite{Citro_Tchoufag_Fabre_Giannetti_Luchini_2016} performed a WNL analysis of the primary instability in the wake of a rotating sphere; inherently, due to the asymmetry induced by the rotation, the result is an imperfect bifurcation. 
\cite{Zampogna_Boujo_2023} and \cite{Chiarini_Boujo_2025} used the multiple-scale method to capture the nonlinear interaction between the two stationary modes $A$ and $B$ in the wakes of Ahmed bodies and rectangular prisms, respectively; in the latter case, the super- or subcritical nature of the bifurcation was found to depend on the aspect ratio of the prism. However, the effect of a geometric imperfection was not considered.

\subsection{Derivation}
\label{sec:WNL_derivation}

In this section, we build upon the weakly nonlinear method developed by \cite{Zampogna_Boujo_2023} for the perfectly aligned case, and extend it to include misalignment induced by pitch and yaw. The specific geometry considered in this study has two stationary modes, referred to as mode $A$ and mode $B$, of symmetries $S_yA_z$ and $A_yS_z$ and  that bifurcate at slightly different critical Reynolds numbers. 
To perform the WNL analysis for both modes at the same reference Reynolds number $Re_c$, we introduce a shift operator $\mathcal{S}$
(\cite{MELIGA_CHOMAZ_SIPP_2009}, \cite{Zampogna_Boujo_2023}).
In general,  $Re_c$ is chosen between the critical Reynolds number of the two modes, but the specific value has little effect on the amplitude equations (\cite{Zampogna_Boujo_2023}).
We quantify the departure from criticality as:
\begin{equation}
    \frac{1}{Re_c} - \frac{1}{Re} = \delta = \epsilon^2\tilde{\delta}.
\end{equation}
We also introduce a small misalignment via pitch and yaw angles,
\begin{equation}
    \alpha = \epsilon^3 \tilde{\alpha}, \quad\quad \beta = \epsilon^3 \tilde{\beta}.
\end{equation}
The choice of the misalignment appearing at order $\epsilon^3$ is motivated by the slow time and the first resonant terms appearing at that order. This allows us to close the problem at order $\epsilon^3$ and construct the coupled amplitude equations (\cite{Citro_Tchoufag_Fabre_Giannetti_Luchini_2016,MIZUSHIMA_SHIOTANI_2000}).
As will become clearer later, we emphasise that a small-amplitude roll (rotation about the $x$-axis shown in figure \ref{fig:geometry}) does not affect modes $A$ and $B$ at the leading order of our WNL expansion.
It is important to note that the weakly nonlinear analysis is rigorously valid in the vicinity of the expansion point, that is, for $\epsilon \ll 1$, $Re$ close to $Re_c$ and  small pitch and yaw angles.
Following the multiple scale approach, we introduce a slow time scale $T_1=\epsilon^2 t$, 
inject in the NS equations an expansion of the flow field,
\begin{equation}
    \boldsymbol{u} = \boldsymbol{u}_0 + \epsilon \boldsymbol{u}_1 + \epsilon^2 \boldsymbol{u}_2 + \epsilon^3 \boldsymbol{u}_3 \dots,
\end{equation}
and collect like-order terms to obtain a series of problems to be solved sequentially.
In general, the rotated geometry depends on the order in the sequence of rotations. However, for small angles, the rotation matrices are commutative. Thus, each point $\boldsymbol{x}_{0,0}$ of the perfectly aligned body corresponds to a point $\boldsymbol{x}_{\alpha,\beta}$ of the rotated body, as follows:
\begin{equation}
    \boldsymbol{x}_{\alpha,\beta} = \mathbf{R}_\alpha\mathbf{R}_\beta \boldsymbol{x}_{0,0},
\end{equation}
where the pitch and yaw rotation matrices read
\begin{equation}
    \mathbf{R}_\alpha = \begin{pmatrix}
                        \cos\alpha & 0 & \sin\alpha\\
                        0 & 1 & 0\\
                        -\sin\alpha & 0 & \cos\alpha
                        \end{pmatrix}, \quad
    \mathbf{R}_\beta = \begin{pmatrix}
                        \cos\beta  & \sin\beta & 0\\
                        -\sin\beta & \cos\beta & 0\\
                        0  & 0  & 1                     
                        \end{pmatrix}.
\end{equation}
For small angles, the combined rotation matrix simplifies to
\begin{equation}
\begin{split}
    \mathbf{R}_\alpha\mathbf{R}_\beta & \simeq \mathbf{R}_\beta\mathbf{R}_\alpha  \simeq               \begin{pmatrix}
                        1 & \epsilon^3\tilde{\beta} & \epsilon^3\tilde{\alpha}\\
                        -\epsilon^3\tilde{\beta} & 1 & 0\\
                        -\epsilon^3\tilde{\alpha} & 0 & 1
                        \end{pmatrix}   + \mathcal{O}(\epsilon^6) 
                 \doteq  \mathbf{I} + \epsilon^3 \tilde{\mathbf{R}} + \mathcal{O}(\epsilon^6).
\end{split}
\end{equation}
The no-slip boundary condition on the wall of the rotated body reads 
\begin{equation}
    \boldsymbol{u}(\boldsymbol{x}_{\alpha,\beta}) = \boldsymbol{0} =  \boldsymbol{u}_0(\boldsymbol{x}_{\alpha,\beta}) + \epsilon \boldsymbol{u}_1(\boldsymbol{x}_{\alpha,\beta})
    +\epsilon^2 \boldsymbol{u}_2(\boldsymbol{x}_{\alpha,\beta}) +\epsilon^3 \boldsymbol{u}_3(\boldsymbol{x}_{\alpha,\beta})  + \mathcal{O}(\epsilon^4),
\end{equation}
which can be ``flattened'' onto the perfectly aligned geometry via a Taylor expansion:
\begin{equation}
\begin{split}
    \boldsymbol{u}(\boldsymbol{x}_{\alpha,\beta}) 
    & 
    = \boldsymbol{u}\left( (\mathbf{I} + \epsilon^3 \tilde{\mathbf{R}} + \mathcal{O}(\epsilon^6)) \boldsymbol{x}_{0,0} \right) 
    \\
    &= \boldsymbol{u}(\boldsymbol{x}_{0,0}) + \epsilon^3 (\nabla \boldsymbol{u})_{|\boldsymbol{x}_{0,0}} (\mathbf{\tilde{R}}\boldsymbol{x}_{0,0}) + \mathcal{O}(\epsilon^6) 
    \\
    & = \left(\boldsymbol{u}_0 + \epsilon \boldsymbol{u}_1 +\epsilon^2 \boldsymbol{u}_2 +\epsilon^3 \boldsymbol{u}_3  + \mathcal{O}(\epsilon^4) \right)_{|\boldsymbol{x}_{0,0}}     
    + \epsilon^3 (\nabla\boldsymbol{u}_0     
    + \mathcal{O}(\epsilon))_{|\boldsymbol{x}_{0,0}}
    (\mathbf{\tilde{R}}\boldsymbol{x}_{0,0}).         
\end{split}
\end{equation}
We can therefore identify the following boundary conditions at each order:
\begin{equation}
\hspace{-6mm}
    \boldsymbol{u}_0(\boldsymbol{x}_{0,0}) = \boldsymbol{u}_1(\boldsymbol{x}_{0,0}) = \boldsymbol{u}_2(\boldsymbol{x}_{0,0}) = \boldsymbol{0},
\end{equation}
\begin{equation}
    \boldsymbol{u}_3(\boldsymbol{x}_{0,0}) = -(\nabla \boldsymbol{u}_0)_{|\boldsymbol{x}_{0,0}}\begin{pmatrix}
                        \tilde{\beta}y_{0,0} + \tilde{\alpha}z_{0,0} \\
                        -\tilde{\beta}x_{0,0}\\
                        -\tilde{\alpha}x_{0,0}
                        \end{pmatrix} \doteq \tilde{\boldsymbol{u}}. 
                        \label{eq:u_tilde} 
\end{equation}
We can now proceed with the WNL analysis.
The detailed derivation at orders $\epsilon^0$, $\epsilon^1$ and $\epsilon^2$ follows the standard WNL methodology and is independent of $\alpha$ and  $\beta$ (see details in appendix \ref{appA}). 
The effect of the imperfection appears at order $\epsilon^3$. 
The third-order field is a solution of forced linearised NS equations,
\begin{equation}
    \partial_t \boldsymbol{u}_3 + \tilde{\mathcal{L}}\boldsymbol{u}_3 = \mathcal{F}_3.
\end{equation}
The forcing terms in $\mathcal{F}_3$ are the same as in the perfectly aligned case; in particular, all of them are either $S_y A_z$-symmetric or  $A_y S_z$-symmetric and are therefore resonant with mode $A$ or $B$, respectively.
Unlike the perfectly aligned case, however, the boundary condition (\ref{eq:u_tilde}) on the body wall is not homogeneous: $\boldsymbol{u}_3 = \tilde{\boldsymbol{u}}$.
As will become clear later, this boundary condition too can resonate with modes $A$ and $B$.
Following the Fredholm alternative, the system must satisfy  a compatibility condition to be invertible. 
Loosely speaking, this condition states that $\mathcal{F}_3$ must be orthogonal to the kernel of the adjoint linearised NS operator, but this statement must be made more precise when boundary conditions are non-homogeneous. We should note that this general expression of the Fredholm alternative for a problem with
non-homogeneous BCs can also be obtained by lifting the problem and applying the Fredholm alternative to the lifted
problem with homogeneous BCs.
Specifically, the solution of $\tilde{\mathcal{L}}\boldsymbol{u}_3 = \mathcal{F}_3$ with wall boundary conditions $\boldsymbol{u}_3 = \tilde{\boldsymbol{u}}$  exists if and only if 
\begin{equation}    \langle\mathcal{F}_3,\boldsymbol{u}^\text{\textdagger}\rangle - \mathcal{W}(\boldsymbol{u}_3,\boldsymbol{u}^\text{\textdagger}) = 0
\label{compatibility_condition}
\end{equation}
for all $\boldsymbol{u}^\text{\textdagger}$ satisfying the  adjoint problem
$\tilde{\mathcal{L}}^\text{\textdagger}\boldsymbol{u}^\text{\textdagger} = \boldsymbol{0}$ with homogeneous wall boundary conditions $\boldsymbol{u}^\text{\textdagger} = \boldsymbol{0}$.
Here, $\mathcal{W}$ is the boundary term arising from integration by parts:
\begin{equation}
    \langle\tilde{\mathcal{L}}\boldsymbol{u},\boldsymbol{u}^\text{\textdagger}\rangle = \langle\boldsymbol{u},\tilde{\mathcal{L}}^\text{\textdagger}\boldsymbol{u}^\text{\textdagger}\rangle + \mathcal{W}(\boldsymbol{u},\boldsymbol{u}^\text{\textdagger}).
\end{equation}
Given the expression (\ref{eq:LNS}) of the linearised NS operator, we obtain
\begin{equation}
    \mathcal{W}(\boldsymbol{u},\boldsymbol{u}^\text{\textdagger}) = \oint [(\boldsymbol{u}\cdot\boldsymbol{u}^\text{\textdagger})(\boldsymbol{u}_0\cdot n) + \boldsymbol{u}\cdot(\sigma^\text{\textdagger} \cdot n)
    -\boldsymbol{u}^\text{\textdagger}\cdot(\sigma \cdot n)]dS,
\end{equation}
where $\sigma(\boldsymbol{u},p) = -pI + \nabla \boldsymbol{u}$ 
and $\sigma^\text{\textdagger}(\boldsymbol{u}^\dag,p^\dag) = p^\text{\textdagger}I + \nabla \boldsymbol{u}^\text{\textdagger}$. 
As the kernel of the linearised NS operator contains two modes, $A$ and $B$, the compatibility condition must be satisfied for both modes.
We will distinguish the contributions from  modes $A$ and $B$ as $\mathcal{W}(\boldsymbol{u}_3,\boldsymbol{u}_1^{A\text{\textdagger}}) \text{ and } \mathcal{W}(\boldsymbol{u}_3,\boldsymbol{u}_1^{B\text{\textdagger}})$, respectively, 
with $\boldsymbol{u}_1^{A\text{\textdagger}}$ and $\boldsymbol{u}_1^{B\text{\textdagger}}$  the standard adjoint modes satisfying homogeneous boundary conditions on the body wall.
Recalling all the  boundary conditions for $\boldsymbol{u}_1^{A\dag}$, $\boldsymbol{u}_1^{B\dag}$ and $\boldsymbol{u}_3$, all terms in $\mathcal{W}$ vanish except on the body wall $\Gamma$:
\begin{equation}
    \begin{split}
        \mathcal{W}(\boldsymbol{u}_3, \boldsymbol{u}_1^{A\dag})
        =  \oint_\Gamma \tilde{\boldsymbol{u}} \cdot (\sigma^{A\text{\textdagger}}_1 \cdot n) dS,
        \quad 
        \mathcal{W}(\boldsymbol{u}_3, \boldsymbol{u}_1^{B\dag})
        =  \oint_\Gamma \tilde{\boldsymbol{u}} \cdot (\sigma^{B\text{\textdagger}}_1 \cdot n) dS.
    \end{split}
\end{equation}
We can first look into the symmetries of $\tilde{\boldsymbol{u}}$ by decomposing the rotation vector into two,
\begin{equation}
    \tilde{\boldsymbol{u}} = -(\nabla \boldsymbol{u}_0)_{|\boldsymbol{x}_{0,0}}\begin{pmatrix}
                        \tilde{\beta}y_{0,0} + \tilde{\alpha}z_{0,0} \\
                        -\tilde{\beta}x_{0,0}\\
                        -\tilde{\alpha}x_{0,0}
                        \end{pmatrix} 
                        = -(\nabla \boldsymbol{u}_0)_{|\boldsymbol{x}_{0,0}}
                        \Bigg(\tilde{\alpha}
                        \begin{pmatrix}
                        z_{0,0} \\
                        0\\
                        -x_{0,0}
                        \end{pmatrix}+\tilde{\beta}
                        \begin{pmatrix}
                        y_{0,0}\\
                        -x_{0,0}\\
                        0
                        \end{pmatrix}
                        \Bigg).
\label{eqab}
\end{equation}
Because $\nabla\boldsymbol{u}_0$ is $S_yS_z$-symmetric like $\boldsymbol{u}_0$, the above decomposition shows that $\tilde{\boldsymbol{u}}$ is the sum of an $A_yS_z$-symmetric term (the $x$ and $z$ components are unchanged under $y\rightarrow-y$) that depends only on pitch, and
a $S_yA_z$-symmetric term (the $x$ and $y$ components are unchanged under $z\rightarrow-z$) that depends only on yaw. 
Next, the stress force $\sigma^{\text{\textdagger}}_1 \cdot n$ has the opposite symmetry as 
the corresponding mode, i.e. $A_yS_z$ for $\sigma^{A\text{\textdagger}}_1 \cdot n$ and $S_yA_z$ for $\sigma^{B\text{\textdagger}}_1 \cdot n$. 
Finally, we get
\begin{equation}
    \mathcal{W}(\boldsymbol{u}_3, \boldsymbol{u}_1^{A\text{\textdagger}}) 
    = -\tilde{\alpha}\oint_\Gamma \Bigg((\nabla \boldsymbol{u}_0)_{|\boldsymbol{x}_{0,0}} 
                        \begin{pmatrix}
                        z_{0,0} \\
                        0\\
                        -x_{0,0}
                        \end{pmatrix} \Bigg) \cdot ( \sigma^{A\text{\textdagger}}_1 \cdot n)dS \doteq -\tilde{\alpha}h_A,
                        \label{hA}
\end{equation}
\begin{equation}
    \mathcal{W}(\boldsymbol{u}_3, \boldsymbol{u}_1^{B\text{\textdagger}}) 
    = -\tilde{\beta}\oint_\Gamma \Bigg((\nabla \boldsymbol{u}_0)_{|\boldsymbol{x}_{0,0}} 
                        \begin{pmatrix}
                        y_{0,0} \\
                        -x_{0,0}\\
                        0
                        \end{pmatrix} \Bigg) \cdot ( \sigma^{B\text{\textdagger}}_1 \cdot n)dS \doteq -\tilde{\beta}h_B. 
                        \label{hB}
\end{equation}
As mentioned earlier, we note that a small-amplitude roll angle would induce an $A_y A_z$ geometry modification in (\ref{eqab}) and, for symmetry reasons, would yield identically zero boundary terms in (\ref{hA})-(\ref{hB}). In other words, and perhaps as expected intuitively, roll does not affect the WNL results. 

The result of the above calculations is the following system of amplitude equations:
\begin{equation}
    \partial_{T_1}A = \tilde{\lambda}_AA - \tilde{\chi}_AA^3-\tilde{\eta}_{A}AB^2 + h_A \tilde{\alpha},
    \label{eq:amp_eq_A}
\end{equation}
\begin{equation}
    \partial_{T_1}B = \tilde{\lambda}_BB - \tilde{\chi}_BB^3-\tilde{\eta}_{B}A^2B + h_B \tilde{\beta}.
    \label{eq:amp_eq_B}
\end{equation}

This system of coupled ordinary differential equations for the amplitudes $A$ and $B$ is the same as in the perfectly aligned case (\cite{Zampogna_Boujo_2023}), except for the two additional imperfection terms $h_A \tilde{\alpha}$ and $h_B \tilde{\beta}$. 
These  terms are independent of $A$ and $B$, and are linear in the pitch and yaw angles $\tilde\alpha$ and $\tilde\beta$. 
Very conveniently, all the coefficients of (\ref{eq:amp_eq_A})-(\ref{eq:amp_eq_B}) only need to be computed once for all on the perfectly aligned geometry.
Computing the imperfection coefficients $h_A$ and $h_B$ only involves evaluating the base flow $\boldsymbol{u}_0$ and eigenmodes $\hat{\boldsymbol{u}}_1^A$, $\hat{\boldsymbol{u}}_1^B$, and does not require solving for any additional field. 
These calculations can be performed on a quarter domain. 

For any $Re$, $\alpha$ and $\beta$, the flow field  can be reconstructed from the solution of the amplitude equations as:
\begin{equation}
    \boldsymbol{q} = \boldsymbol{q}_0 
    + \epsilon(A\hat{\boldsymbol{q}}_1^A + B\hat{\boldsymbol{q}}_1^B) 
    + \epsilon^2(\tilde{\delta}\hat{\boldsymbol{q}}_2^\delta + A^2\hat{\boldsymbol{q}}_2^{A^2} + B^2\hat {\boldsymbol{q}}_2^{B^2} + AB\hat{\boldsymbol{q}}_2^{AB}) 
    + \mathcal{O}(\epsilon^3).
\end{equation}
The values of the WNL coefficients and the amplitudes depend on the choice of $\epsilon$ and the normalisation of the direct/adjoint modes, while the reconstructed flow field is independent of these choices.
The WNL analysis is rigorously valid in the vicinity of $Re=Re_c$ and $\alpha=\beta=0$. 
Therefore, we expect the fully nonlinear bifurcation diagram to depart from the WNL analysis away from this point.
We  also note that the WNL analysis is built with the bifurcating stationary eigenmodes $A$ and $B$, and does not account for the secondary instabilities (Hopf) that occur on the deflected wake. 
However, we should appreciate that it is  accurate and computationally cheap  in the vicinity of $Re=Re_c$ and $\alpha=\beta=0$. In particular, it allows us to construct bifurcation diagrams for simultaneously non-zero $\alpha$ and $\beta$, while it is much more expensive to compute fully non linear base flows deflected in both horizontal and vertical directions and solve for the eigenmodes, or to run fully nonlinear DNS.

\subsection{Results}
We now investigate the solutions of the amplitude equations (\ref{eq:amp_eq_A})-(\ref{eq:amp_eq_B}) for various attitudes $(\alpha,\beta)$. We can solve the amplitude equations at different $Re$ values and reconstruct the velocity at sensor $M$.
Comparing the weakly nonlinear results (black dots) with the fully nonlinear solution (blue spline) in figure~\ref{WNLpitch}, which shows the vertical or lateral velocity  at sensor $M$, we  observe good agreement, especially near the reference Reynolds number. 
In particular, the location of the saddle-node is well captured.
\begin{figure}[h]
        \centering
\begin{subfigure}{0.48\textwidth}
\begin{overpic}[width=1\linewidth]{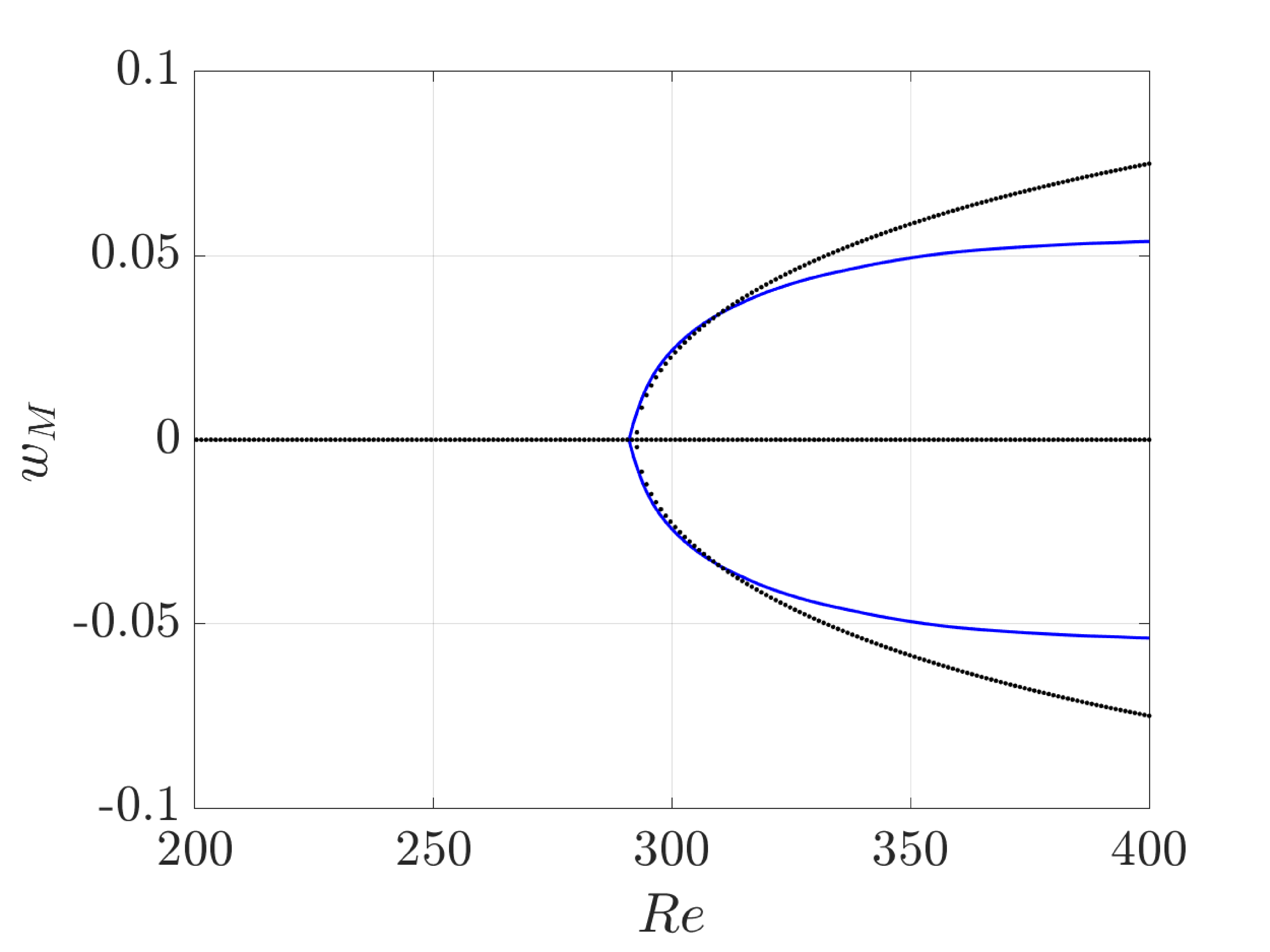} 
\put(-5,70){\small (a)}
\end{overpic}
\end{subfigure}
\begin{subfigure}{0.48\textwidth}
\begin{overpic}[width=1\linewidth]{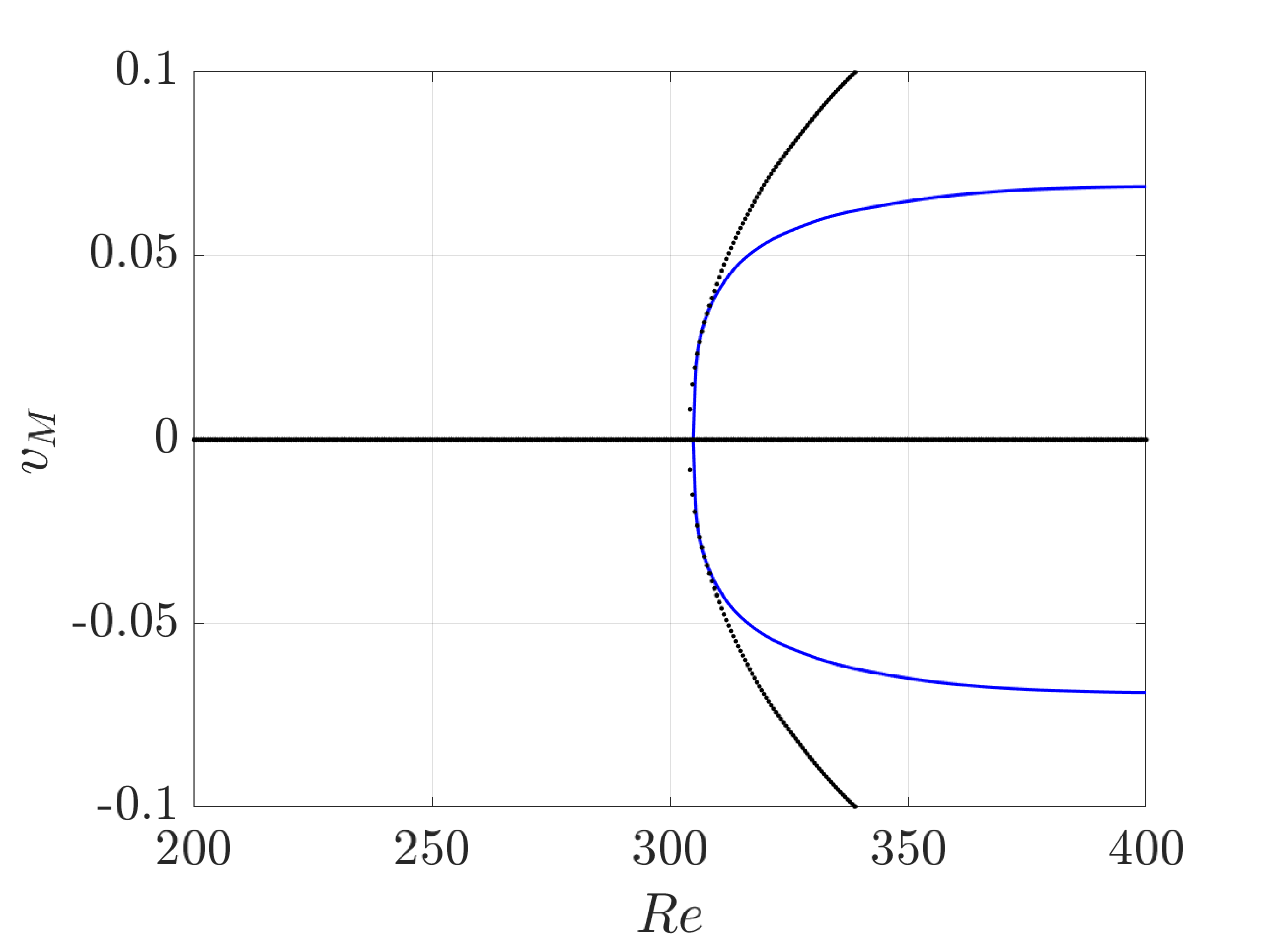} 
\put(-5,70){\small (b)}
\end{overpic}
\end{subfigure}
\begin{subfigure}{0.48\textwidth}
\begin{overpic}[width=1\linewidth]{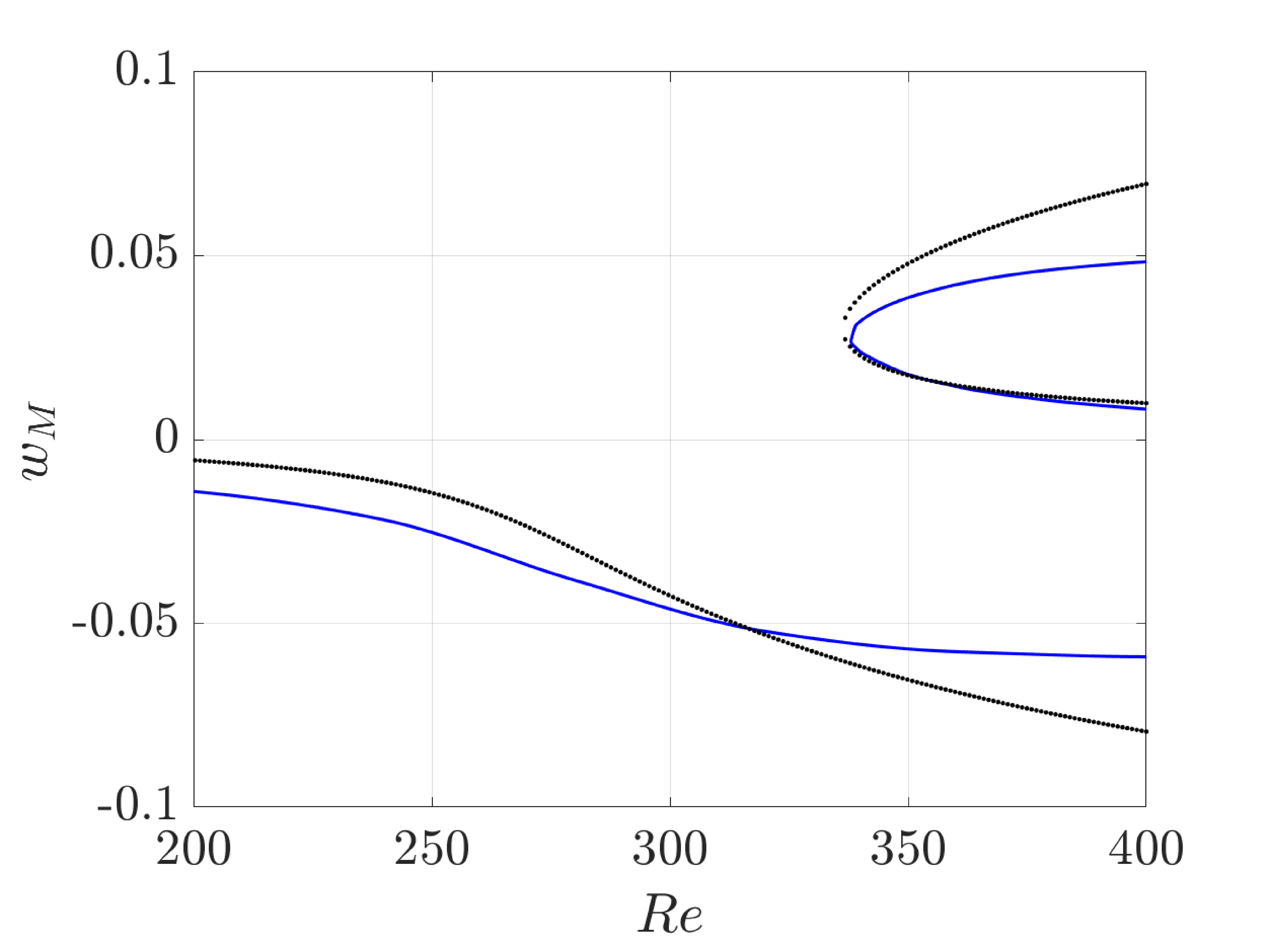}
\put(-5,70){\small (c)}
\end{overpic}
\end{subfigure}
\begin{subfigure}{0.48\textwidth}
\begin{overpic}[width=1\linewidth]{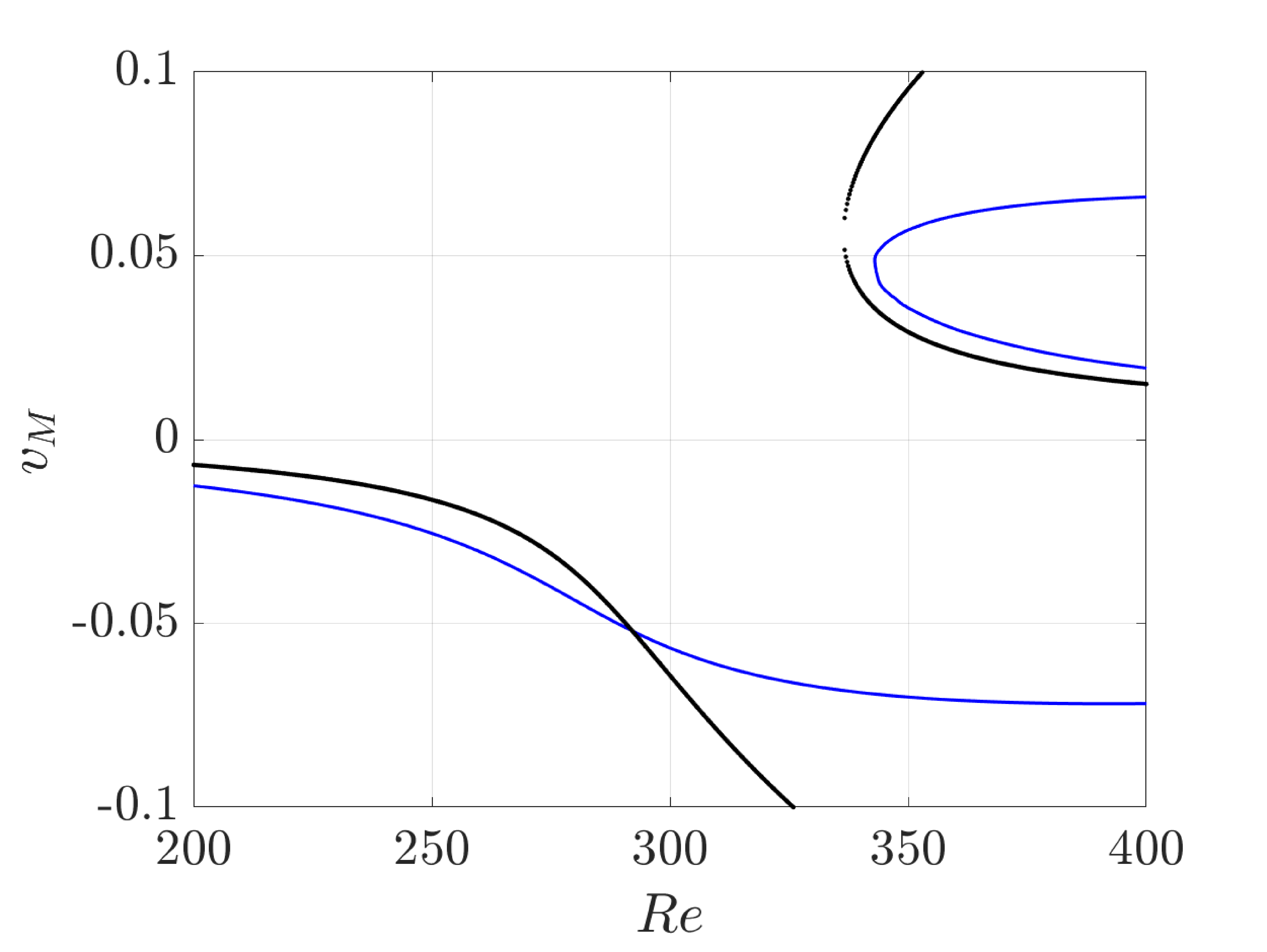}
\put(-5,70){\small (d)}
\end{overpic}
\end{subfigure}
\caption{Bifurcation diagrams from fully nonlinear calculations (blue)  and the WNL analysis (black dots). 
(a)~Vertical deflection in state $(A,0)$ for the perfectly aligned case $(\alpha,\beta) = (0^\circ,0^\circ)$, 
(b)~horizontal deflection in state $(0,B)$ for the perfectly aligned case $(\alpha,\beta) = (0^\circ,0^\circ)$, 
(c)~vertical deflection for a pure pitch case, $(\alpha,\beta) = (0.5^\circ,0^\circ)$, and 
(d)~horizontal deflection for a pure yaw case, $(\alpha, \beta) = ( 0^\circ,0.5^\circ)$.}
\label{WNLpitch}
\end{figure}
\begin{figure}[h]
    \centering
    \includegraphics[width=1\linewidth]{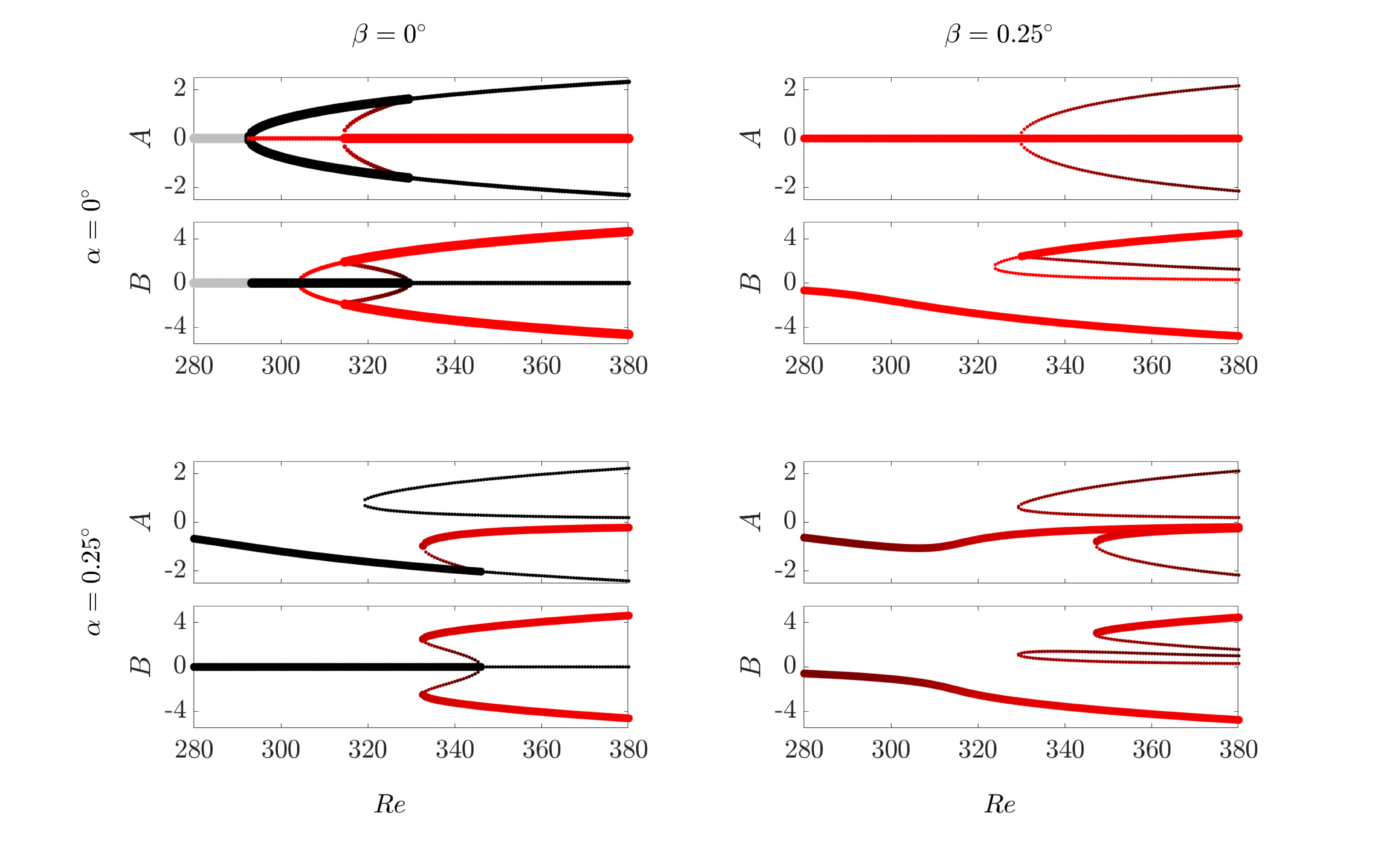  } 
    \includegraphics[width=0.3\linewidth]{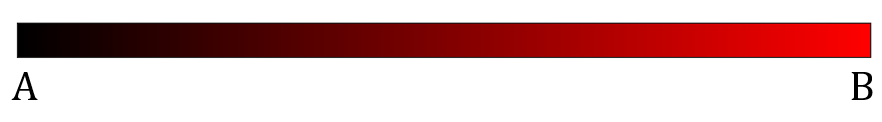}     
    \caption{WNL amplitude-based bifurcation diagram with $Re$ as the control parameter. Top, bottom: $\alpha = 0^{\circ}$ and $ 0.25^{\circ}$. Left, right: $\beta = 0^{\circ} $ and $ 0.25^{\circ}$. Smaller and larger points correspond to linearly unstable and stable solutions, respectively. Black, red,  and intermediate shades correspond to states $(A,0)$, $(0,B)$, and $(A,B)$, respectively, with the intermediate shades denoting the contribution of each amplitude.}
    \label{WNL}
\end{figure} 
We  now compute the amplitudes for a range of pitch and yaw values, along with the linear stability of each solution by computing the two eigenvalues of the Jacobian of (\ref{eq:amp_eq_A})-(\ref{eq:amp_eq_B}).

Figure~\ref{WNL} shows bifurcation diagrams $(A,B)$ as a function of $Re$ for four attitudes with  $\alpha=0^\circ$ (top row) or $\alpha=0.25^\circ$ (bottom row), and with $\beta=0^\circ$ (left column) or $\beta=0.25^\circ$ (right column). 
Colours range from black (pure vertical deflection $(A,0)$) to red (pure horizontal deflection $(0,B)$).
Smaller and larger points correspond to linearly stable and unstable solutions, respectively.
For the angles considered here, the bistability observed in the perfectly aligned case between vertically and horizontally deflected states persists under pitch but disappears under yaw.
We  also note that, similar to fully nonlinear results, imperfection breaks the vertical or horizontal symmetry and preselects one of two states at low Reynolds number. 
\begin{figure}[h]
    \centering
    \begin{subfigure}[t]{0.328\textwidth}
        \centering
        \begin{overpic}[width=\linewidth]{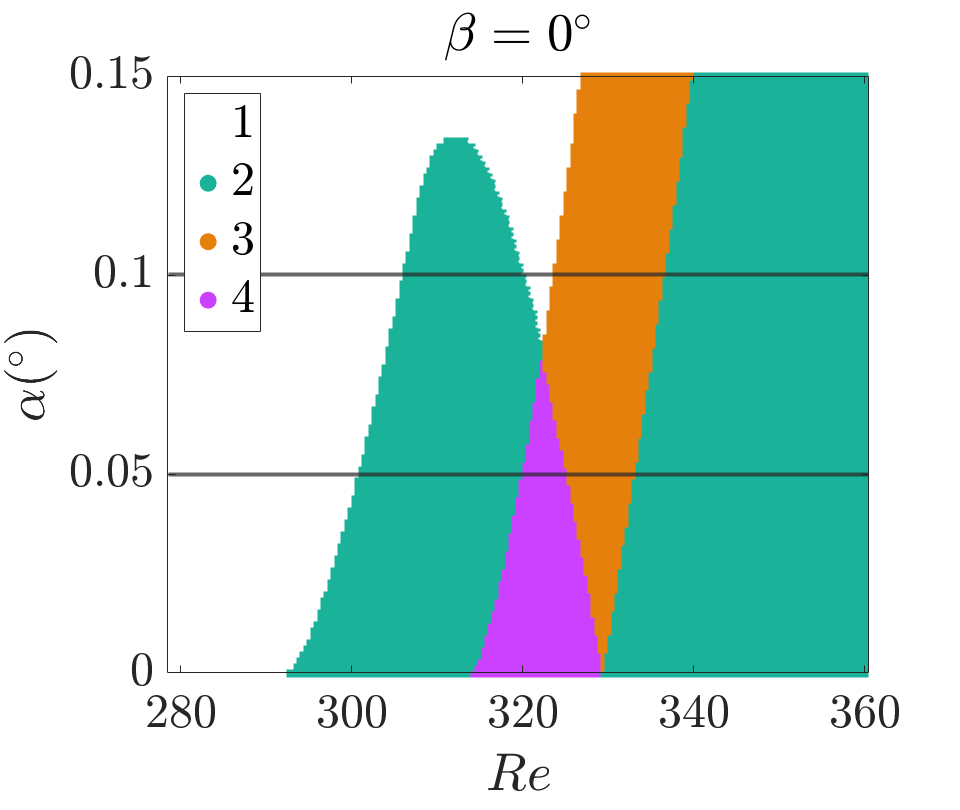}
        \put(-5,70){\small (a)}
        \end{overpic}
    \end{subfigure}
    \begin{subfigure}[t]{0.328\textwidth}
        \centering
        \begin{overpic}[width=\linewidth]{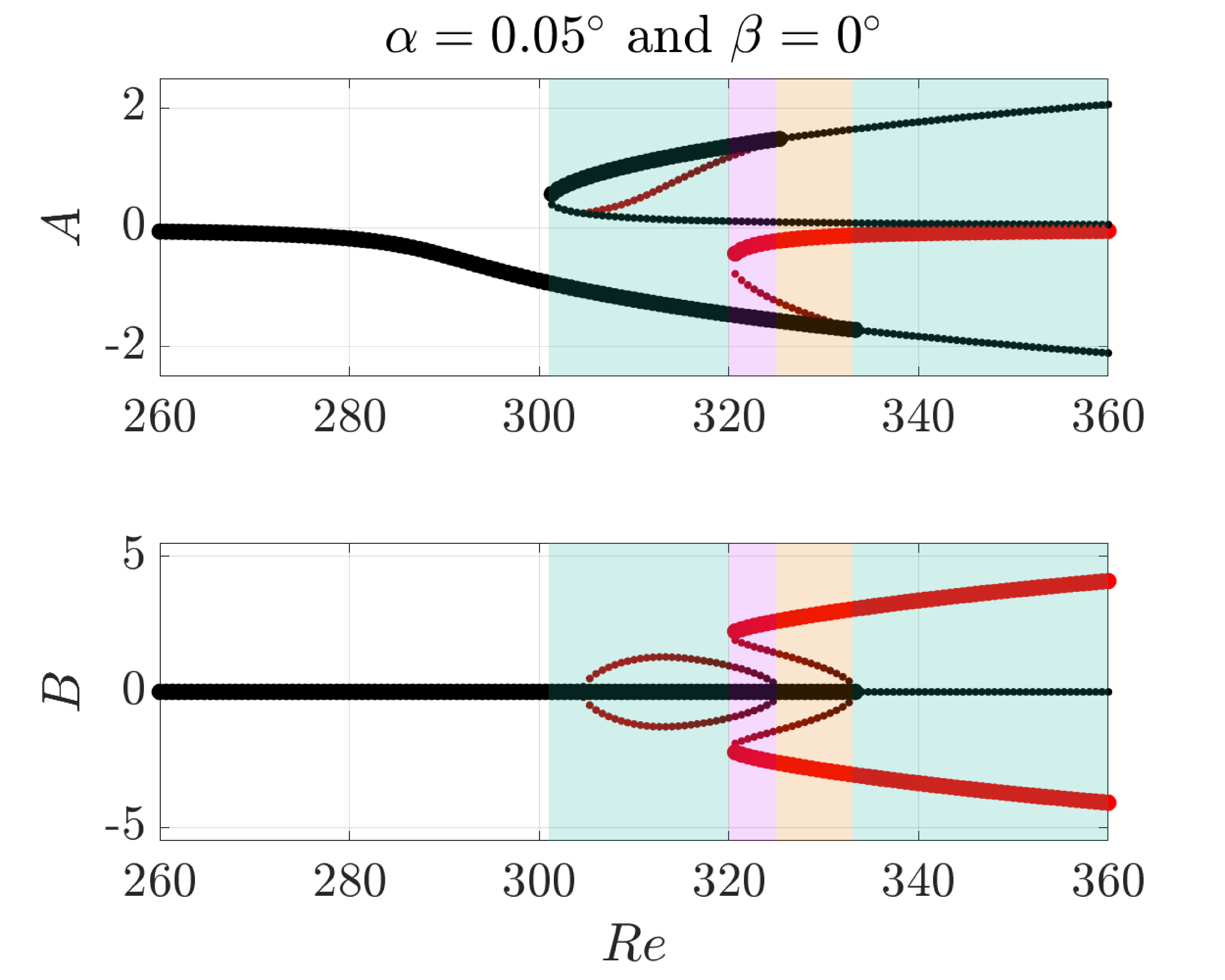}
        \put(-5,70){\small (b)}
        \end{overpic}
    \end{subfigure}
    \begin{subfigure}[t]{0.328\textwidth}
        \centering
        \begin{overpic}[width=\linewidth]{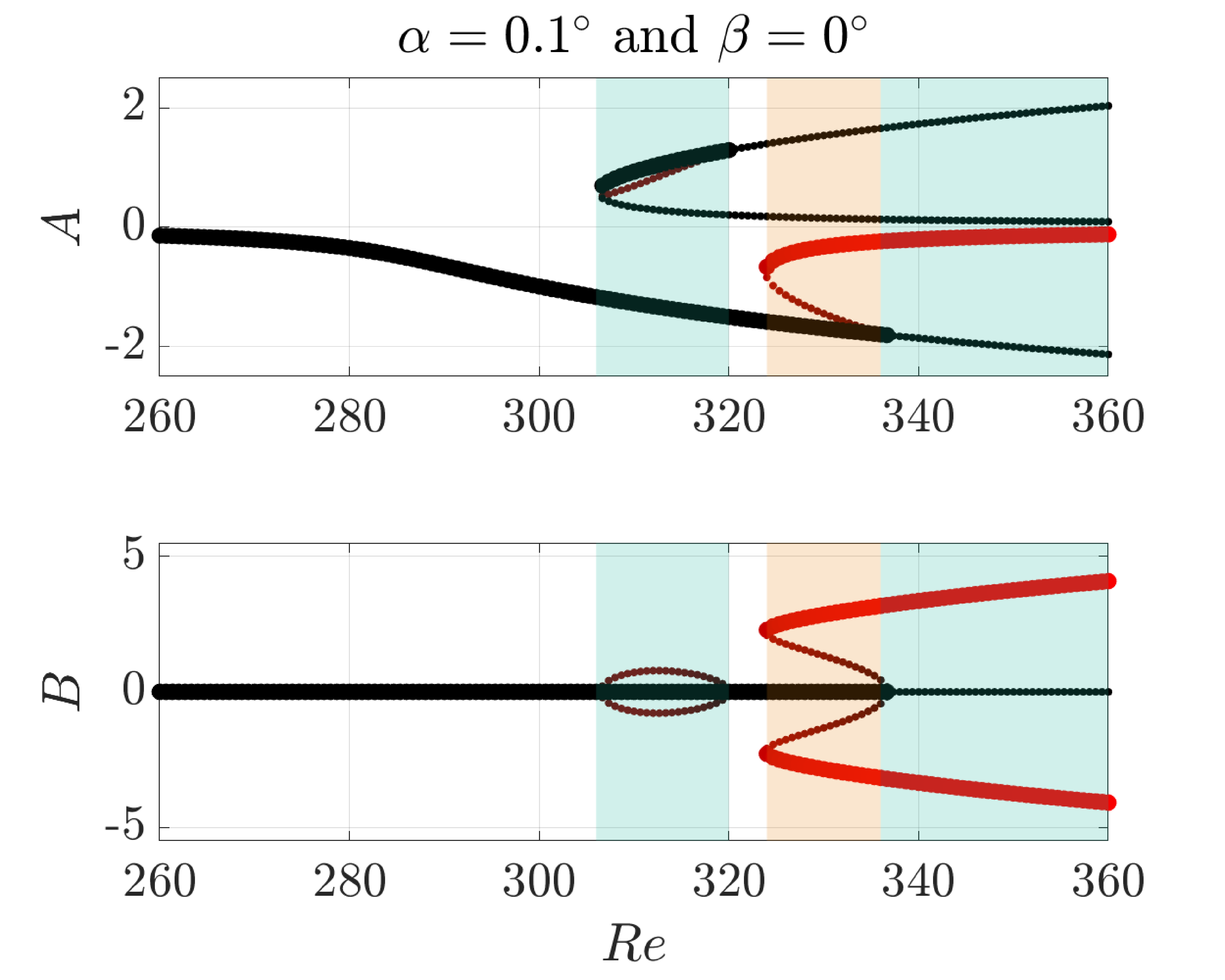}
        \put(-5,70){\small (c)}
        \end{overpic}
    \end{subfigure}
    \caption{
     $(a)$ Number of stable WNL solutions coded by colour, as a function of pitch angle and $Re$,  for pure pitch. Horizontal lines correspond to the pitch angles of panels $(b,c)$.
    $(b,c)$~Bifurcation diagrams for $\alpha=0.05^\circ$ and $\alpha=0.1^\circ$, respectively.} 
    \label{WNLstablesols_pitch_with_demonstration}
\end{figure}
\begin{figure}[h]
    \centering
    \begin{subfigure}[t]{0.325\textwidth}
        \centering
        \begin{overpic}[width=1\linewidth]{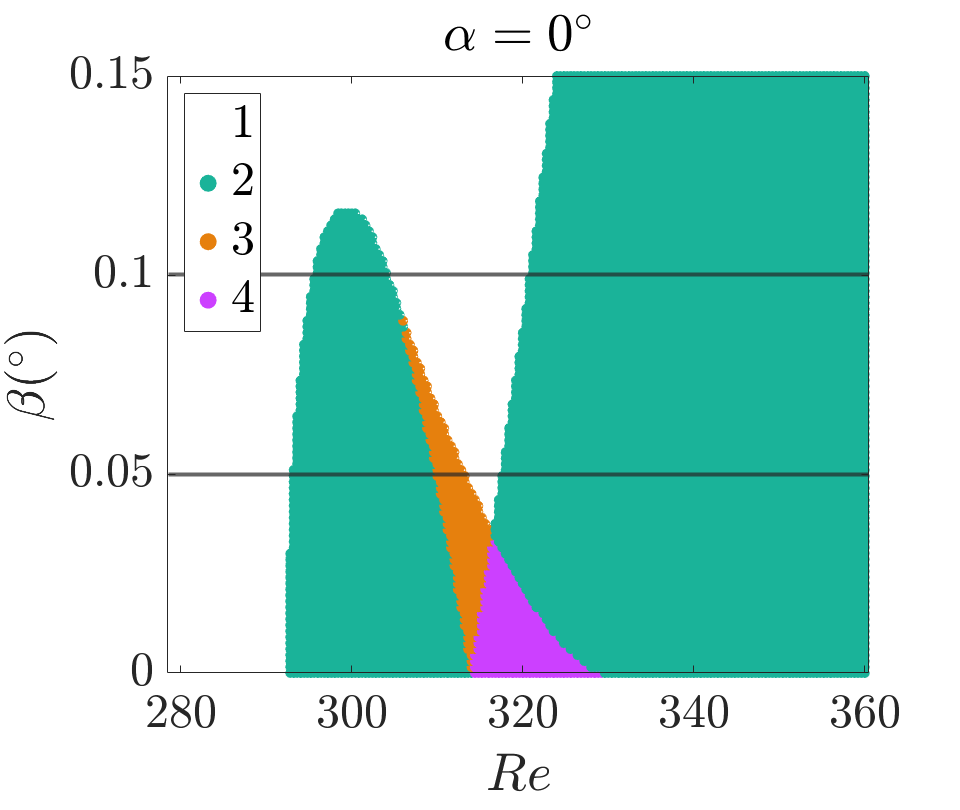}
        \put(-5,70){\small (a)}
        \end{overpic}
    \end{subfigure}
    \hfill
    \begin{subfigure}[t]{0.325\textwidth}
        \centering
        \begin{overpic}[width=1\linewidth]{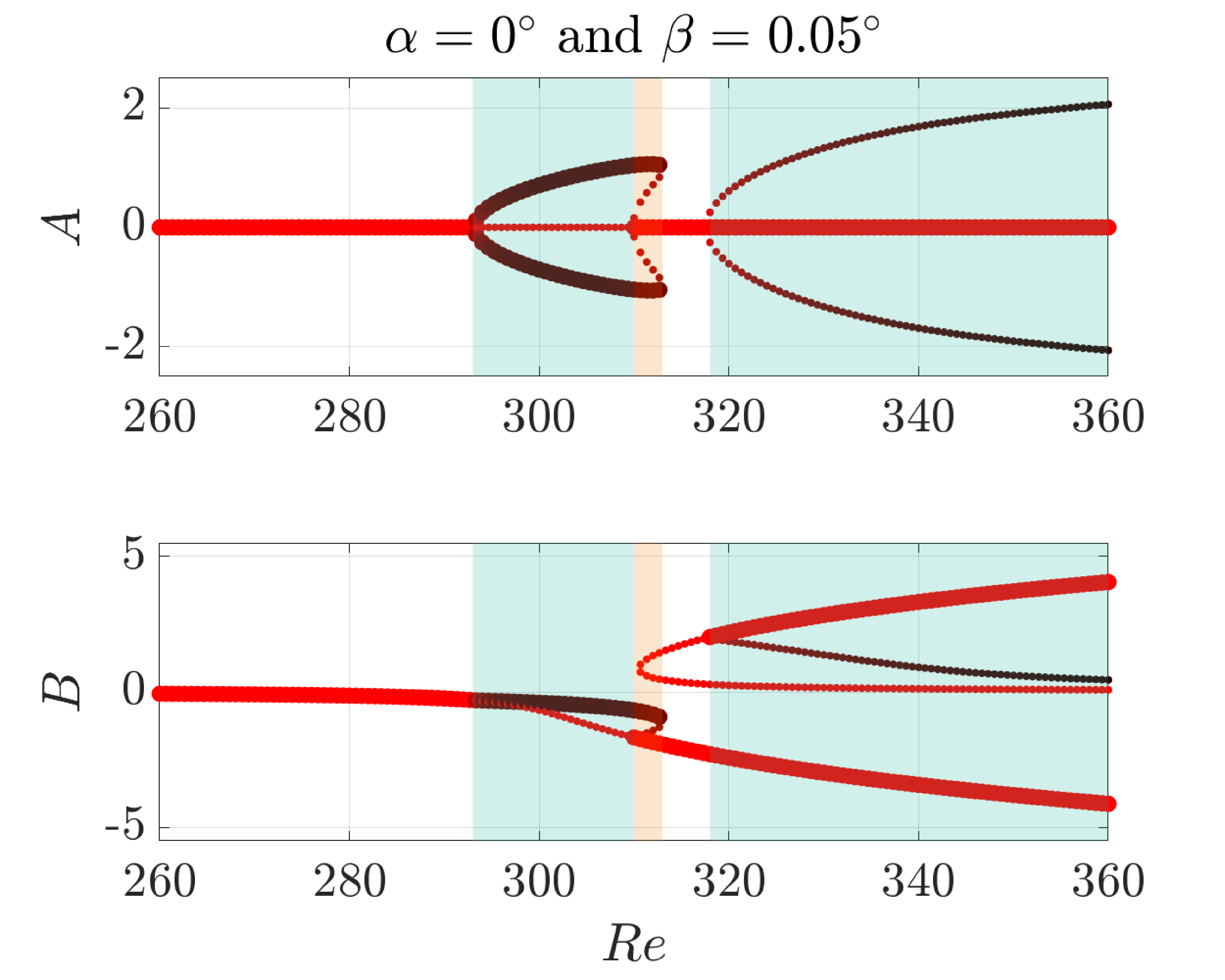}
        \put(-5,70){\small (b)}
        \end{overpic}
    \end{subfigure}
    \hfill
    \begin{subfigure}[t]{0.325\textwidth}
        \centering
        \begin{overpic}[width=1\linewidth]{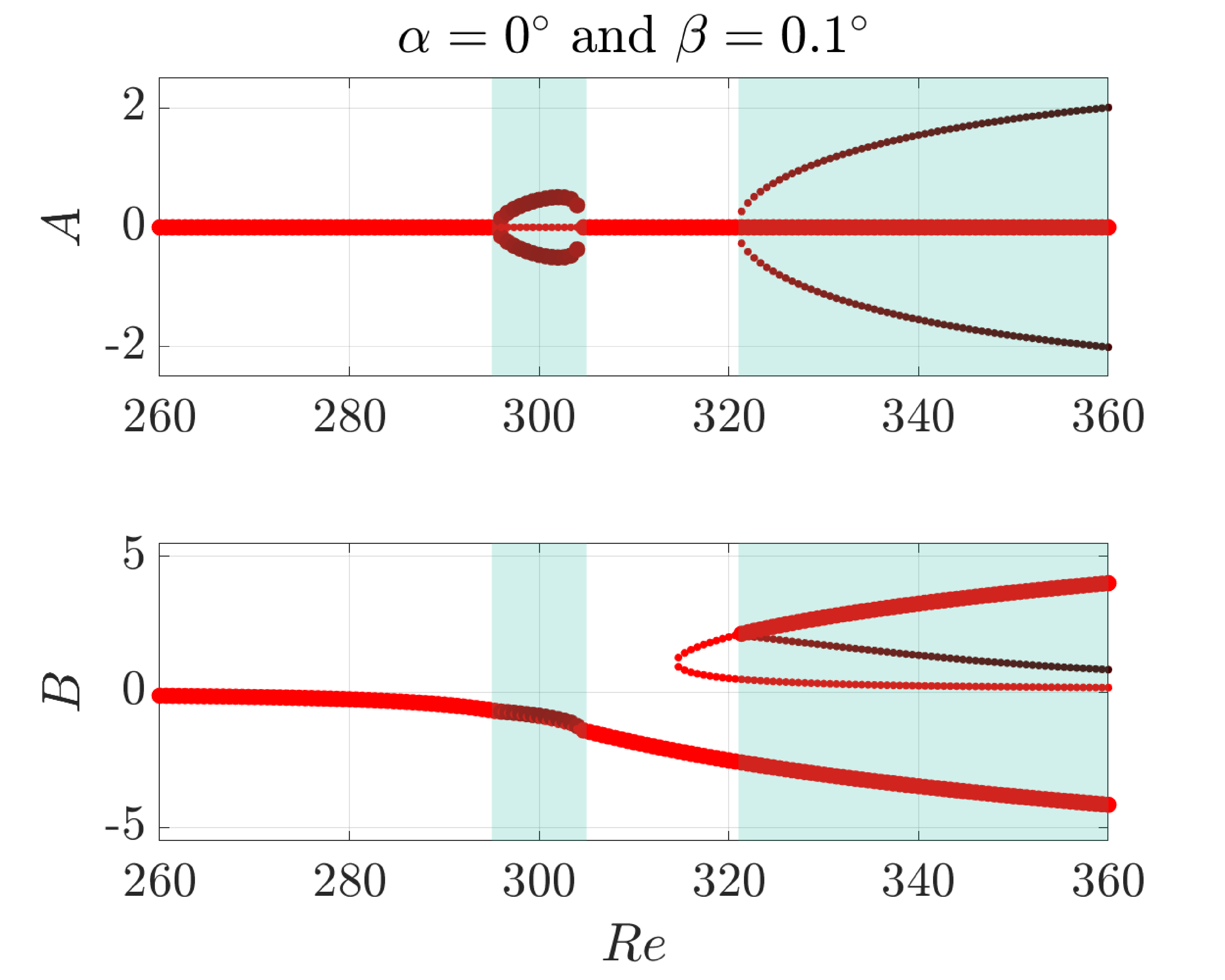}
        \put(-5,70){\small (c)}
        \end{overpic}
    \end{subfigure}
    \caption{   
    Same as figure~\ref{WNLstablesols_pitch_with_demonstration} as a function of yaw angle and $Re$, for pure yaw.
    }  
    \label{WNLstablesols_yaw_with_demonstration}
\end{figure}

\begin{figure}[h!]
        \centering
\begin{subfigure}{0.32\textwidth}
\includegraphics[width=1\linewidth]{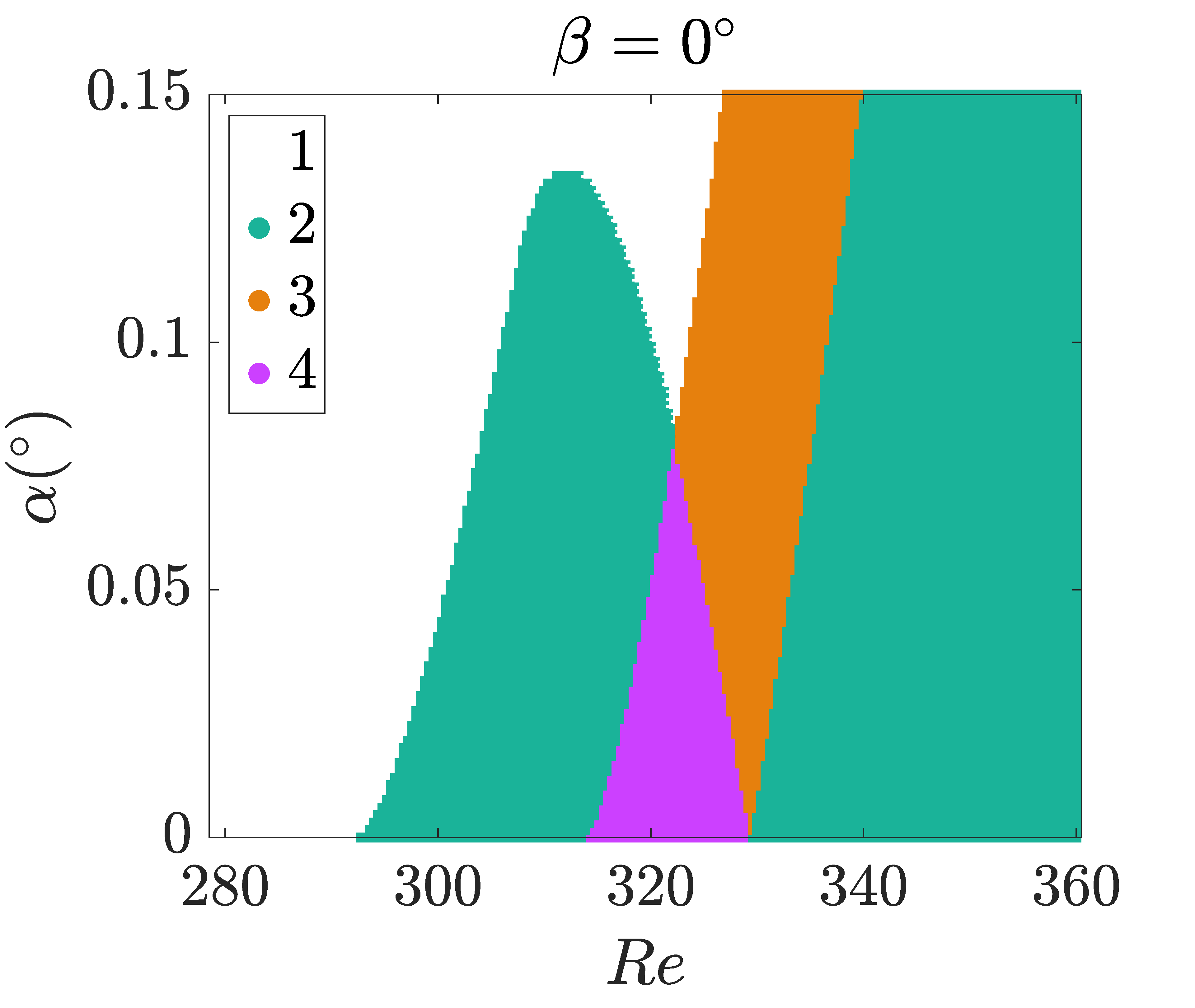}
\end{subfigure}
\begin{subfigure}{0.32\textwidth}
\includegraphics[width=1\linewidth]{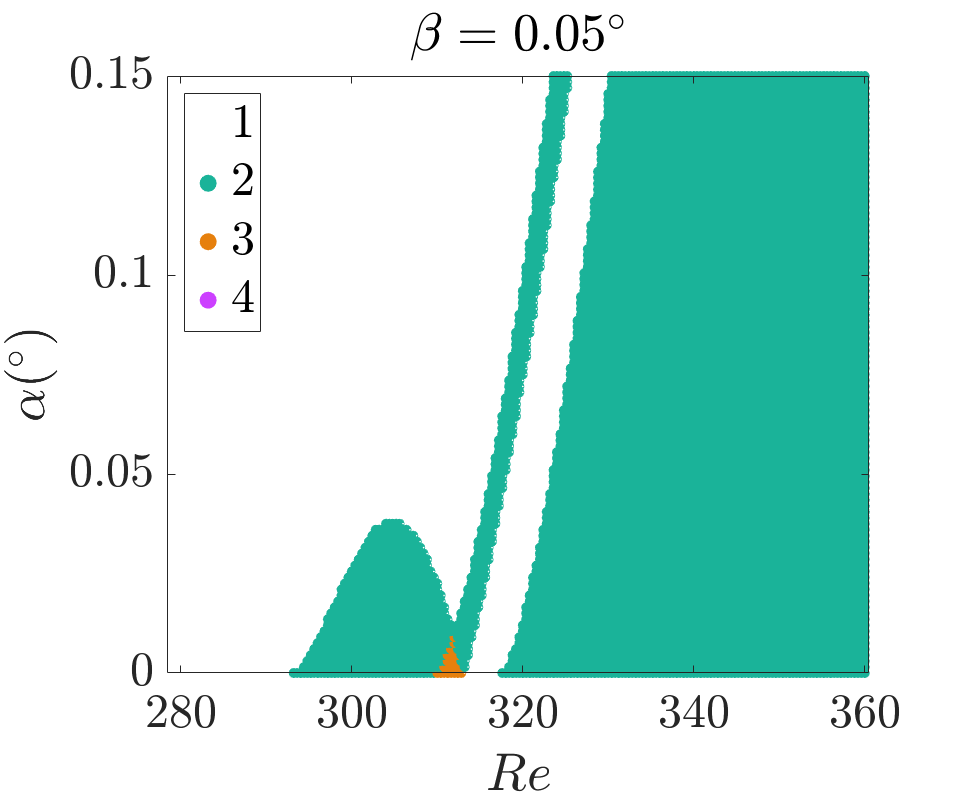}
\end{subfigure}
\begin{subfigure}{0.32\textwidth}
\includegraphics[width=1\linewidth]{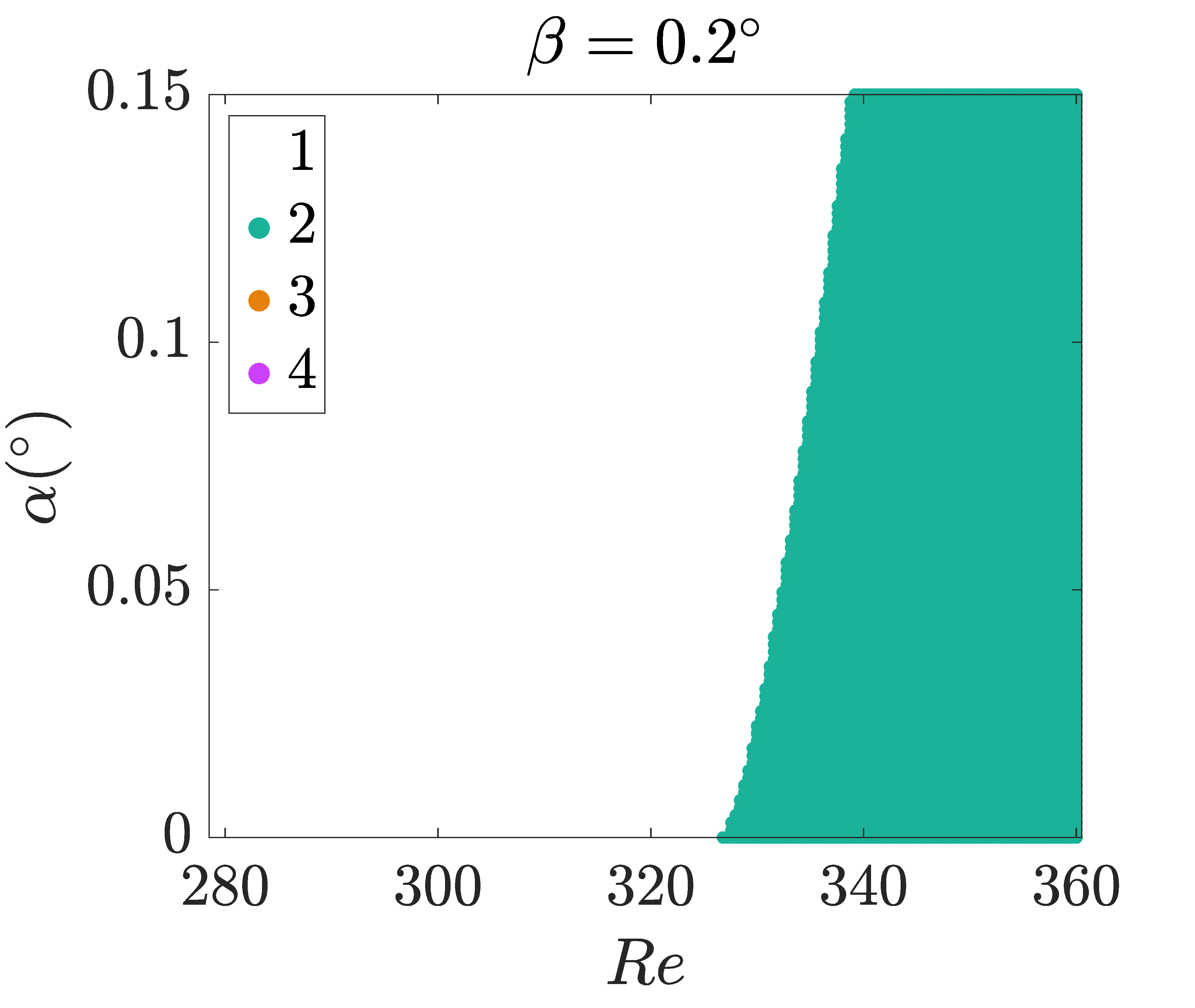}
\end{subfigure}
\begin{subfigure}{0.32\textwidth}
\includegraphics[width=1\linewidth]{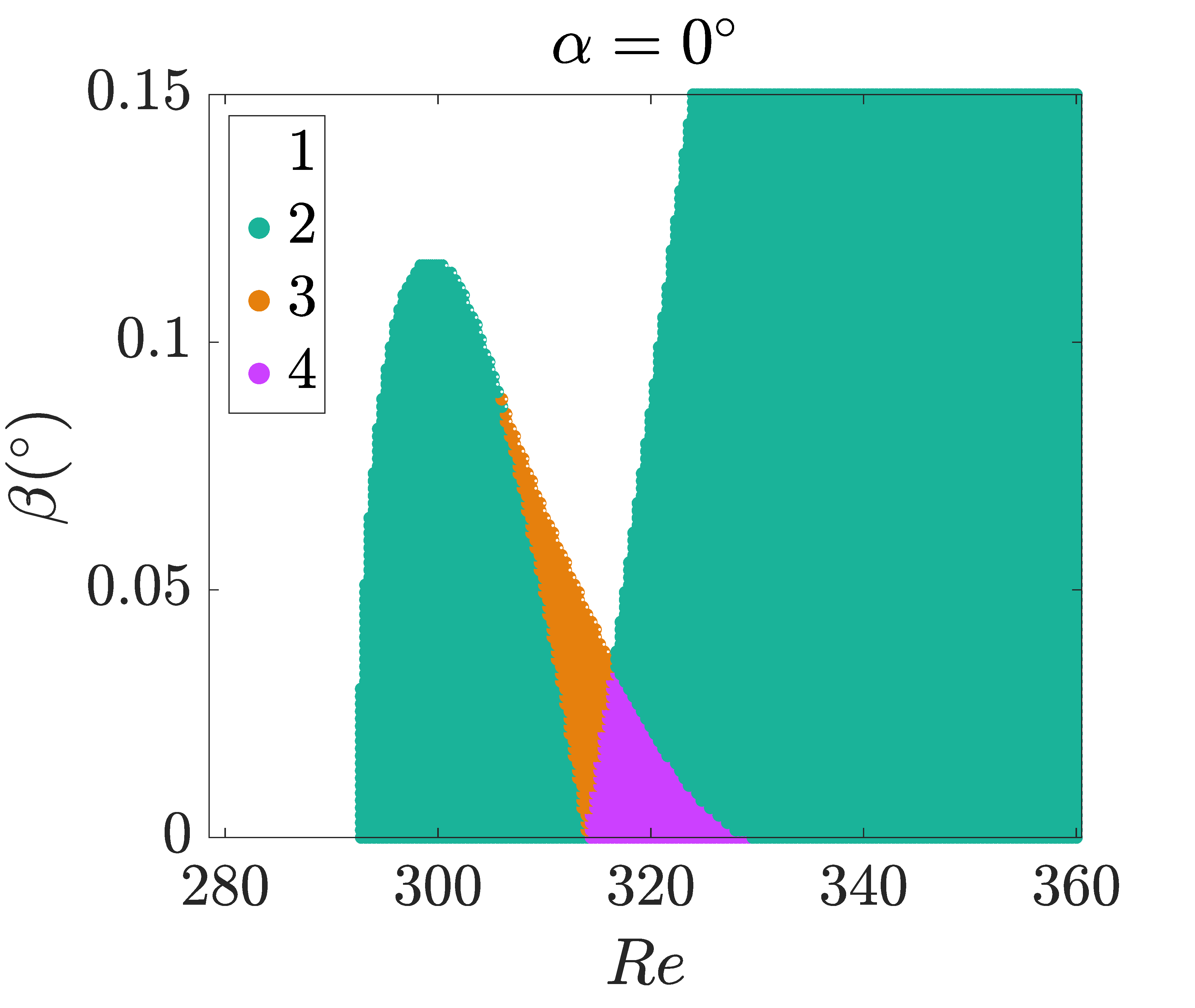} 
\end{subfigure}
\begin{subfigure}{0.32\textwidth}
\includegraphics[width=1\linewidth]{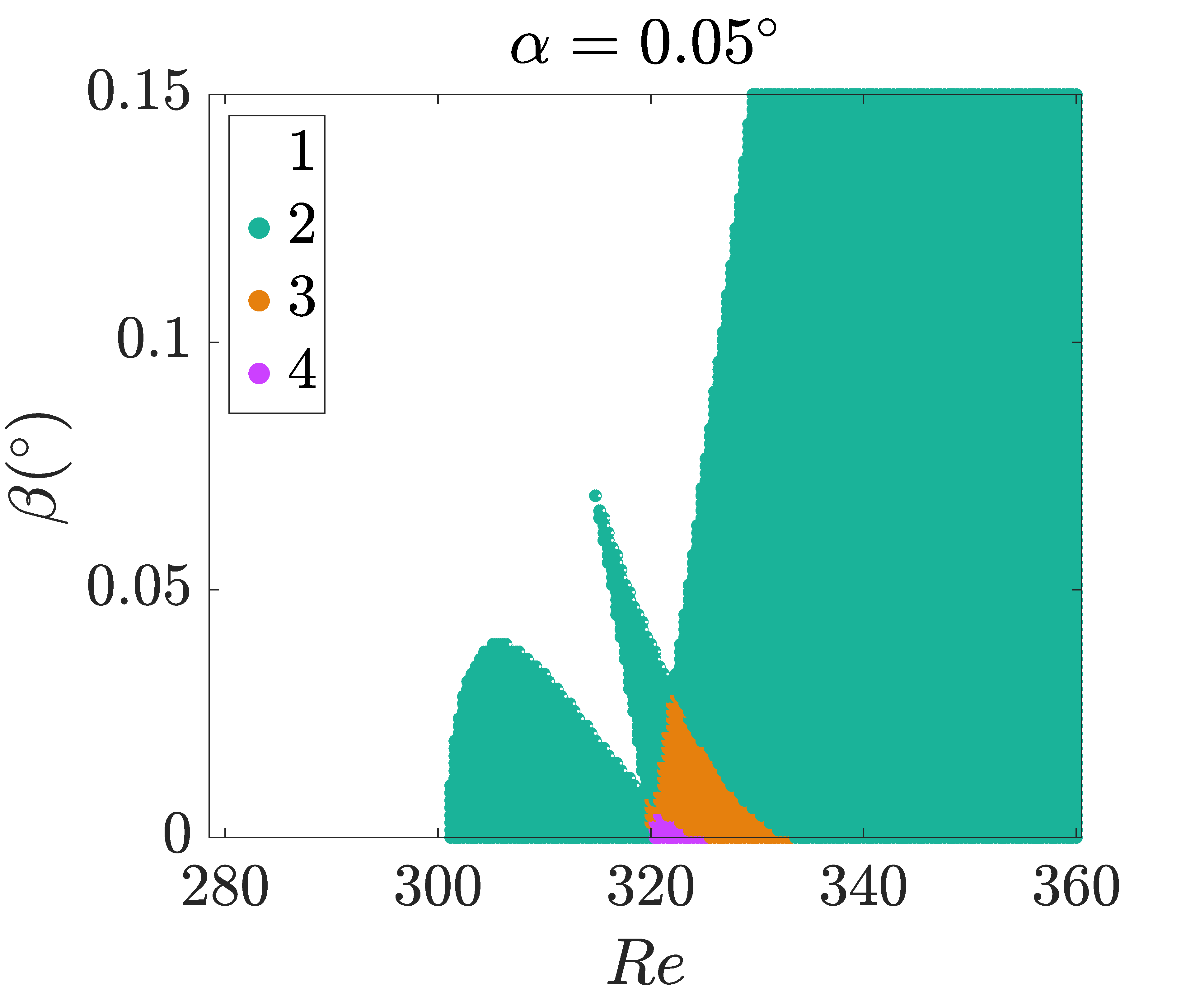}
\end{subfigure}
\begin{subfigure}{0.32\textwidth}
\includegraphics[width=1\linewidth]{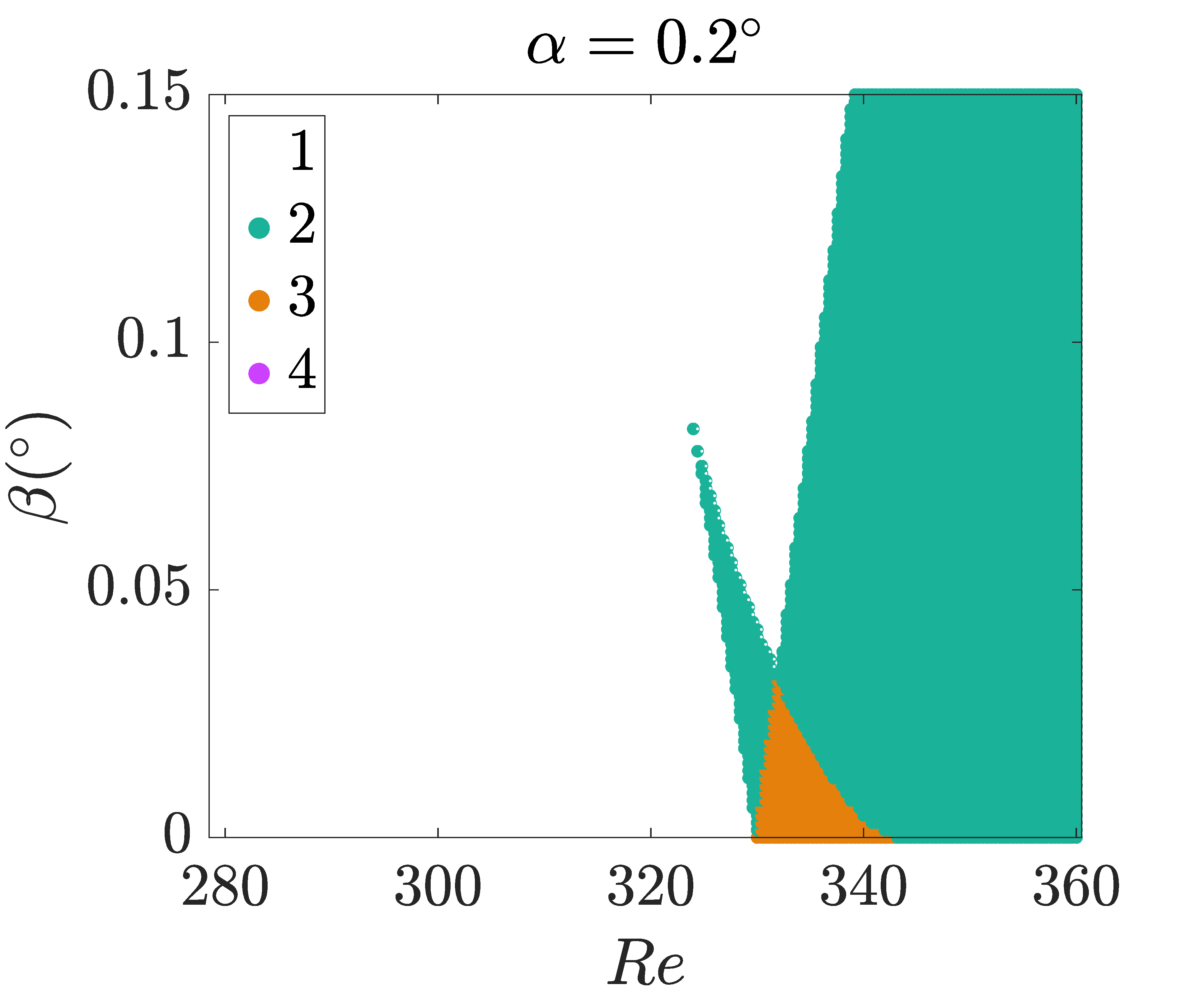}
\end{subfigure}
\caption{Number of stable WNL solutions for oblique attitudes, with both non-zero pitch and yaw angles.}
\label{WNLstablesols_yaw2}
\end{figure}
\begin{figure}[h!]
        \centering
\begin{subfigure}{0.32\textwidth}
\begin{tikzpicture}
    \node[anchor=south west](img) at (0,0)
        {\includegraphics[trim=1.2cm 0cm 1.5cm 0cm, clip=true, width=\textwidth]{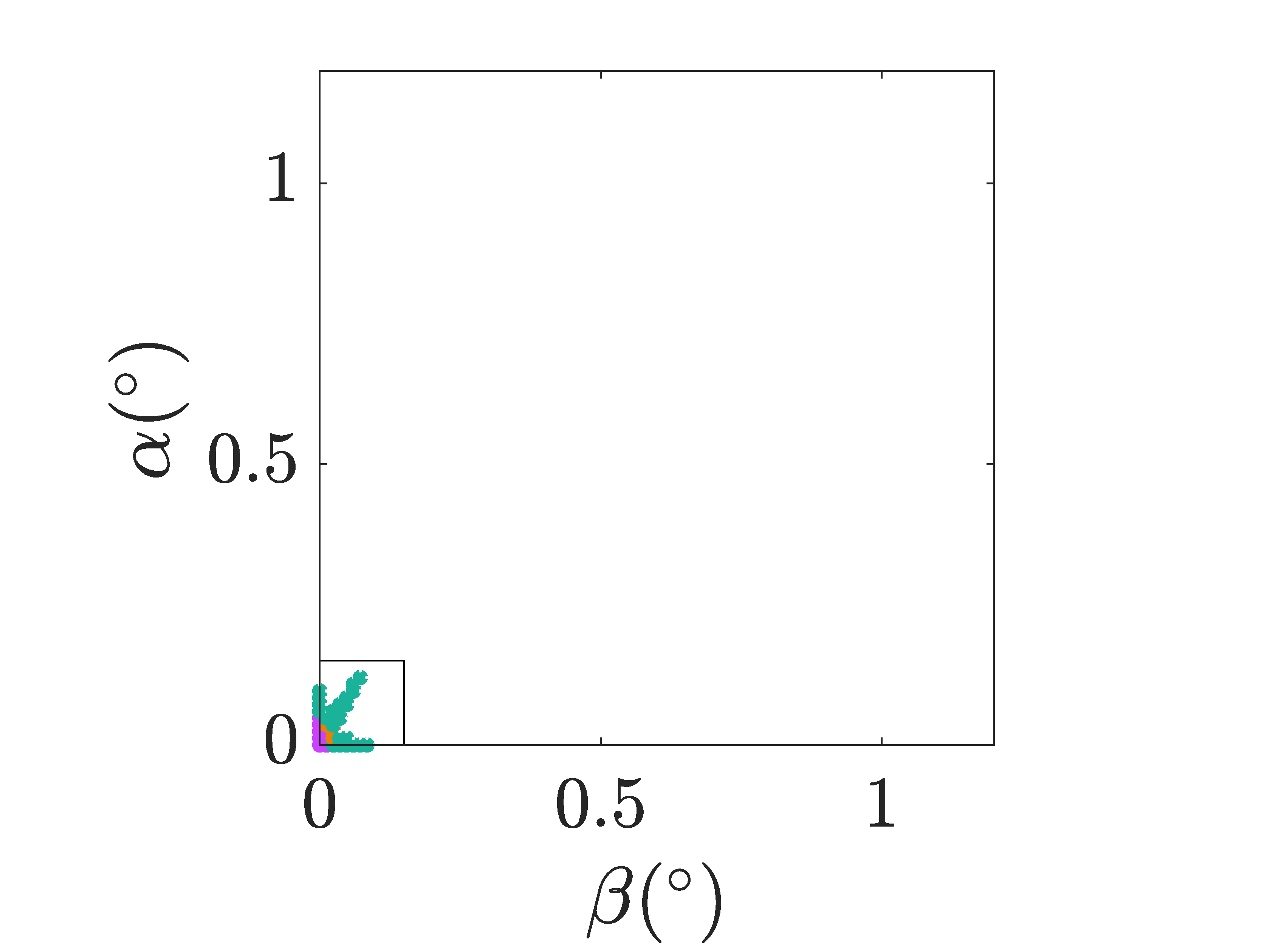}};
    \node[anchor=north west] at (img.north west) {\small (a)};
    \node[anchor=north east] at (3.6cm,3cm)
        {\includegraphics[width=0.45\textwidth]{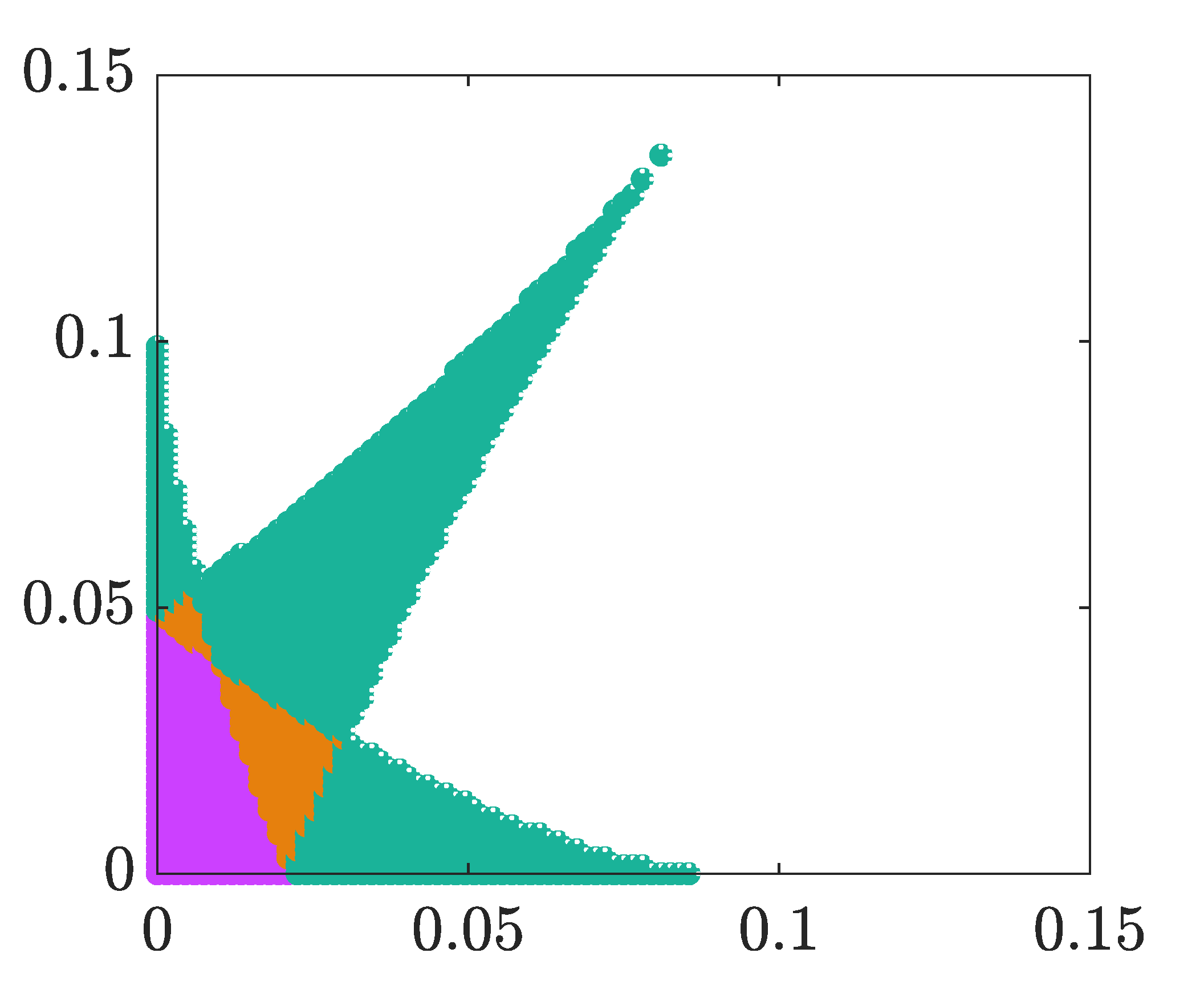}};
\end{tikzpicture}
\end{subfigure}
\begin{subfigure}{0.32\textwidth}
\begin{tikzpicture}
    \node[anchor=south west](img) at (0,0)
        {\includegraphics[trim=1.2cm 0cm 1.5cm 0cm, clip=true, width=\textwidth]{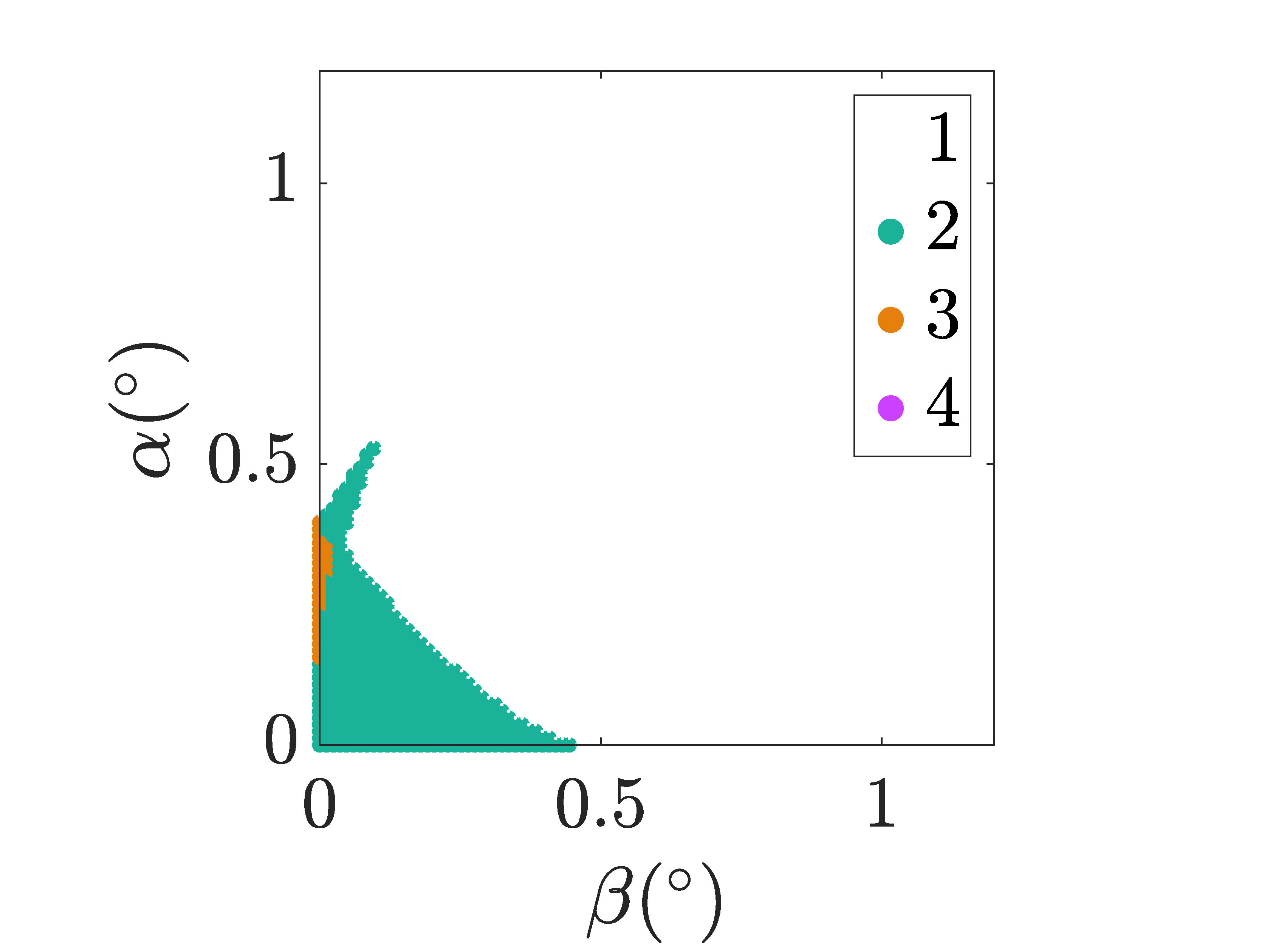}};
    \node[anchor=north west] at (img.north west) {\small (b)};
\end{tikzpicture}
\end{subfigure}
\begin{subfigure}{0.32\textwidth}
\begin{tikzpicture}
    \node[anchor=south west](img) at (0,0)
        {\includegraphics[trim=1.2cm 0cm 1.5cm 0cm, clip=true, width=\textwidth]{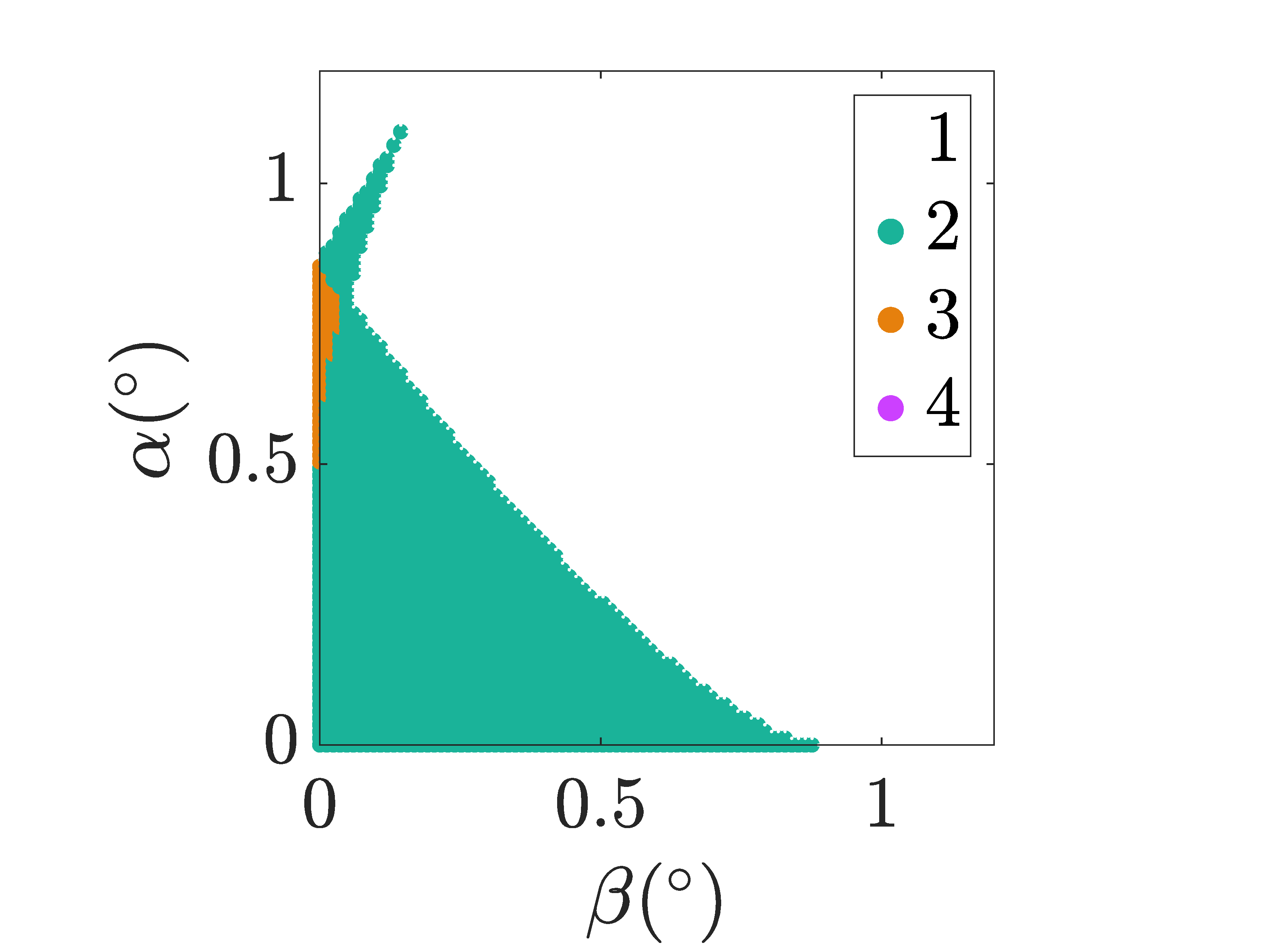}};
    \node[anchor=north west] at (img.north west) {\small (c)};
\end{tikzpicture}
\end{subfigure}
\caption{Number of stable WNL solutions for varying pitch and yaw at $(a)$~$Re = 320$ with zoomed inset, $(b)$~$Re = 340$, and $(c)$~$Re = 360$.}
\label{WNLstablesol}
\end{figure}

Based on the linear stability of the WNL solutions, we can also look at the number of stable solutions  encountered for each combination of pitch and yaw.
Figures~\ref{WNLstablesols_pitch_with_demonstration}-\ref{WNLstablesols_yaw2} compile the bifurcation diagrams obtained when systematically varying $\alpha$, $\beta$ or both.
In each of these figures, panel $(a)$ gives the number of stable solutions, while panels $(b,c)$ show bifurcation diagrams  highlighting the intervals of $Re$ with 1, 2, 3 or 4 stable solutions at two specific attitudes. Note that panel $(a)$ is symmetric with respect to $\alpha=0$ or $\beta=0$.
As seen in figures ~\ref{WNLstablesols_pitch_with_demonstration}-\ref{WNLstablesols_yaw_with_demonstration}, in the perfectly aligned case $\alpha=\beta=0^\circ$ we recover the bifurcation sequence of \cite{Zampogna_Boujo_2023}: 
1 stable solution (white) $(A,B)=(0,0)$ at low $Re$ when the doubly symmetric flow is stable; 
2 stable solutions (green) $(A,0)$ for the wake deflected upward or downward;
4 stable solutions (magenta) $(A,0)$ and $(0,B)$ as the wake can be deflected vertically or horizontally;
and finally 2 stable solutions $(0,B)$ as the vertically deflected wake becomes unstable. 

As $|\alpha|$ increases (figure~\ref{WNLstablesols_pitch_with_demonstration}), solutions remain even in $B$ (the flow remains horizontally symmetric), and one of the two $(A,0)$ branches is preselected.
All the multistable regions described above start at increasingly larger $Re$, following the motion of the corresponding saddle-nodes.
Simultaneously, however, the destabilisation of the secondary $(A,0)$ branch occurs at smaller $Re$, so  the first region with 2 stable solutions eventually disappears, and the region with 4 stable solutions eventually contains only 3 stable solutions: the two symmetric $(0,B)$ horizontally deflected wakes and the primary $(A,0)$ vertically deflected wake.
At large $Re$,  vertically deflected states $(A,0)$ cannot be sustained and only vertically deflected states $(0,B)$ remain stable, similar to the purely aligned case (\cite{Zampogna_Boujo_2023}).
At a fixed Reynolds number $Re \gtrsim 330$ and for increasing pitch angle, the number of stable solutions increases from 2 to 3, before decreasing to 1. This is due to the persistence of the stable solution $(A,0)$ until a Reynolds number larger than that of the saddle-node of the two symmetric $(0,B)$ branches. 

Broadly speaking, the scenario is qualitatively similar when adding yaw instead of pitch (figure~\ref{WNLstablesols_yaw_with_demonstration}). 
Here, solutions remain even in $A$ (the flow remains vertically symmetric), and one of the two $(0,B)$ branches is preselected.
As $|\beta|$ increases, the first region with 2 stable solutions and the regions with 3 and 4 stable solutions eventually disappear, as vertically deflected solutions  $(A,0)$ can only be sustained in an increasingly smaller interval of $Re$ and, again, only vertically deflected states $(0,B)$ remain stable, similar to the purely aligned case (\cite{Zampogna_Boujo_2023}).
At a fixed Reynolds number $Re \gtrsim 330$ and for increasing yaw angle, the number of stable solutions $(0,B)$ decreases from 2 to 1, as the saddle-node of the secondary $B$ branch moves to larger Reynolds numbers. 

Note that if the flow underwent only one pitchfork bifurcation, for example the bifurcation of mode $A$ (resp. mode $B$), figure~\ref{WNLstablesols_pitch_with_demonstration}$(a)$   (resp. \ref{WNLstablesols_yaw_with_demonstration}$(a)$) would look like a cusp in $Re=Re_c$, $\alpha=0$ (resp. $\beta=0$), with 1 stable solution at low $Re$ / large angle, and 2 stable solutions at large $Re$ / small  angle. 
This can also be understood as the projection of the 3D surface of figure~\ref{splots}$(a)$ onto the plane of control parameter ($Re$) and imperfection ($\alpha$ or $\beta$).
As already mentioned in section \ref{BaseFlows}, an imperfect pitchfork bifurcation is sometimes called a ``cusp catastrophe'' after the shape of this projection.
Here, the picture is richer due to the presence of two pitchfork bifurcations. 

Figure~\ref{WNLstablesol} offers an alternative representation in the $(\alpha,\beta)$ plane at fixed $Re$. Again, results are symmetric with respect to $\alpha=0$ and $\beta=0$.
As already seen in figures~\ref{WNLstablesols_pitch_with_demonstration}-\ref{WNLstablesols_yaw2}, the number of stable solutions varies rapidly near the reference Reynolds number, due to the bifurcation of mode $A$.
For $Re \gtrsim 330$, however, the multistable region  keeps the same qualitative shape while growing in size.
At these Reynolds numbers, there are either one or two solutions, with a deflection that is mostly horizontal, except for a small cusp at large $|\alpha|$ and small $|\beta|$, where an additional state exists  with a deflection that is mostly vertical.

\begin{figure}
\vspace{0.5cm}
    \centerline{
        \begin{overpic}[width=1\linewidth]{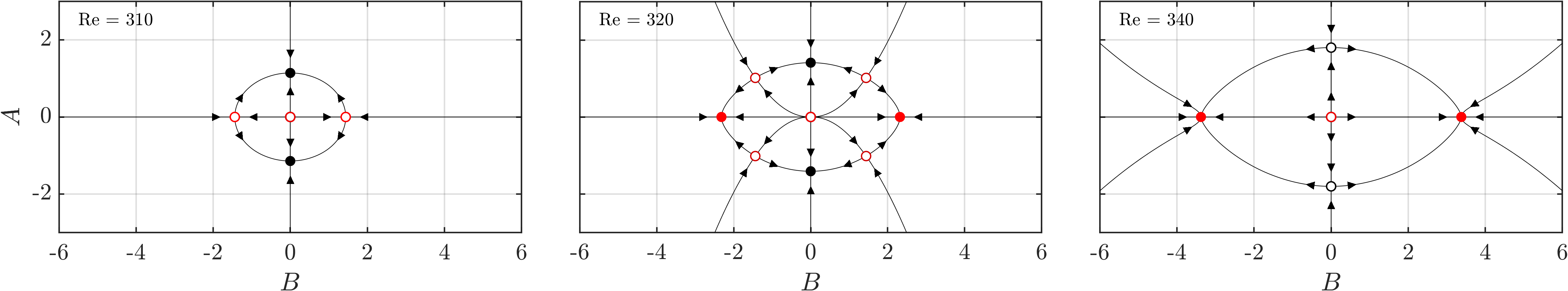}
        \put(0,20){\small (a)}
        \end{overpic}
    }
\vspace{0.4cm}
    \centerline{
        \begin{overpic}[width=1\linewidth]{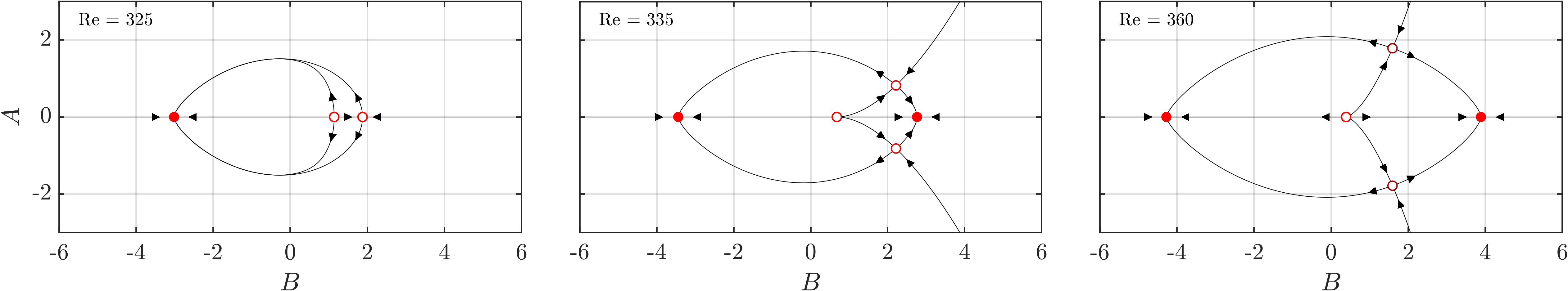}
        \put(0,20){\small (b)}
        \end{overpic}
    }
\vspace{0.4cm}
    \centerline{
        \begin{overpic}[width=1\linewidth]{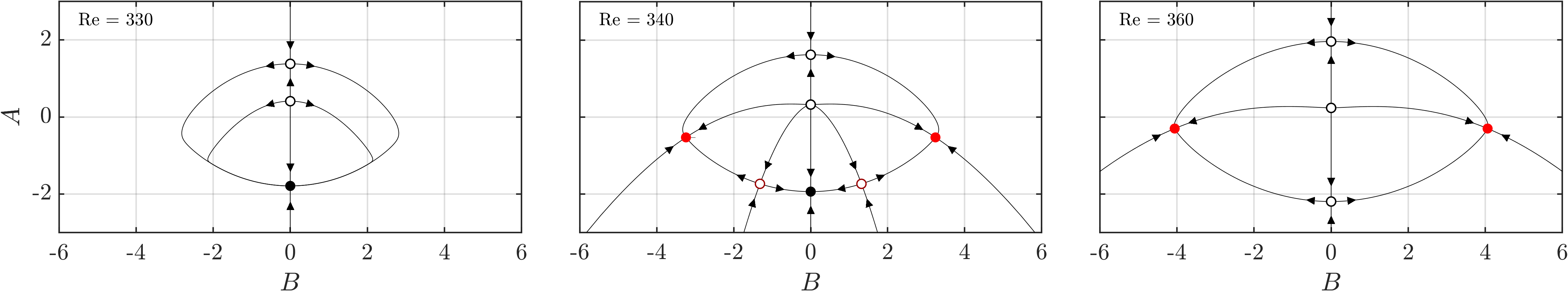}
        \put(0,20){\small (c)}
        \end{overpic}
    }
\vspace{0.4cm}
    \centerline{
        \begin{overpic}[width=1\linewidth]{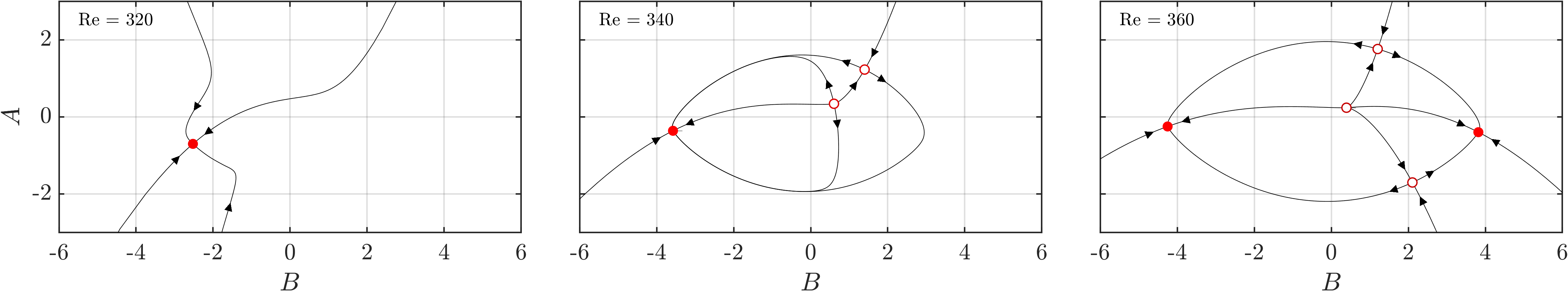}
        \put(0,20){\small (d)}
        \end{overpic}   
    }
\caption{WNL based trajectories in $A-B$ phase space for 
    (a)~$(\alpha,\beta) = (0^\circ,0^\circ)$, 
    (b)~$(\alpha,\beta) = (0^\circ,0.25^\circ)$, 
    (c)~$(\alpha,\beta) = (0.25^\circ,0^\circ)$ and 
    (d)~$(\alpha,\beta) = (0.25^\circ,0.25^\circ)$.
}
\label{combined_traj}
\end{figure}
Figure \ref{combined_traj} shows the phase-space solutions (fixed points) and some phase-space trajectories predicted from the WNL amplitude equations (\ref{eq:amp_eq_A})-(\ref{eq:amp_eq_B}) for different combinations of pitch and yaw, providing some insight into the emergence and disappearance of the different solutions and changes in their stability. 

Interestingly, the phase space is a direct analogue to the position of the wake, since $A$ and $B$ are the amplitudes of modes $A$ and $B$ responsible for vertical and horizontal deflection, respectively.
As already observed, misalignment breaks one or both planar symmetries, preselecting one state at low $Re$, and affecting the whole phase space at larger $Re$ regardless of subsequent bifurcations. 
For example, in figure~\ref{combined_traj}(d), pitch induces a slight vertical deflection on top of the horizontal deflection induced by the imperfect pitchfork bifurcation of mode $B$. 

It is also worth noting that the presence of some fixed points is imprinted on the phase space  before they appear or after they disappear via a saddle-node bifurcation. In figure~\ref{combined_traj}(c), for example, new fixed points appear between $Re=330$ and 340 but are already ``felt'' by the dynamics at $Re=330$, as observed in the shape of the trajectories leaving the unstable points with large $B\neq0$ excursions. These are typical examples of ``saddle-node remnants'' or ``ghosts'' \citep{Strogatz_2015}. 

Finally, we return to the study of \cite{Fan_2022}, who investigated the effect of combined pitch and yaw in the turbulent flow ($Re=2.1 \times 10^5$) past a taller-than-wide Ahmed body in the presence of ground. 
For small angles $|\alpha|,|\beta| \lesssim 2^\circ$, they observed  that the primary static deflection is always vertical across the larger dimension of the body. This is consistent with existing results for perfectly aligned bodies:  turbulent experiments without ground \citep{Legeai20} as well as laminar experiments and numerical calculations~\citep{Grandemange12PRE, Evstafyeva17, Zampogna_Boujo_2023, Chiarini_Boujo_2025}.
That this deflection persists under pitch and yaw shows that it is a robust feature. 
This is also consistent with our present laminar results, in particular with the WNL predictions: as $Re$ increases, the stable state predominantly corresponds to $|B| \gg A$, i.e. the static deflection of the wake is across the larger dimension of the body. 
\section{Conclusion}\label{Conclusion}
In this study, we investigated the stability of a flat-back Ahmed body $(L = 3H, W =1.2H, R = 0.347H)$ wake in presence of a misalignment to the incoming flow, namely by varying pitch and yaw angles. Linear stability analysis for pure pitch $(\alpha\neq0^\circ, \beta = 0^\circ)$ and pure yaw $(\alpha=0^\circ, \beta \neq 0^\circ)$ cases unveiled a complicated and rich bifurcation scenario, with different modes competing and arising. The stationary modes $A$ and $B$ break the vertical and horizontal symmetry respectively. We observe that the presence of misalignment intrinsically pre-selects a preferential position for the wake, with it being deflected either vertically (state $A$ in case of pitch) or horizontally (state $B$ in case of yaw). We also observe interesting regions and competing modes at low values of pure yaw and pure pitch. At misalignment values, we always directly encounter a Hopf bifurcation, with oscillations about the deflected wake. We observe a curious case of crossover of the critical $Re$ curves for the pitchfork and Hopf bifurcations, approximately around $\alpha= 1°$ on the primary branch.
DNS studies conducted around the critical Reynolds number cross-over point shed light on the competition between the pitchfork and Hopf bifurcation. Slow dynamics, reminiscent of the experimental observations in the presence of ground as seen by \cite{Grandemange12PRE}, are observed. 

Next, we developed a weakly nonlinear expansion around the critical $Re$ for modes $A$ and $B$, using the perfectly aligned body as the reference point. We then introduced the imperfection at lower orders and as a result get a coupled amplitude equation for the two modes, with the imperfection parameter only appearing as a constant in the corresponding mode's amplitude equation. The resulting WNL bifurcation diagrams showcase a useful method to understand the bifurcation scenario for both non-zero pitch and yaw, which can be computationally expensive nonlinearly. We then compute the stability of the WNL solution, using which we can quantify the number of stable solutions one might expect for different pairs of pitch and yaw. 

Future natural extensions to the work may include (i)~exploring the effect of noise on transitions between coexisting states,
and (ii)~studying the effect of ground.
Regarding~(i), \cite{NoiseYves} derived a rigorous amplitude equation describing transitions between two symmetry-breaking states induced by a stochastic forcing near a bifurcation, and applied the method  to a sudden-expansion flow near the first pitchfork bifurcation.
Investigating noise-induced transitions between  two or more multiple stable states in bluff-body wakes will serve as a stepping-stone to understand rare transitions in the turbulent regime triggered by fluctuations in the incoming flow. 
Regarding~(ii), \cite{MCT}  reported exhaustive results on the flow past cylinders and spheres moving parallel to or rolling along a wall. 
Studying the effect of a fixed or moving ground on the wake stability of simplified ground vehicles remains to be explored and is of interest to guide the development of methods to subdue or trigger  instabilities.

\begin{bmhead}[Declaration of interests] The authors report no conflict of interest.
\end{bmhead}
\begin{bmhead}[Funding] This work was supported by the Swiss National Science
Foundation (SNSF) under Grant No.~10001059.
\end{bmhead}

\begin{appen}

\section{}\label{appA}

We give here some intermediate steps of the weakly nonlinear derivation of section~\ref{sec:WNL_derivation}.
We introduce the scaled growth rates 
\begin{equation}
    \tilde{\sigma}_A = \frac{\sigma_A}  {\epsilon^2}, \quad 
    \tilde{\sigma}_B = \frac{\sigma_B}{\epsilon^2},
\end{equation}
the shift operator ${\mathcal{S}}$ such that
\begin{equation}
    \begin{split}
\mathcal{S} \hat{\boldsymbol{q}}_1^A = \tilde{\sigma}_A(Re_c) \mathcal{B} \hat{\boldsymbol{q}}_1^A , 
\\
\mathcal{S} \hat{\boldsymbol{q}}_1^B = \tilde{\sigma}_B(Re_c) \mathcal{B} \hat{\boldsymbol{q}}_1^B,  \\
    \end{split}
\end{equation}
and $\mathcal{S} \hat{\boldsymbol{q}}_1 = \boldsymbol{0}$ for all other modes, and the  shifted linearised NS operator $\tilde{\mathcal{L}} = \mathcal{L} + \epsilon^2 \mathcal{S}$ such that
\begin{equation}
    \begin{split}
        \tilde{\mathcal{L}} \hat{\boldsymbol{q}}_1^A &= 
        {\mathcal{L}} \hat{\boldsymbol{q}}_1^A + 
        \sigma_A(Re_c)  \mathcal{B} \hat{\boldsymbol{q}}_1^A, 
        \\
        \tilde{\mathcal{L}} \hat{\boldsymbol{q}}_1^B &= 
        {\mathcal{L}} \hat{\boldsymbol{q}}_1^B + \sigma_B(Re_c)  \mathcal{B} \hat{\boldsymbol{q}}_1^B.  \\
    \end{split}
\end{equation}
By construction, $\tilde{\mathcal{L}}$ is  singular for both $\hat{\boldsymbol{q}}_1^A$ and $\hat{\boldsymbol{q}}_1^B$  at $Re=Re_c$.
Recall that  we also define the departure from criticality as
\begin{equation}
    \frac{1}{Re_c} - \frac{1}{Re} = \delta = \epsilon^2\tilde{\delta},
\end{equation}
the small misalignment as
\begin{equation}
    \alpha = \epsilon^3 \tilde{\alpha}, \quad\quad 
    \beta = \epsilon^3 \tilde{\beta},
\end{equation}
and  the slow time scale $T_1=\epsilon^2 t$. 
We then inject these and the flow field expansion 
\begin{equation}
    \boldsymbol{u} = \boldsymbol{u}_0 + \epsilon \boldsymbol{u}_1 + \epsilon^2 \boldsymbol{u}_2 + \epsilon^3 \boldsymbol{u}_3 \dots
\end{equation}
in the NS equations, collect like-order terms, and obtain the following series of problems.

\subsubsection{Order $\epsilon^0$ and $\epsilon^1$}
At order $\epsilon^0$, we recover the steady nonlinear NS equations \eqref{NS}, with the base flow being the solution at the reference Reynolds number $Re_c$,
\begin{equation}
    \boldsymbol{u}_0\cdot\nabla \boldsymbol{u}_0 + \nabla p_0 - \frac{1}{Re_c}\nabla^2 \boldsymbol{u}_0 = \boldsymbol{0},
\end{equation}
with the homogeneous wall boundary condition $\boldsymbol{u}_0(\boldsymbol{x}_{0,0}) = \boldsymbol{0}$.

At order $\epsilon^1$, we recover the eigenvalue problem 

\begin{equation}
    \partial_t \boldsymbol{u}_1 + \tilde{\mathcal{L}}\boldsymbol{u}_1  = \boldsymbol{0},
\end{equation}
with the homogeneous wall boundary condition $\boldsymbol{u}_1(\boldsymbol{x}_{0,0}) = \boldsymbol{0}$.
We assume that $\boldsymbol{u}_1$ is a superposition of the two bifurcating stationary modes, mode $A$ which preserves the left-right symmetry and mode $B$ which preserves the top-down symmetry,
\begin{equation}
    \boldsymbol{u}_1 = A(T_1) \hat{\boldsymbol{u}}_1^A + B(T_1) \hat{\boldsymbol{u}}_1^B,
\end{equation}
where $A$ and $B$ are slowly-varying amplitudes, yet to be determined.

\subsubsection{Order $\epsilon^2$}

At order $\epsilon^2$, we obtain a linearised NS problem with forcing,
\begin{equation}
    \partial_t \boldsymbol{u}_2 + \tilde{\mathcal{L}}\boldsymbol{u}_2 = \mathcal{F}_2,
\end{equation}
with the homogeneous wall boundary condition $\boldsymbol{u}_2(\boldsymbol{x}_{0,0}) = \boldsymbol{0}$. 
Here, we have
\begin{equation}
\begin{split}
    \mathcal{F}_2 &= -\tilde\delta \nabla^2\boldsymbol{u}_0  - \boldsymbol{u}_1\cdot\nabla \boldsymbol{u}_1 
    \\
    &= -\tilde\delta \nabla^2\boldsymbol{u}_0 - A^2 \hat{\boldsymbol{u}}_1^A \cdot \nabla \hat{\boldsymbol{u}}^A_1 - \mathcal{C}(\hat{\boldsymbol{u}}_1^A,\hat{\boldsymbol{u}}_1^B)AB - B^2 \hat{\boldsymbol{u}}_1^B \cdot \nabla \hat{\boldsymbol{u}}^B_1.
\end{split}
\end{equation}
We should note that the forcing terms are non-resonant with the eigenmodes, as they do not contain terms with $S_yA_z$ or $A_yS_z$ symmetries. Hence we can invert the system and find a solution of the form
\begin{equation}
    \boldsymbol{u}_2 =  -\tilde\delta \nabla^2\boldsymbol{u}_2^\delta - A^2\boldsymbol{u}_2^{A^2} + AB\boldsymbol{u}_2^{AB}- B^2\boldsymbol{u}_2^{B^2},
\end{equation}
where the fields are the solution of
\begin{equation}
    \begin{split}
        \tilde{\mathcal{L}}\boldsymbol{u}_2^\delta &= -\nabla^2 \boldsymbol{u}_0, \\
        \tilde{\mathcal{L}}\boldsymbol{u}_2^{A^2} &= -\hat{\boldsymbol{u}}_1^A\cdot\nabla \hat{\boldsymbol{u}}_1^A, \\ 
        \tilde{\mathcal{L}}\boldsymbol{u}_2^{AB} &= - \mathcal{C}(\hat{\boldsymbol{u}}_1^A,\hat{\boldsymbol{u}}_1^B), \\
        \tilde{\mathcal{L}}\boldsymbol{u}_2^{B^2} &= -\hat{\boldsymbol{u}}_1^B\cdot\nabla \hat{\boldsymbol{u}}_1^B.
    \end{split}
\end{equation}
Each of the above problems has boundary conditions similar to that of the eigenvalue problem. As expected, we  observe that $\boldsymbol{u}_2^\delta,\boldsymbol{u}_2^{A^2}$ and $\boldsymbol{u}_2^{B^2}$ are $S_yS_z$ doubly-symmetric and $\boldsymbol{u}_2^{AB}$ is $A_yA_z$ doubly-antisymmetric, confirming their non-resonant nature. 

\subsubsection{Order $\epsilon^3$}

At order $\epsilon^3$, we obtain again a forced linear NS problem, but now all the forcing terms are resonant with mode $A$ or mode $B$, and the wall boundary condition $\boldsymbol{u}_3(\boldsymbol{x}_{0,0}) = \tilde{\boldsymbol{u}}$ is non-homogeneous and resonant too.
Enforcing the compatibility condition (\ref{compatibility_condition}),

we get the following amplitude equations,
\begin{equation}
    \partial_{T_1}A = \tilde{\lambda}_AA - \tilde{\chi}_AA^3-\tilde{\eta}_{A}AB^2 + h_A \tilde{\alpha},
\end{equation}
\begin{equation}
    \partial_{T_1}B = \tilde{\lambda}_BB - \tilde{\chi}_BB^3-\tilde{\eta}_{B}A^2B + h_B \tilde{\beta},
\end{equation}
with the imperfection terms given by (\ref{hA})-(\ref{hB}) and the other coefficients defined as
\begin{equation}
    \tilde{\lambda}_A = \tilde{\sigma}_A + \tilde{\delta} \langle \boldsymbol{u}^{A\text{\textdagger}}_1, -\mathcal{C}(\hat{\boldsymbol{u}}^{A}_1,\boldsymbol{u}^{\delta}_2) -\nabla^2 \hat{\boldsymbol{u}}^{A}_1 \rangle,
    \label{app:start}
\end{equation}
\begin{equation}
    \tilde{\chi}_A = \langle \boldsymbol{u}^{A\text{\textdagger}}_1, \mathcal{C}(\hat{\boldsymbol{u}}^{A}_1,\boldsymbol{u}^{A^2}_2) \rangle, \qquad\qquad\quad \hspace{15pt}
\end{equation}    
\begin{equation}
    \tilde{\eta}_A = \langle \boldsymbol{u}^{A\text{\textdagger}}_1, \mathcal{C}(\hat{\boldsymbol{u}}^{B}_1,\boldsymbol{u}^{AB}_2) + \mathcal{C}(\hat{\boldsymbol{u}}^{A}_1,\boldsymbol{u}^{B^2}_2) \rangle, \hspace{6pt}
\end{equation}    
\begin{equation}
    \tilde{\lambda}_B = \tilde{\sigma}_B + \tilde{\delta} \langle \boldsymbol{u}^{B\text{\textdagger}}_1, -\mathcal{C}(\hat{\boldsymbol{u}}^{B}_1,\boldsymbol{u}^{\delta}_2) -\nabla^2 \hat{\boldsymbol{u}}^{B}_1 \rangle,
\end{equation}
\begin{equation}
    \tilde{\chi}_B = \langle \boldsymbol{u}^{B\text{\textdagger}}_1, \mathcal{C}(\hat{\boldsymbol{u}}^{B}_1,\boldsymbol{u}^{B^2}_2) \rangle, \qquad\qquad\quad \hspace{15pt}
\end{equation}    
\begin{equation}
    \tilde{\eta}_B = \langle \boldsymbol{u}^{B\text{\textdagger}}_1, \mathcal{C}(\hat{\boldsymbol{u}}^{A}_1,\boldsymbol{u}^{AB}_2) + \mathcal{C}(\hat{\boldsymbol{u}}^{B}_1,\boldsymbol{u}^{A^2}_2) \rangle. \hspace{6pt}
    \label{app:end}
\end{equation}

\end{appen}

\bibliographystyle{jfm}
\bibliography{jfm}

\end{document}